\documentclass[twocolumn, floatfix, pra, longbibliography, superscriptaddress]{revtex4-2}
\usepackage{graphicx}
\usepackage{bm}
\usepackage[usenames,dvipsnames]{xcolor}
\usepackage[colorlinks=true,urlcolor=blue,citecolor=blue,linkcolor=blue]{hyperref}
\usepackage{amssymb,amsmath}
\usepackage[normalem]{ulem}
\usepackage{braket}
\usepackage[T1]{fontenc}
\usepackage[utf8]{inputenc}
\usepackage[english]{babel}
\usepackage{xspace}
\usepackage[final]{microtype}
\usepackage{enumitem}

\newcommand{\deriv}[2]{\frac{\mathrm{d}#1}{\mathrm{d}#2}}
\newcommand{\meanv}[1]{\langle #1 \rangle}

\begin{document}
\title{All-optical switching at the two-photon limit with interference-localized states}

\author{Ville A. J. Pyykk\"onen}
\affiliation{Department of Applied Physics, Aalto University School of Science,
FI-00076 Aalto, Finland}

\author{Grazia Salerno} 
\affiliation{Department of Applied Physics, Aalto University School of Science, FI-00076 Aalto, Finland}

\author{Jaakko K\"ah\"ar\"a}
\affiliation{Department of Applied Physics, Aalto University School of Science, FI-00076 Aalto, Finland}
\affiliation{Institute for Atmospheric and Earth System Research/Physics, Faculty of Science, University of Helsinki, FI-00014 Helsinki, Finland}

\author{P\"aivi T\"orm\"a}
\email{paivi.torma@aalto.fi}
\affiliation{Department of Applied Physics, Aalto University School of Science, FI-00076 Aalto, Finland}

\begin{abstract}
We propose a single-photon-by-single-photon all-optical switch concept based on interference-localized states on lattices and their delocalization by interaction. In its `open' operation, the switch stops single photons while allows photon pairs to pass the switch. Alternatively, in the `closed' operation, the switch geometrically separates single-photon and two-photon states. We demonstrate the concept using a three-site Stub unit cell and the diamond chain. The systems are modeled by Bose-Hubbard Hamiltonians, and the dynamics is solved by exact diagonalization with Lindblad master equation. We discuss realization of the switch using photonic lattices with nonlinearities, superconductive qubit arrays, and ultracold atoms. We show that the switch allows arbitrary `ON'/`OFF' contrast while achieving picosecond switching time at the single-photon switching energy with contemporary photonic materials. 
\end{abstract}

\maketitle
\section{Introduction}
\label{sec:introduction}
All-optical devices have the potential to change data transmission and processing, having faster speeds and lower power consumption compared to their electronic counterparts~\cite{almeida2004all, reed2010silicon, chai2017ultrafast}. 
The switch is an essential component that allows the control and the manipulation of signals in circuits, representing a logical AND gate, namely allowing a {\it signal} to pass only if a {\it control signal} is present. The implementation of all-optical switches can lead to the development of integrated and compact circuits~\cite{sasikala2018all}. 

Performing switching of light with only a single control photon allows operation at minimal energy, but it requires strong photon-photon interaction, a long-standing goal of quantum optics ~\cite{chang2014quantum}. Non-linear elements in an optical cavity can confine light and prolong the interaction time, allowing for larger effective photon-photon coupling~\cite{lukin2001controlling}. Examples of such cavity systems realising switching of many-photon light beams using only few control photons include organic molecules \cite{pscherer2021single-molecule,zasedatelev2021}, rubidium-87 atoms \cite{shomroni2014all,tiecke2014nanophotonic,hacker2016photon}, quantum dots \cite{volz2012ultrafast,giesz2016coherent,dietrich2016gaas,sun2018}, ultracold Rydberg or Cesium atoms \cite{peyronel2012quantum, chen2013all, baur2014single, gorniaczyk2014single,chai2017ultrafast} among others. Switching a single-photon signal with a control light made of many photons is a complementary challenge, which was realized by exciton depletion at semiconducting quantum dots \cite{munoz2020all}, for single-photon transport of transmission-line-resonator arrays \cite{liao2009}, and plasmon in nanowires~\cite{chang2007single}.
Switching a single-photon signal with a single control photon represents a fundamental quantum limit, which has been proposed using a cavity QED system \cite{bermel2006}.

We approach single-photon-by-single-photon switching using interference-induced single-particle localized states, which have received interest in recent years in the context of flat bands~\cite{leykam2018}.
In these systems, the single-particle kinetic energy is suppressed while other energy scales, such as interaction, are relevant even if they are weak compared to single-particle ones. Importantly, the interaction can allow propagation of many-particle states while single particles remain localized~\cite{leykam2018,tovmasyan2018preformed, torma2018quantum, pyykkonen2023suppression}, which can lead to  superconductivity in flat band systems \cite{kopnin2011high, peotta2015superfluidity, julku2016geometric, liang2017band, torma2022}.

The interference-localized states are also at play in the context of Aharanov-Bohm cages, where the (artificial) magnetic flux localizes the single particles and leads to the emergence of a flat dispersion. The dice lattice \cite{vidal1998aharonov} and the diamond chain \cite{vidal2000interaction} are examples of Aharonov-Bohm cages that have been studied and realized in superconducting nanowires \cite{abilio1999magnetic}, circuit QED lattices \cite{alaeian2019creating, hung2021quantum, martinez2023interaction, chase2023compact}, photonic lattices~\cite{longhi2014aharonov,mukherjee2015observation,mukherjee2018experimental,diliberto2019nonlinear,caceres2022controlled} and also in ultracold atoms~\cite{creffield2010coherent, li2022aharonov}.

\begin{figure}
    \centering
    \includegraphics[width=0.47\textwidth]{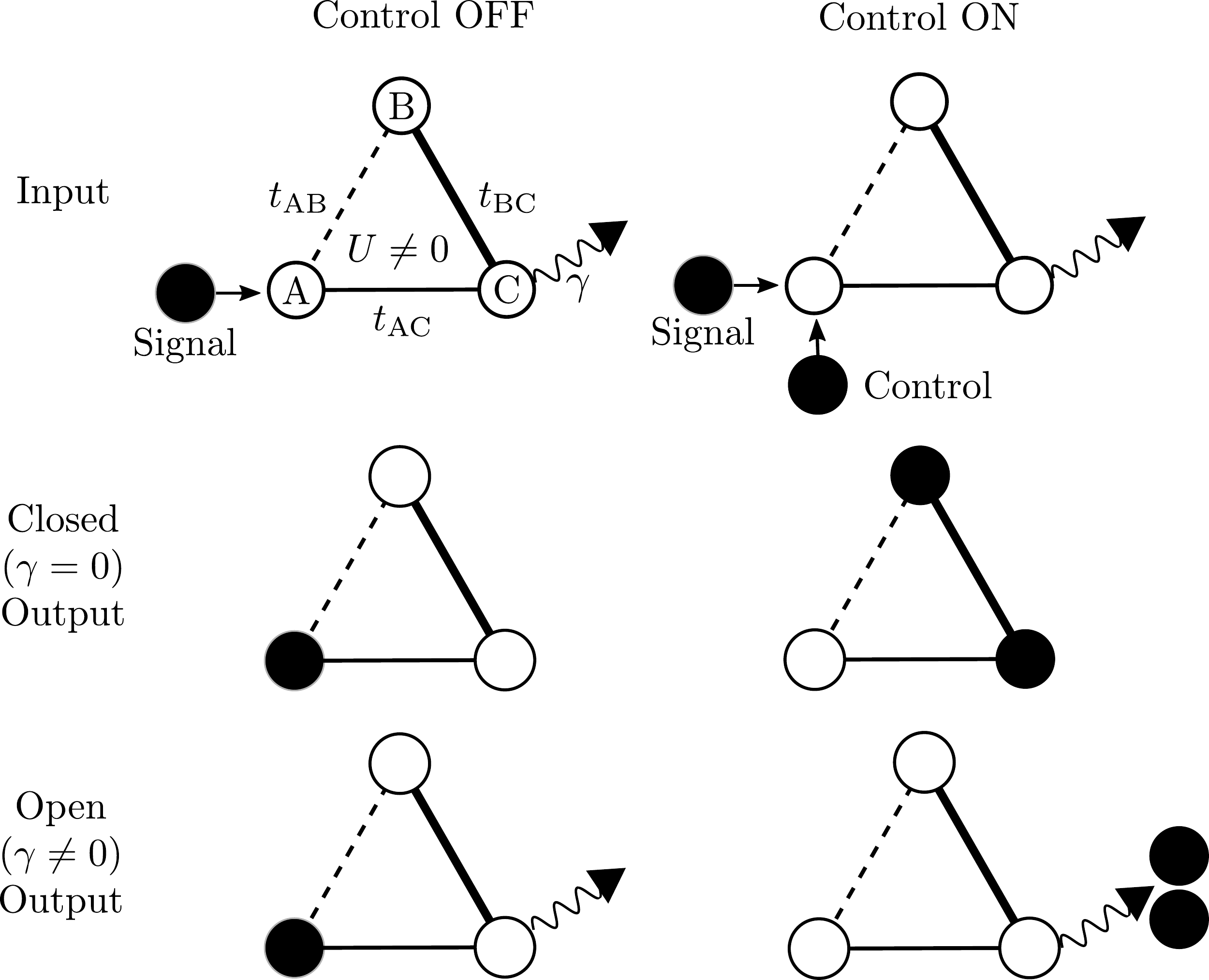}
    \caption{The switching concept illustrated with a system consisting of three sites, labeled A, B, and C. The wiggly arrow from the C site represents the presence of a sink with a decay rate $\gamma$. Empty circles denote empty sites while filled circles correspond to photon presence. The three-site system has an interference-localized state around the A site. If the {\it  control photon} is `OFF', the {\it signal photon} is trapped in the localized state, as shown in the left column representing the switch being `OFF'. If the {\it  control photon} is `ON', the two photons can delocalize due to interactions, as shown in the right column representing the switch being `ON'. The first row represents the input of the switch while the second and third rows depict the output in the closed ($\gamma=0$) and open ($\gamma\neq0$) operation, respectively, see main text for more information.}
    \label{fig:Fig1}
\end{figure}

In this article, we propose a switching concept based on single-particle localization and interaction-induced delocalization of pairs of particles in photonic lattice models. Figure~\ref{fig:Fig1} illustrates the concept using a three-site model described with a Bose-Hubbard model. The {\it signal photon} enters the system and remains localized in a state around the input site, representing the `OFF' state of the switch. 
If also {\it a control photon} is present, the photon pair can delocalize due to interaction, representing the `ON' state of the switch. The switch can be operated as an {\it closed} or a {\it open} system. In the closed operation of the switch, the dynamics of the system is conservative, i.e. particle number is conserved, and solved using exact diagonalization techniques. The photons are geometrically separated from the initial state and can be collected once the maximal separation is reached. In the open operation of the switch, a sink is introduced to continuously deplete the photons from the system, and the corresponding dynamics is modeled by a Lindblad master equation.
We consider the unit cell of a Stub lattice and a diamond chain as notable models offering localized single-particle states with alternative advantages.

The proposed switch operates inherently at the single-photon-by-single-photon limit meaning that the input and control signals consist of no more than a single photon, respectively. In other words, the operation is at the purely quantum mechanical limit at the minimal switching energy in terms of the control signal, and does not work in the classical (mean-field) level. We show that the switching can be realized in principle at arbitrarily small interaction strengths between photons at the cost of increasing the switching time. Given the state-of-the-art Kerr nonlinearities in photonic materials, we show that the switching time can be as fast as picoseconds. 

The article is structured as follows. Firstly, in Sec.~\ref{sec:three_site}, we discuss general conditions for localization in a three-site model, and demonstrate the switching concept by using the notable special case of a single Stub lattice unit cell. We compare the single and the two-particle dynamics in both the closed and the open operation of the switch. Section~\ref{sec:diamond} illustrates the switch in the diamond chain, which is known for its Aharonov-Bohm cages, and Sec.~\ref{sec:experiments} discusses potential experimental platforms to realize the switching concept. In Sec.~\ref{sec:comparison}, we compare the switch to non-linear Mach-Zehnder interferometer and other switching paradigms. We discuss the results and conclude in Sec.~\ref{sec:conclusions}.

\section{Switching with a three-site model}
\label{sec:three_site}
A minimal switching scheme can be realized using the localized eigenstates of a three-site model.
The general non-interacting Hamiltonian corresponding to the three-site model, shown in Fig.~\ref{fig:Fig1} with sites labeled A, B and C, is $\hat{H}_{0,\mathrm{three~site}} = \sum_{ij} H_{0,\mathrm{three~site},ij}\hat{b}_i^\dagger\hat{b}_j$, where $\hat{b}_{i},\hat{b}_i^\dagger$ are annihilation and creation operators of particles at sites $i\in\{\mathrm{A,B,C}\}$ and the matrix is
\begin{equation}
H_{0,\mathrm{three~site}} =
    \begin{pmatrix}
        \epsilon_\mathrm{A} & -t_{\mathrm{AB}} & -t_{\mathrm{AC}} \\
        -t_{\mathrm{BA}} & \epsilon_\mathrm{B} & - t_{\mathrm{BC}} \\
        -t_{\mathrm{CA}} & -t_{\mathrm{CB}} & \epsilon_\mathrm{C}
    \end{pmatrix}~.
    \label{eq:three_loc}
\end{equation}
Without loss of generality, we set $\epsilon_B=0.$
Firstly, we see that single-site localization, for instance at the A site, yields disconnected sites, i.e. $t_{\mathrm{BA}}=t_{\mathrm{CA}}=0$.
Secondly, we find that the two-site localized state has the form $\ket{\mathrm{loc}} \doteq (\alpha_{\mathrm{loc}},\beta_{\mathrm{loc}},0)$, if 
\begin{equation}
    \epsilon_\mathrm{A} = t_{AB}\left(\frac{t_{\mathrm{BC}}}{t_{\mathrm{AC}}} - \frac{t_{\mathrm{AC}}}{t_{\mathrm{BC}}}\right)~,
    \label{eq:three_loc_cond}
\end{equation}
and the amplitudes fulfill the condition
\begin{equation}
    r_{\mathrm{AB}}\equiv \frac{\alpha_{\mathrm{loc}}}{\beta_{\mathrm{loc}}}
    = -\frac{t_{\mathrm{BC}}}{t_{\mathrm{AC}}}~.
\end{equation}
Such localized state satisfies $H_{0,\mathrm{three~site}}\ket{\mathrm{loc}} = E_{\mathrm{loc}} \ket{\mathrm{loc}}$ with energy
\begin{equation}
    E_{\mathrm{loc}} = \frac{t_{\mathrm{AB}}t_{\mathrm{BC}}}{t_{\mathrm{AC}}}~.
    \label{eq:loc_energy}
\end{equation}
Similar processes to construct localized states are used in flat-band lattice construction~\cite{calugaru2022general}. Destructive interference on a three-state system is also behind the formation of dark states in the electromagnetically induced transparency phenomenon~\cite{fleischhauer2005electromagnetically}, that is relevant for all-optical switching by slowing light down or enhancing optical Kerr nonlinearities~\cite{schmidt1996giant, lukin2001controlling}.

\begin{figure*}[ht]
    \centering
    \includegraphics[width=0.95\textwidth]{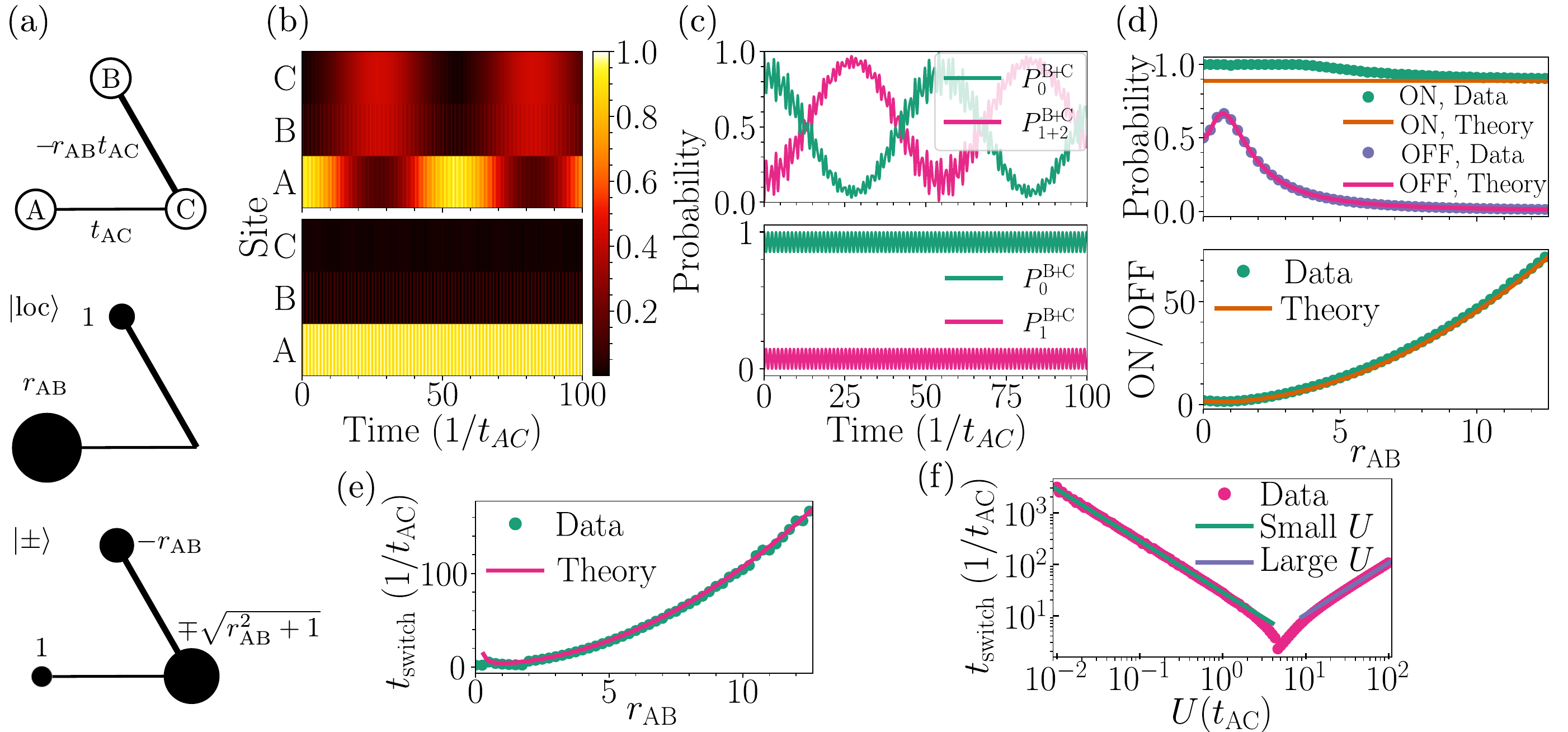}
    \caption{Switching in the closed operation with the Stub unit cell. (a) The Stub unit cell, where A is the input site, and B and C are the output sites. The single-particle eigenstates $\ket{\text{loc}}$ and $\ket{\pm}$ are also represented, with the sizes of the filled circles corresponding to the relative sizes of the wavefunciton amplitudes. (b) The site-resolved normalized particle number as function of time with two and one initial photon at the input site respectively in the upper and lower panel. Single photon remains stuck in the localized state but the interacting photon pair undergoes a Rabi-type oscillation between the input and the output sites. Parameters are $r_{\mathrm{AB}} = -5$ and $U= t_\mathrm{AC}$. (c) The probability to observe $n=\{0,1,2\}$ photons at the output sites with respect to time.  
    Parameters as in (b). (d) The probability of finding one or two photons at the output sites at time $t_{\mathrm{switch}}$ in the `ON' and `OFF' state with respect to $r_{\mathrm{AB}}$ for $U= t_\mathrm{AC}$ (upper panel), and the switching contrast, defined as the `ON' versus `OFF' signal (lower panel). (e) The switching time $t_{\mathrm{switch}}$ with respect to the ratio $r_{\mathrm{AB}}$ at $U= t_\mathrm{AC}$ . (f) The interaction strength $U$ dependence of $t_{\mathrm{switch}}$ at $r_{\mathrm{AB}}=-5$.
    }
    \label{fig:Fig2}
\end{figure*}

The three-site model contains notable solutions. The unit cell of the sawtooth lattice is obtained by setting $\epsilon_A = t_{AC} = t_{AB}/\sqrt{2} = t_{BC}/\sqrt{2}$. The sawtooth lattice case is discussed in Appendix \ref{sec:sawtooth}. Furthermore, the unit cell of the Stub lattice is obtained by setting $\epsilon_\mathrm{A}=0$ and $t_{\mathrm{AB}}=0$, i.e., A and B are disconnected. This case and its eigenstates are illustrated in Fig. \ref{fig:Fig2} (a) and further discussed in the following due to its simplicity, positive hoppings and zero on-site energies. 

\subsection{The closed-operation dynamics}
\label{sec:three_sites_closed}
The system is described by a Bose-Hubbard model
\begin{equation}
    \hat{H} = \sum_{ij} H_{0,\mathrm{three~site}}\hat{b}_{i}^\dagger \hat{b}^{}_j + \sum_i U \hat{n}_i(\hat{n}_{i}-1)/2~,
\label{eq:Hamiltonian_tbbh}
\end{equation}
where $U$ is the on-site interaction strength.

The closed-operation dynamics of the switch is demonstrated with the Stub unit cell in Fig. \ref{fig:Fig2}. The input is located at site A while the output is comprised of the B and C sites. The single-particle localized state $\ket{\mathrm{loc}}$ has energy $E_\mathrm{loc}=0$ and the two delocalized states $\ket{\pm}$ have energies $E_\pm=\pm t_{\mathrm{AC}}\sqrt{r_{\mathrm{AB}}^2+1} = \pm\Delta$, see Fig.~\ref{fig:Fig2}(a). 

We start by analysing the `ON' state of the switch, that is when the initial state has two photons on the input A site. 
The initial state, expanded in the single-particle states, is
$\ket{\mathrm{A,A}}
    = \frac{1}{r_{\mathrm{AB}}^2+1}
    \big(r_{\mathrm{AB}}^2\ket{\mathrm{loc,loc}}
    + r_{\mathrm{AB}}\ket{\mathrm{loc},+} + r_{\mathrm{AB}}\ket{\mathrm{loc},-} + \frac{1}{2}\ket{+,+} + \frac{1}{2}\ket{-,-} + \frac{1}{\sqrt{2}}\ket{+,-}\big)~.$
For interactions $U$ that are small compared to the energy gap $\Delta$, we obtain that $\ket{\mathrm{loc, loc}}$ and $\ket{+,-}$ describe a two-level system, with other eigenstates being stationary. In the limit of large $r_{\mathrm{AB}}$, the state $\ket{\mathrm{loc,loc}}$ is a good approximation of the input state, while $\ket{+,-}$ is the output state. 
From this two-level description, the state dynamics consists of Rabi oscillations with frequency $\Omega = \frac{3}{2} \frac{U r_{\mathrm{AB}}^2}{(1+r_{\mathrm{AB}}^2)^2},$ see Appendix \ref{sec:two_state_model}. 

The Rabi oscillations of the photon pair is visible the upper panel of Figs.~\ref{fig:Fig2}(b) and (c). The half-period of such oscillation defines the switching time
\begin{equation}
t_{\mathrm{switch}} = \frac{\pi (1+r_{\mathrm{AB}}^2)^2}{3 U r_{\mathrm{AB}}^2}.
\label{eq:switch_time_Stub_low_U}
\end{equation} 
We notice that the switching time is only dependent on the interaction strength $U$ and the ratio $r_{\mathrm{AB}}$.

The probability to find one or two photons at the output sites in the output state $\ket{+,-}$, i.e. the `ON' state of the switch, is given by $P_\mathrm{ON}=8/9$, see Appendix~\ref{sec:two_state_model} for details.
This theoretical prediction is shown in Fig.~\ref{fig:Fig2}(d) as orange solid line, together with numerical data as green dots.

We now analyse the non-interacting one-photon case, i.e. the `OFF' state of the switch. 
The dynamics of a single photon starting from the input A site, shows clear localization, see  lower panel of Figures~\ref{fig:Fig2}(b) and (c). There is however a finite probability over time to find the photon at the output B and C sites, given by 
\begin{equation}
    \meanv{P_{\mathrm{OFF}}} = (1/2+2r_{\mathrm{AB}}^2)/(r_{\mathrm{AB}}^2+1)^2.
    \label{eq:probability_stub_off}
\end{equation}

This expression gives the switch's `OFF' state leakage probability, which is shown as a pink solid line in the upper panel of Fig.~\ref{fig:Fig2}(d), together with numerical data as violet dots. Importantly, from Eq.~\eqref{eq:probability_stub_off}, we see that the ratio $r_{\text{AB}}$ can be used to minimize the false `OFF' signal. In this way, the `ON' versus `OFF' switching contrast can be increased, see lower panel in Fig.~\ref{fig:Fig2}(d). 

In fact, by increasing the ratio $r_{\mathrm{AB}}$, the localized eigenstate is more weighted on the A than on the B site. However, in this case also the switching time $t_{\mathrm{switch}}$ increases, see Eq.~\eqref{eq:switch_time_Stub_low_U} and Fig.~\ref{fig:Fig2}(e). 

The behavior of the switching time as a function of $U$ is further shown Fig.~\ref{fig:Fig2}(f), where we see that $t_{\mathrm{switch}}$ is optimal for intermediate values of the interaction $U\approx t_\mathrm{AC}$. A more thorough dependence of $t_{\mathrm{switch}}(U)$ at large interactions is given in Appendix~\ref{sec:two_state_model}.

\begin{figure*}[ht!]
    \centering
    \includegraphics[width=0.90\textwidth]{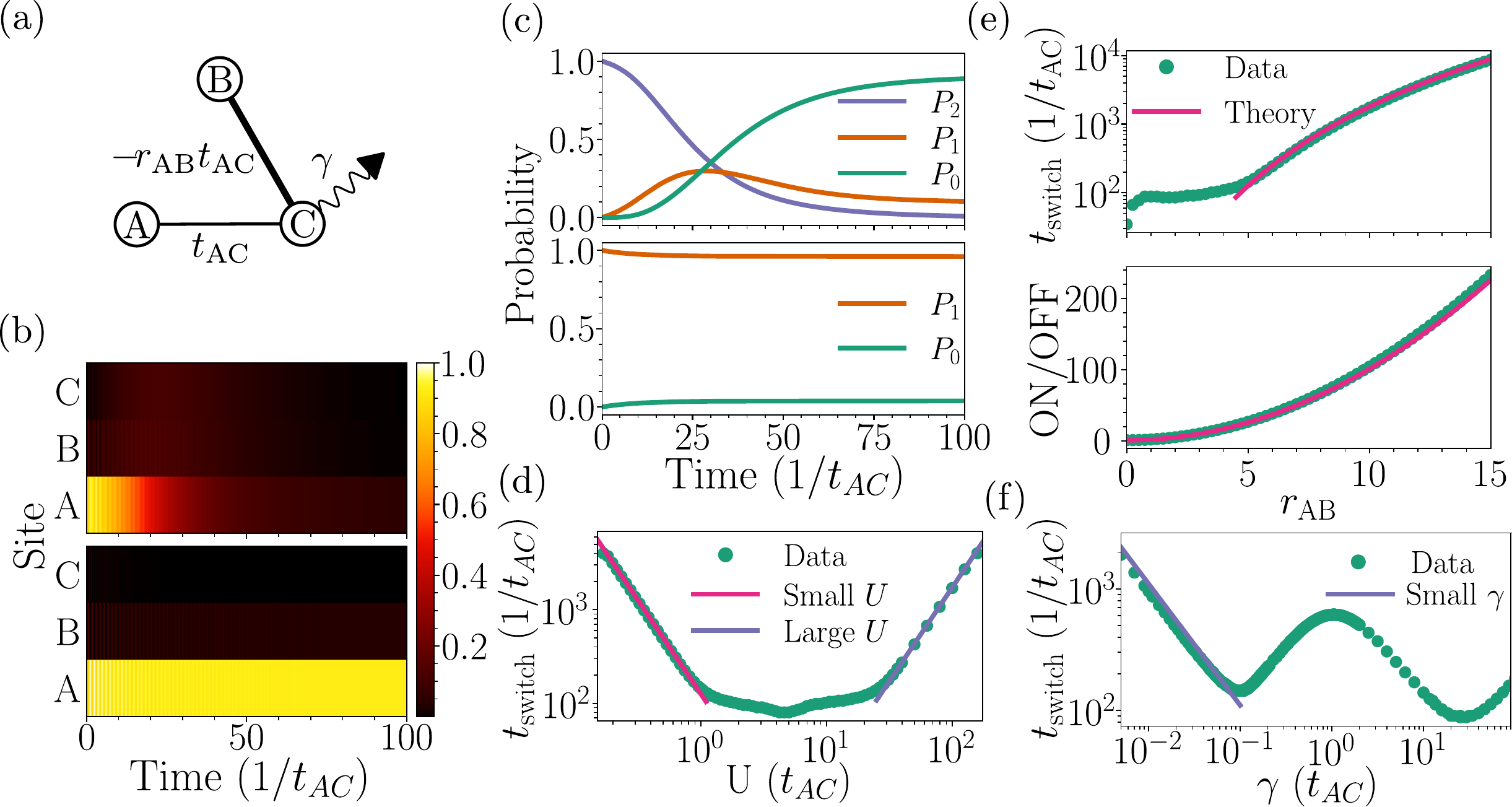}
    \caption{Switching in the open operation with the Stub unit cell. (a) The switch input is at site A and the output is at site C with losses $\gamma$. (b) The site-resolved normalized particle number with two (upper panel) and one (lower panel) initial photons at the input site with respect to time. Parameters are: $\gamma=0.1t_{\mathrm{AC}}$, $r_{AB} = -5$ and $U=t_{\mathrm{AC}}$. (c) The probability to observe $n=\{0,1,2\}$ photons in the system for one (upper panel) and two (lower panel) photons with respect to time. Parameters are the same as in (b). (d) The switching time with respect to the interaction strength $U$. We define $t_\text{switch}$ as the time after which the probability to observe two particles in the system is below a certain threshold $\epsilon=10^{-3}$. Parameters are as in (b). (e) Switching time and the `ON' versus `OFF' contrast as a function of $r_{\mathrm{AB}}$. (f) The switching time dependency on the decay rate $\gamma$.}
    \label{fig:Fig3}
\end{figure*}

\subsection{The open-operation dynamics}
The open-operation dynamics is modelled with a Lindblad master equation
\begin{equation}
    \dot{\hat{\rho}} = -\frac{i}{\hbar}[\hat{H},\hat{\rho}]
    + \sum_n \gamma_n \left(
    \hat{L}_n\hat{\rho}\hat{L}_n^\dagger - 
    \frac{1}{2}\left\{\hat{L}_n^\dagger \hat{L}_n,\hat{\rho}\right\}\right)~,
\end{equation}
where $\hat{L}_n$ are the ladder operators and $\gamma_n$ are the rates corresponding to the connection to an external reservoir. For a sink on the C site, $\hat{L} = \hat{b}_\mathrm{C}$, and $\gamma$ is the decay rate, and A is the input site, see Fig.~\ref{fig:Fig3}(a).

The open-operation dynamics of the switch with the Stub unit cell is illustrated in Figs.~\ref{fig:Fig3}(b) and (c). Similarly to the closed-operation dynamics, in the `OFF' state the single photon remains mostly confined around the A site and only slightly at B, see lower panel in Fig.~\ref{fig:Fig3}(b). Therefore, there is a very high probability $P_1$ of the single photon being found in the system, as it does not reach the sink, see lower panel in Fig.~\ref{fig:Fig3}(c). 

In the `ON' state, the pair delocalizes due to the interaction, evolving according to the two-state dynamics, as in the close-operation case, where Rabi oscillations are dampened by the decay rate $\gamma$. However, when the pair reaches the sink, the system gets depleted, as shown in the upper panel of Fig.~\ref{fig:Fig3}(b). Consequently, the probability $P_2$ of finding two photons in the switch goes to zero over time, while there is obviously an increasing probability $P_0$ of finding zero photons in the system, i.e. the switch is empty, see upper panel of Fig.~\ref{fig:Fig3}(c). We also notice that probability $P_1$ of finding a single photon in the system plateaus due to a finite possibility of being stuck in the localized state.

In contrast to the closed operation, the success of the switching, i.e., a single photon exiting the system in the presence of the control photon, can be made as high as wanted. The `OFF' signal can be limited by utilizing the ratio $r_{\mathrm{AB}}$ to focus the initial state to the localized state. In fact, we observe that the `ON' versus `OFF' contrast increases with increasing $r_{\mathrm{AB}}$, as illustrated in the lower panel of Fig.~\ref{fig:Fig3}(e). 

In Figs.~\ref{fig:Fig3}(d), (e) and (f), we show the dependence of the switching time on the interaction strength $U$, the ratio $r_{\mathrm{AB}}$, and the decay rate $\gamma$ respectively. 
The switching time $t_{\mathrm{switch}}$ is defined as the time it takes for a photon to exit the system with probability above a certain threshold $1-\epsilon$, with $\epsilon= 10^{-3}$. In this open operation, the switching time depends quite a lot on the decay rate $\gamma$, opposed to the closed operation, where $t_\text{switch}$ is determined by the Rabi oscillations.

It can be shown that $t_{\mathrm{switch}}$ has two opposing behaviors depending on the decay rate $\gamma$ being small or large in comparison to the delocalization time-scale, see Appendix~\ref{sec:two_state_losses}.

In the small $\gamma$ limit, it can be shown that the switching time is independent on the interactions
\begin{equation}
    t_{\mathrm{switch}} 
= -9\ln(\epsilon)/4\gamma\,,
\end{equation} 
see Appendix~\ref{sec:two_state_losses} for details. 

When $\gamma$ is large compared to the two-particle delocalization time scale, one can approximate the sink site to be empty at all times. In Appendix~\ref{sec:two_state_losses}, we show that this approximation leads to the following expression for the switching time, valid for small interactions
\begin{equation}
    t_{\mathrm{switch}} = -\frac{\ln(\epsilon) \gamma (1+r_{\mathrm{AB}}^2)^4}{8r_{\mathrm{AB}}^4 U^2}.
    \label{eq:switching_time_large_gamma_smallU}
\end{equation} 
This result is shown as the solid pink line in Fig.~\ref{fig:Fig3} (d) and in Fig.~\ref{fig:Fig3} (e). We see that increasing $\gamma$ at this limit actually increases the switching time. 
The reason is that the sink operates so fast that all the photons ending up at the sink site get directly out from the system, while also the coherence between the two states decays, slowing the delocalization in direct proportion to $\gamma$.

When analysing the switching time behavior as a function of the interaction, we find that $t_\text{switch}$ is inversely proportional to $U^2$ at small interactions, see Eq.~\eqref{eq:switching_time_large_gamma_smallU}, but is directly proportional to $U^2$ at large interactions, see Fig.~\ref{fig:Fig3} (d) and Appendix \ref{sec:two_state_losses}.
Thus, the optimum is in the intermediate interaction range and the decay rate $\gamma$ is the limiting factor.

\begin{figure}
    \centering
    \includegraphics[width=0.48\textwidth]{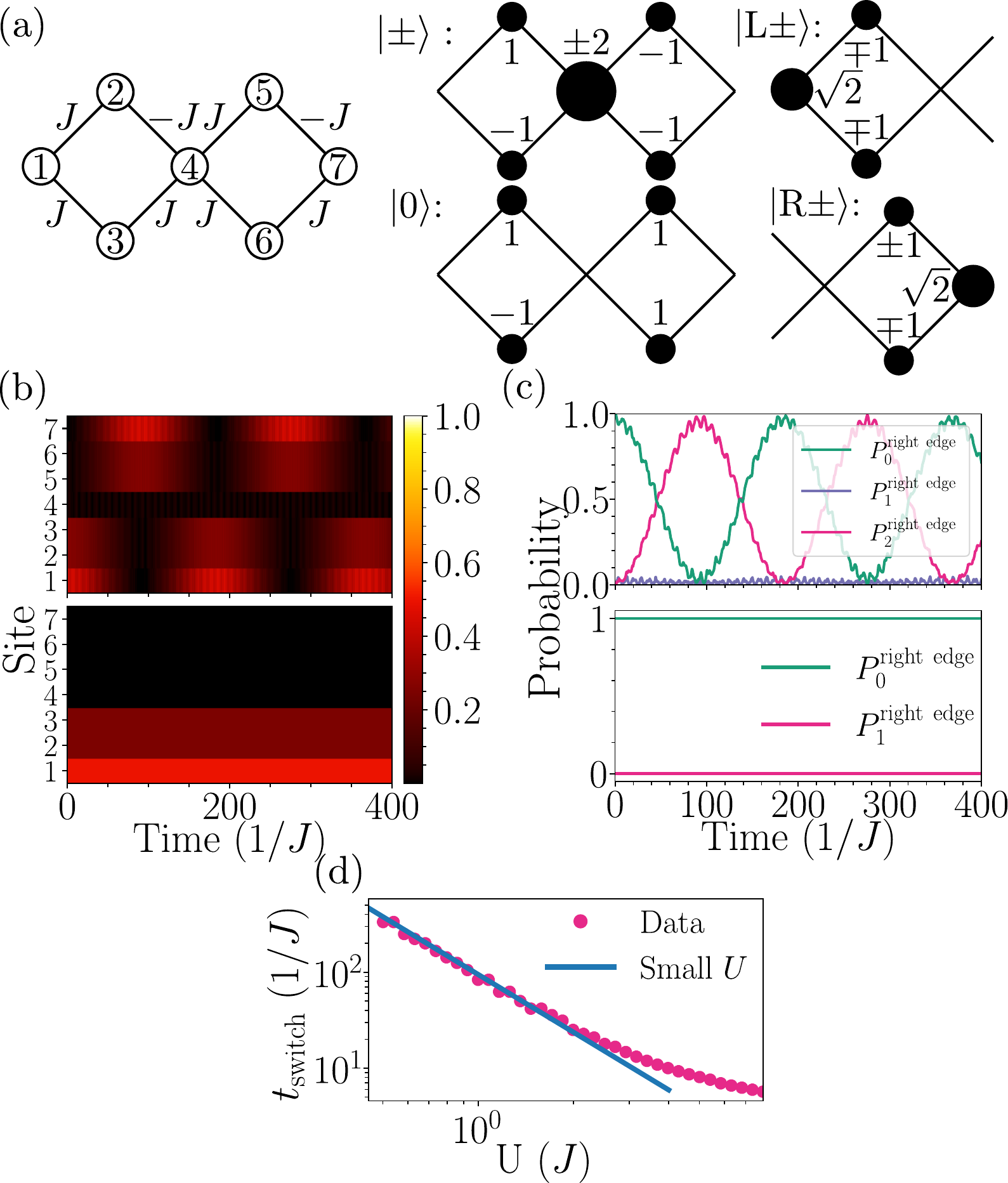}
    \caption{Closed operation of the switch with a diamond chain. (a) The diamond chain, consisting of two rhombi with a $\pi$ flux. We show the bulk eigenstates $\ket{\pm}$ and $\ket{0}$ with energies $E_{\pm}=\pm 2J$ and $E_0=0,$ respectively, while the edge states $\ket{\mathrm{L/R} \pm}$ have energies $E_{\mathrm{L/R}\pm} = \pm \sqrt{2} J$. 
    (b) The site-resolved normalized particle number with two (upper panel) or one (lower panel) initial photons at the left edge state $\ket{\mathrm{L}\pm}.$  Here $U = J$. (c) The probability of finding $n=\{0,1,2\}$ photons at the right edge, consisting of sites 5, 6, and 7, over time. Parameters are the same as in panel (b). (d) The interaction strength $U$ dependence of the switching time $t_{\mathrm{switch}}.$ One sees monotonous decrease with increasing $U$.}
    \label{fig:Fig4}
\end{figure}

\section{Switching with the Diamond chain}
\label{sec:diamond}
We now briefly consider the switch in the diamond chain, as it is a notable model that has been extensively studied for its interference-localized states known as Aharonov-Bohm cages \cite{longhi2014aharonov,mukherjee2015observation,mukherjee2018experimental,diliberto2019nonlinear,gligoric2019nonlinear, kremer2020square, pelegri2020interaction,danieli2021nonlinear, danieli2021quantum,  li2022aharonov, caceres2022controlled, martinez2023interaction, kolovsky2023conductance}. 
The diamond chain is illustrated in Fig.~\ref{fig:Fig4}(a) together with its eigenstates. When a $\pi$ flux is inserted through the plaquette, the system supports only purely localized eigenstates. The bulk eigenstates are $\ket{\pm}$ and $\ket{0}$, with energies $E=\pm 2J$ and $E=0$ respectively, where $J$ is the hopping amplitude between the sites in the chain. In addition, due to finite length of the chain, there are edge states $\ket{L\pm}$ and $\ket{R\pm}$ at energies $E=\pm\sqrt{2}J$.

The closed operation of the switching for the diamond chain is illustrated in Figs.~\ref{fig:Fig4}(b) and (c).
For a successful switching, the input is prepared in the left edge state $\ket{L\pm}$. The single photon remains perfectly localized in the left edge state while the photon pair oscillates between the opposite edge states. Two rhombi are needed to obtain perfect geometric separation of the edge states.
Note that, in the diamond lattice, the switching contrast is perfect: the `OFF' state does not have any contribution at the output. However, in comparison to the three-site system discussed above, the switching time is longer as the chain is longer. Fig.~\ref{fig:Fig4}(d) depicts the switching time dependence on the interaction strength. The switching time is decreasing with interaction, being proportional to $U^{-2}$.

\begin{figure*}
    \centering
    \includegraphics[width=0.85\textwidth]{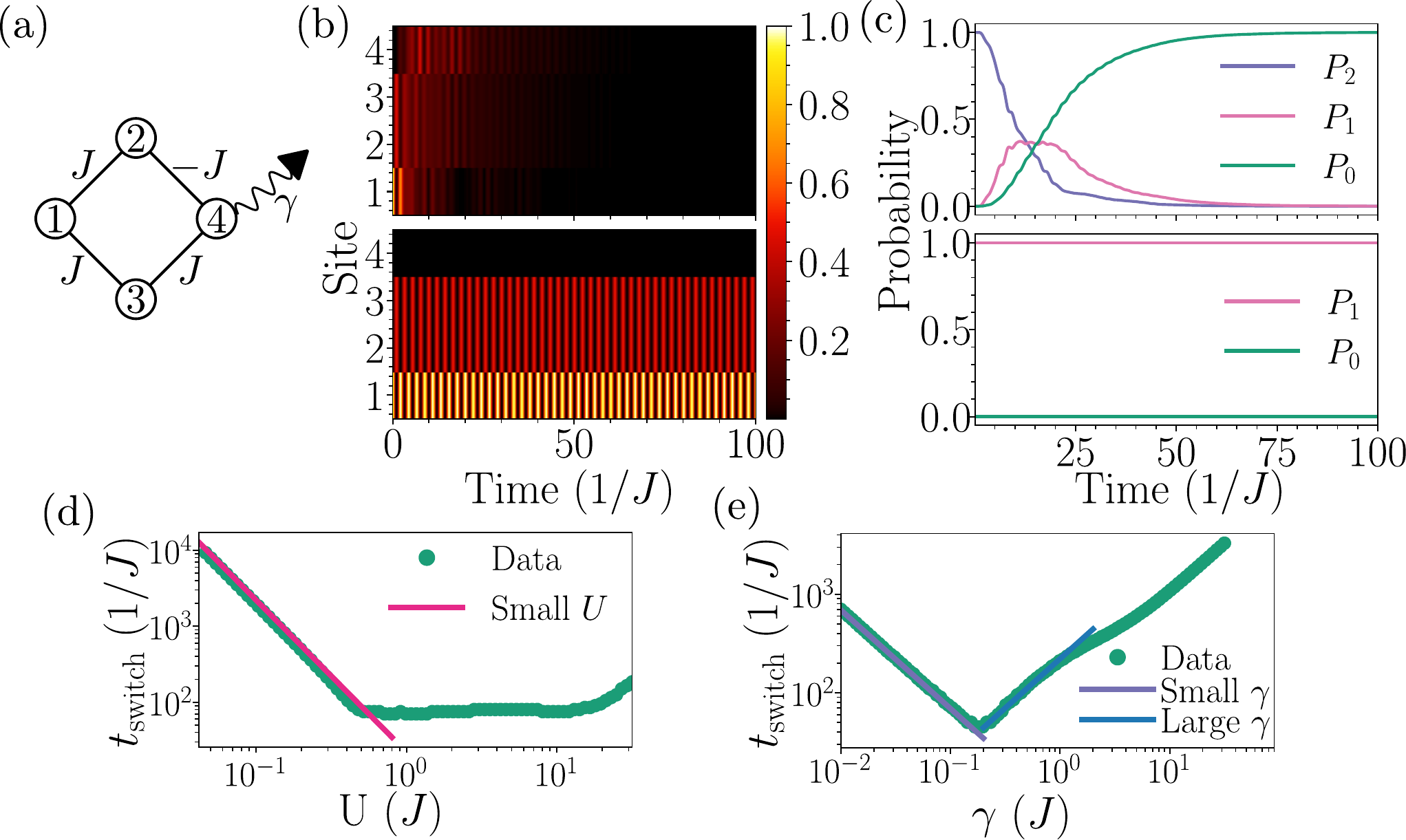}
    \caption{The open operation of the switch with a single rhombus of the diamond chain. (a) The site 1 acts as the input, and the site 4 as the output, where the sink with decay rate $\gamma$ is located. (b)  The site-resolved normalized particle number with two (upper panel) and one (lower panel) initial photon at the input site with respect to time. Here $U = \gamma = J.$ (c) The probability to observe $n=\{0,1,2\}$ photons in the system with respect to time for two initial photons (upper panel) and one initial photon (lower panel). (d) The switching time $t_{\mathrm{switch}}$ dependence on the interaction strength $U$. The threshold is set at $\epsilon=10^{-3}.$ (e) The decay rate $\gamma$ dependence of the switching time.}
    \label{fig:Fig5}
\end{figure*}

The open operation of the switching with the diamond chain is shown in Fig.~\ref{fig:Fig5}. In contrast to the closed operation, the open operation needs only a single rhombus. Furthermore, the switching is successful already if the photons enter from a single site, e.g. site 1, rather than the edge state. This is due to the fact that the site only overlaps with the edge state localized around the site, and the photon remains stuck. This is seen in Fig.~\ref{fig:Fig5}(b) as an oscillation between site 1 and sites 2 and 3, while site 4 remains empty. With two photons, instead, the photons delocalize towards the right edge, where they enter the sink and exit the system. A noteworthy difference to the Stub unit cell behavior shown in Fig.~\ref{fig:Fig3} is that both photons exit the system and do not remain stuck in the localized states. This highlights that, in the presence of interactions, photons move in pairs from one localized state to another. The switching contrast is perfect, see Fig.~\ref{fig:Fig5}(c). The switching time is inversely proportional to the interaction strength and the decay rate in the small $U$ and $\gamma$ limit, respectively, as shown in Figs.~\ref{fig:Fig5}(d) and (e). For large $U$ and $\gamma$, the switching time increases with $U$ and $\gamma$, thus the optimal $t_\text{switch}$ is found for theintermediate values.

\section{Experimental realization}
\label{sec:experiments}
Main limiting factors in the realizations of a switch are the accurate realization of the interacting models and the achievement of sufficiently large optical nonlinearities.  Pioneering works on how to enhance photon-photon interactions for switching include the use of dipolar gases and Rydberg blockade~\cite{dudin2012strongly, peyronel2012quantum, firstenberg2016nonlinear, murray2017coherent} as well as electromagnetic-induced transparency~\cite{schmidt1996giant, fleischhauer2005electromagnetically, bajcsy2009efficient, chen2013all} and ways to interface atoms with optical fibers~\cite{vetsch2010optical, oshea2013fiber}. While extensive efforts have been made in the ultracold gases community, miniaturized on-chip realizations in all-optical devices are in high demand and are only emerging. The switching concept proposed in this work can be realized in all-optical systems that allow simulating the Bose-Hubbard physics with various interaction strengths in the two-photon limit. 

In order to utilize tight-binding models based on the interference-localized state, small deviations from the exact flux condition have a detrimental effect on the single-photon states, as shown in Appendix \ref{sec:sensitivity}. 
Advances in the realization of synthetic fluxes have made possible to study and realize Aharonov-Bohm caging models in circuit-QED lattices \cite{yanay2020two-dimensional, alaeian2019creating, hung2021quantum, martinez2023interaction}, photonic lattices \cite{amo2016exciton,ozawa2019topological, mukherjee2018experimental, caceres2022controlled} and ultracold lattices \cite{creffield2010coherent, rey2015synthetic, cooper2019topological, li2022aharonov}, 
However, the requirement of a flux can be avoided by using e.g.~the Stub unit cell lattice instead of the diamond chain. Nevertheless, a challenge with the Stub unit cell realization of the switch is to eliminate the hopping amplitude between the A and B sites, as illustrated further in Appendix \ref{sec:sensitivity}. However, the Stub unit cell is not sensitive to other hopping amplitude imperfections. 

On one hand, control of single microwave photons in circuit-QED devices is already a standard, and Fock state photon dynamics have been observed in Bose-Hubbard models \cite{roushan2017chiral, yanay2020two-dimensional, martinez2023interaction}.  
On the other hand, various down-conversion techniques exist for producing single photons in the optical domain and their performance is constantly improving~\cite{eisaman2011invited, meyer2020single}. These techniques can be used to realize the single-photon-by-single-photon switch for example in cavity-QED arrays \cite{tomadin2010manybody}, coupled waveguides~\cite{kang2023topological}, or micropillar lattices \cite{amo2010exciton,amo2016exciton}. 

In the long run, provided that the single-photon manipulation is within reach, the switching concept presented in our work could be attainable in photonic crystals at arbitrarily low interaction strengths. In fact, since the localized states strongly confine the single photons, the switching is successful at any interaction if the interaction-induced delocalization time is shorter than the lifetime of the photons in the system. 
Moreover, the single-particle hopping parameters and the lattice geometry can be freely tuned in micro- and nanostructured optical materials. Therefore these systems offer a promising and flexible platform for implementing the models for the all-optical switch based on single-particle localized states. 
Besides photonic systems, ultracold atoms in optical lattices~\cite{torma2014quantum, gross2017quantum, schafer2020tools, li2022aharonov} can also implement the switch, as the dynamics of single atoms and interacting atom pairs can be observed and manipulated by a quantum gas microscope \cite{bakr2009quantum, sherson2010single}.

To estimate the reachable switching times, let us evaluate the Hubbard interaction parameter using state-of-the-art Kerr non-linear refractive index values. 
The Hamiltonian corresponding to the Kerr nonlinearity has the same form of the interaction term in Eq.~\eqref{eq:Hamiltonian_tbbh} and is given by $\chi\hat{n}(\hat{n}-1)$,
where $\hat{n}$ is the input photon number operator at the Kerr non-linearity. The constant $\chi$, equivalent to $U/2$ in Eq.~\eqref{eq:Hamiltonian_tbbh}, is given by~\cite{carusotto2013}
\begin{equation}
    \chi = \frac{3 \hbar^2 \omega^2 \chi^{(3)}}{4\epsilon_0\epsilon_r^2 V_{\mathrm{eff}}}
    \label{eq:kerr_coeff}
\end{equation}
where $\omega$ is the angular frequency of the photons, $\chi^{(3)}$ is the third-order non-linearity of the Kerr material, $\epsilon_0$ is the permittivity of vacuum, $\epsilon_r$ is the relative permittivity of the material, and $V_{\mathrm{eff}}$ is the mode volume in the cavity. The connection between the third-order non-linearity $\chi^{(3)}$ and the non-linear refractive index $n_2$ is given by $n_2 = \chi^{(3)}/(\epsilon_r\epsilon_0c),$ where $c$ is the speed of light.

Non-linear refractive indexes up to $n_2=10^{-9}~\mathrm{m^2/W}$ have been reported for GaAs/GaAlAs quantum wells~\cite{miller1983degenerate} and graphidyne/graphene heterostructures~\cite{zhang2022nonlinear}.
Using Eq.~\eqref{eq:kerr_coeff} with the value of the GaAs/GaAlAs quantum wells at $\lambda=1343$~nm laser with relative permittivity of 13 \cite{miller1983degenerate} and mode volume $V_{\mathrm{eff}}=\lambda^3$, we obtain $\chi \approx 10^{-3}$ eV. For the Stub lattice unit cell in the closed operation where $t_\text{switch} \approx 1/U$ in Eq.~\eqref{eq:switch_time_Stub_low_U}, this translates into a switching time $t_{\mathrm{switch}} \approx 10^{-12}$ seconds, i.e.~in the order of picoseconds. This value is substantially faster than the rate of single-photon sources \cite{eisaman2011invited, meyer2020single, tomm2021bright}.
Similar estimates for the switching time from the interaction $U$ can be made for circuit QED and ultracold atom systems.

\section{Comparison to other types of switches}
\label{sec:comparison}
The Mach-Zehnder interferometer (MZI) is a renowned switch concept based on the destructive interference between two different optical paths~\cite{o2010compact}. Due to its widespread usage, we now compare our switch concept to the MZI one, and illustrate the key differences.

The core principle of the MZI relies in using a control beam of intensity $I$ to induce a phase shift between two arms of the interferometer via Kerr nonlinearities.
Such phase shift is quantified as $\Delta \phi = kLn_2I$ for light of wavevector $k$ travelling along an arm of length $L$ with nonlinear refractive index $n_2$, relatively to the other arm. The switch is \textit{on} when $\Delta \phi = \pi$, because the signal is not transmitted due to destructive interference. Conversely, when the control beam is \textit{off}, no phase difference is acquired and the signal is transmitted. 

There are fundamental limitations for operating a MZI switch in the few-photon limit with coherent light~\cite{sanders1992quantumPRA}. In fact, the intensity fluctuations of coherent light, relative to the mean intensity, are significantly larger in the few-photon limit. These fluctuations have dramatic consequences on the phase matching and hence the switching mechanism~\cite{sanders1992quantum}. In the low mean photon-number limit, the non-linearity only allows for a phase-shift that is at maximum equal to the mean photon number~\cite{sanders1992quantumPRA, sanders1992quantum}.

The MZI switch could work in the few-photon limit with Fock states of light, provided that the necessary $\pi$ phase shift is achievable. We now illustrate such principle, calculating the action of the MZI on Fock states.
We model each stage of the MZI as quantum operators; the photon can either be in arm 1  or 2, and the corresponding annihilation operators are $\hat{b}_1$ and $\hat{b}_2$.  The beam-splitter is described by the unitary operator $\hat{B} = \exp(i\pi[\hat{b}_1^\dagger\hat{b}_2+\hat{b}_2^\dagger\hat{b}_1])$~\cite{sanders1992quantum}. The non-linear Kerr element operator is $\hat{K} = \exp(-i\chi'[\hat{n}_2^2 +\hat{n}_c^2]-4i\chi' \hat{n}_2\hat{n}_c)$, where $\chi' = \chi L/\hbar n c$ and $\chi$ is given by Eq. \eqref{eq:kerr_coeff}, $\hat{n}_2=\hat{b}_2^\dagger\hat{b}_2$ and $\hat{n}_c=\hat{b}_c^\dagger\hat{b}_c$ is the photon number operator of the control beam~\cite{sanders1992quantum}. The allowed Fock states are $\ket{1,0,0}$ and $\ket{0,1,0}$ for a photon in arm 1 and 2, respectively, when the control photon is absent. In the presence of the control photon, the states are $\ket{1,0,1}$ and $\ket{0,1,1}$ for a photon in arm 1 and 2, respectively. In the basis spanned by these four states, the beam-splitter operator $\hat{B}$ is represented by
\begin{equation}
    \hat{B}
    =\frac{1}{\sqrt{2}}
    \begin{pmatrix}
        1 & 1 & 0 & 0 \\
        -1 & 1 & 0 & 0 \\
        0 & 0 & 1 & 1 \\
        0 & 0 & -1 & 1 \\
    \end{pmatrix}
\end{equation}
and similarly, the Kerr non-linearity operator is given by
\begin{equation}
    \hat{K}
    =
    \begin{pmatrix}
        1 & 0 & 0 & 0 \\
        0 & 1 & 0 & 0 \\
        0 & 0 & 1 & 0 \\
        0 & 0 & 0 & e^{-4i\chi}\\
    \end{pmatrix},
\end{equation}
where the phase shift caused by the self-modulation terms $\chi \hat{n}_2^2, \chi \hat{n}_c^2$ is not included as it can be compensated by the other arm.
The overall effect of the MZI, where the phase shift is $4\chi=\pi$, is given by the following operator
\begin{equation}
    \hat{B} \hat{K} \hat{B} = 
     \begin{pmatrix}
        0 & 1 & 0 & 0 \\
        -1 & 0 & 0 & 0 \\
        0 & 0 & 1 & 0 \\
        0 & 0 & 0 & -1\\
    \end{pmatrix}~.
\end{equation}
By acting this operator onto the Fock states $\ket{1,0,0}$ and $\ket{0,1,0}$, the MZI switches the output paths, while in the presence of the control photon for $\ket{1,0,1}$ and $\ket{0,1,1}$, the output ports are the same, up to a phase.
Since the evolution is necessarily conserving, one can treat one of the outputs as a scrap-collector while the other acts as the primal output of the switch.

The remaining question is what would the required features of the Kerr non-linearity be
to achieve the appropriate phase-shift. In the Kerr operator $\hat{K}$ given above, we have required the effective phase-shift $\Delta \phi = 4\chi'$ to be $\pi$.  Using $\chi \approx 10^{-3}$ obtained above for the heterostructure material GaAs/AlGaAs with the high Kerr parameter $n_2 = 10^{-9}$ m$^{2}$/W~\cite{miller1983degenerate}, the necessary length $L= \hbar n c \pi/(4 \chi)$ is in the order of millimetres. Such lengthscale represents, for state-of-the-art materials, a fundamental limit for on-chip all-optical switches based on MZI.

Both the MZI and the single-particle localized switch discussed in this work are based on using destructive interference, but in a different way. In the MZI, one uses light-light effective interaction to attain a relative phase between its arms that realizes a destructive interference. However, in the localized-state switch, the interactions are used to break the destructive interference that localized the photons. On one hand, due to the localization, even a weak non-linearity allows switching in the single-particle localized switch if the delocalization time is faster than the system-losses by other means, since the interaction time is limited only by the losses. On the other hand, in the MZI the interaction time is limited by the length of the bulk non-linear component and the speed of light, requiring instead large intensity and aligning of the control pulse with the signal to sufficient precision. These facts render the MZI at the single-photon limit unpractical with currently existing materials, while the localized state switch can be realized as such with circuit QED systems and microstructured optical materials, as discussed above.

Other types of single-photon switch discussed in the Introduction rely on different realizations of the photon-photon interaction, usually enhancing the effect by placing the non-linear system in an optical cavity. However, our proposed switch paradigm is not dependent on the particular realization of the photon-photon non-linearity but is more a way to use these non-linearities in an effective way. Instead of having a non-linear element in a cavity and increasing the interaction time, one can realize the localized states of the flat-band system and allow photons to delocalize in a controlled way by interaction.

Our proposed switch is inherently single-photon-by-single-photon, since the input and the control signals can consist of no more than a single photon, respectively. If the signal or control beam consists of many photons, the interaction would delocalize them from the lattice, resulting in a false signal. Even if the interaction could be turned on or off by other means, the delocalization time-scale would be much longer for many particles, see Appendix \ref{sec:multiphoton}. With the Stub unit cell, the photons delocalize to the output but do not perform a full population oscillation. Similarly, the seven-site diamond chain with odd photon number does not show population oscillation, while the even photon numbers do to some extend. However, even partial delocalization would lead to successful switching in the open operation, but with the technical problem that a control photon is not anymore effective in operating the switch. At the classical limit of many photons, considered in Appendix \ref{sec:classical}, the delocalization does not occur or occurs very weakly, depending on the model.

\section{Conclusion}
\label{sec:conclusions}
In this article, we have introduced a switching concept based on the single-particle localization in a lattice by destructive interference with coincident delocalization of correlated two-particle state. In other words, single photons remain trapped in a part of the system while a pair delocalizes. The proposed switching concept is purely quantum mechanical and operates only at the quantum limit of few photons. 
One can either continuously extract the photons from outside the localized state via a sink, which we call the open operation of the switch, or one can wait for photons to geometrically separate from the initial state and collect them instantaneously, which we call the closed operation. 

We have demonstrated the switching scheme using the minimal three-sites Stub unit cell model , and the diamond chain.
Based on analytical two-level models, we have found expressions for the switching time in terms of the Rabi oscillation frequency of the two interacting photons between opposite edges.
Furthermore, we have found that the `ON' versus `OFF' contrast in the Stub unit cell is limited, whereas it is perfect for the diamond chain. However, the contrast in the Stub unit cell can be improved by tuning a geometric parameter that increases the overlap of the localized state on the input site.

We discussed how to realize the switching concept in various experimental platforms and compared with existing switches at the single-photon limit. 
We have estimated the achievable switching times in photonic systems to be in the order of picoseconds. This estimate is at par with most of the existing all-optical single-photon switch proposals, being limited by the photon-production rate of the existing single-photon sources. Nonetheless, as most proposals use light of many photons either as the control or the signal, there are limited direct comparisons for our single-photon-by-single-photon switch concept.

\begin{acknowledgments}
We acknowledge the computational resources provided by the Aalto Science-IT project. This work was supported by Academy of Finland under Projects No. 349313. VAJP acknowledges financial support by
the Jenny and Antti Wihuri Foundation. GS has received funding from the European Union's Horizon 2020 research and innovation programme under the Marie Sk\l{}odowska-Curie grant agreement No 101025211 (TEBLA).
\end{acknowledgments}

\appendix
\section{Subspace projections of the two-photon Hamiltonian}
\label{sec:projection}
In the simple lattice systems we are considering, the {\it non-interacting} two-particle eigenstates appear in degenerate sets separated from other states by finite energy gaps. 
For instance, 
the Stub unit cell has the single-particle eigenenergies $E_\mathrm{loc}=0$ and $E_{\pm} =\pm t_{AC}\sqrt{r^2+1}$, which result in two
degenerate two-particle eigenstates $\ket{\mathrm{loc},\mathrm{loc}}$ and $\ket{+-}$ and other states at separate energies.
At the low interaction limit, to first order in the interaction, the dynamics can be understood by dividing the initial state to its various non-interacting two-particle eigenstate components and considering evolution of the components independently. 

\subsection{Schrieffer-Wolff transformation}
\label{sec:sw_trans}
The projection to a 
non-interacting Hamiltonian subeigenspace is accomplished by the Schrieffer-Wolff transformation. The following review is based on Ref.~\cite{bravyi2011schrieffer-wolff}.
Let us assume that the system Hamiltonian is written as $\hat{H}=\hat{H}_0+\epsilon\hat{H}_{\mathrm{pert}}$ for some non-perturbed
Hamiltonian $\hat{H}_0$ and a perturbation $\hat{H}_{\mathrm{pert}}$. Furthermore, let us assume that we have a subspace of the
unperturbed Hamiltonian $\hat{H}_0$ with a collection of eigenenergies separated from others by a gap $\Delta$. We denote the projection
to this subspace by $P_0$ $(P_0^2 = P_0$, the subscript $0$ is to differentiate from the projection with respect to the total Hamiltonian) and the projection to the orthogonal subspace by $Q_0=1-P_0.$ Furthermore, for a general operator $\hat{X}$ in the Hilbert space, we define operators $\mathcal{O}(\hat{X}) = P_0\hat{X}Q_0+Q_0\hat{X}P_0$ and $\mathcal{D}(\hat{X}) = P_0\hat{X}P_0 + Q_0\hat{X}Q_0$, which give the components of the operator $X$ between the orthonormal subspaces and within them, respectively. 

The central result behind the Schrieffer-Wolff transformation is the existence of a one-to-one correspondence between unperturbed and perturbed states, and the subspaces they span,
when $\epsilon||\hat{H}_{\mathrm{pert}}|| < \Delta/2$. Here $||\hat{H}_\mathrm{pert}||$ is the norm of the perturbation operator defined as the maximal factor by which it scales the norm of a state it operates on. The result is physically intuitive: if the perturbation is small enough, it cannot mix the states belonging to the orthogonal subspaces separated by a larger energy than twice the perturbation. The one-to-one correspondence is given as a unitary transformation $\hat{U}$, called the Schrieffer-Wolff transformation, from perturbed to unperturbed subspace,
such that the perturbed subspace Hamiltonian can be expressed exactly in the basis of the respective non-perturbed eigenstates by
\begin{equation}
    \hat{H}_{\mathrm{eff}} = P_0 \hat{U}(\hat{H}_0+\epsilon \hat{H}_{\mathrm{pert}})\hat{U}^\dagger P_0~.
    \label{eq:sw_effective}
\end{equation}

Oftentimes, the exact transformation $\hat{U}$ is difficult to determine but one can express it using a perturbation expansion.
It can be shown that the Schrieffer-Wolff transformation can be written as $\hat{U}=\exp(i\hat{S})$, where $\hat{S}$ is a
block-off-diagonal, Hermitian operator, that is, $\hat{S} = \mathcal{O}(\hat{S})$ and $\hat{S}^{\dagger}=\hat{S}.$ One develops the perturbation series by 
expanding $\hat{S}$ as a power series $\hat{S} = \sum_{n} \epsilon^n \hat{S}_n$ and utilizing Baker-Campbell-Hausdorff formula $\exp(A)B\exp(-A) = \sum_n \frac{1}{n!}[A,^n B]$, where $[A,^nB] \equiv [A,[A,\dots,[A,B]\dots]]$  denotes that the commutator of $B$ with $A$ is taken $n$ times in succession.
We obtain 
\begin{equation}
\begin{split}
    &\exp(i\hat{S})(\hat{H}_0+\epsilon \hat{H}_{\mathrm{pert}})\exp(-i\hat{S})\\
    =& \sum_{n} \frac{i^n}{n!} [\hat{S},^n\hat{H}_0+\epsilon \hat{H}_{\mathrm{pert}}]\\
     =& \sum_{nm} \frac{i^n}{n!} \epsilon^{nm}[\hat{S}_m,^n\hat{H}_0]+\frac{i^n}{n!}\epsilon^{nm+1}[\hat{S}_m,^n\hat{H}_{\mathrm{pert}}]~.\\
    \end{split}
\end{equation}
The operators $\hat{S}_n$ are solved recursively by putting $\hat{S}_0=0$ and demanding at every order of $\epsilon$ that
the expression is block-diagonal. Block-diagonality means that the sub-manifolds are not mixed --- the very requirement of the Schrieffer-Wolff transformation. The zeroth order satisfies the condition automatically. At the first order, we
obtain $i[\hat{S}_1,\hat{H}_0] + \mathcal{O}(\hat{H}_{\mathrm{pert}}) = 0$ where by taking the matrix elements with respect to
the $\hat{H}_0$ eigenbasis $\ket{n}$, we find $\braket{n|\hat{S}_1|m} = i\braket{n|\mathcal{O}(\hat{H}_{\mathrm{pert}})|m}/(E_n-E_m)$.
Inserting this expression back to the expansion, we have $\hat{H}_{\mathrm{eff},1} = P_0\hat{H}_{\mathrm{pert}}P_0$. Thus, the first order perturbation is just the perturbation Hamiltonian projected to the subspace. Similarly, we have
$\hat{H}_{\mathrm{eff},2} =  \frac{1}{2}P_0 [i\hat{S}_1,\mathcal{O}(\hat{H}_{\mathrm{pert}})]P_0$, which, represented in the basis of the subspace $P_0$ eigenstates, is  
\begin{align}
    \braket{n|\hat{H}_{\mathrm{eff},2}|m}
    = -\frac{1}{2}&\sum_{n'\in Q_0}\left(\frac{1}{E_n-E_{n'}}-\frac{1}{E_{n'}-E_m}\right) \nonumber \\
    &\braket{n|\hat{H}_{\mathrm{pert}}|n'}\braket{n'|\hat{H}_{\mathrm{pert}}|m}~,
\end{align}
where $n'$ is summed over the orthogonal complement of the subspace $P_0.$
In this work, we do not use higher order than two terms but they can be obtained in a similar manner.

\subsection{Two-particle basis}
\label{sec:2p_basis}
In a system with single-particle quantum numbers $\alpha$ (e.g. orbital, location in an array of sites or single-particle eigenstate index), the two-particle states can be labeled by a list of the occupied quantum numbers and taking into account the indistinguishability by
$\ket{\alpha\beta} = N_{\alpha\beta}\hat{b}_\alpha^\dagger\hat{b}_{\beta}^\dagger \ket{0}$, where $\ket{0}$ is the vacuum state, $\hat{b}_{\alpha}^\dagger,\hat{b}_{\alpha}$ are creation and annihilation operators, and $N_{\alpha\beta} = 1$ if $\alpha\neq\beta$ and $N_{\alpha\beta} = 1/\sqrt{2}$ if $\alpha=\beta.$ We are interested in a second-quantized Hamiltonian $\hat{H} = \hat{H}_0+\hat{H}_{\mathrm{int}}$, 
where $\hat{H}_0=-\sum_{\alpha}t_{\alpha\beta}\hat{b}_{\alpha}^\dagger\hat{b}_\beta$
is a single-particle term and $\hat{H}_{\mathrm{int}}= \sum_{\alpha\beta\gamma\delta} U_{\alpha\beta\gamma\delta} \hat{b}_\alpha^\dagger\hat{b}_\beta^\dagger \hat{b}_\gamma\hat{b}_\delta$ is
is a two-particle interaction term.
By using the Wick's theorem, one finds the matrix elements of single-particle term as
\begin{equation}
\begin{split}
    &H_{0,\alpha\beta\gamma\delta}\\
    =&\braket{\alpha\beta|\hat{H}_0|\gamma\delta}\\
    =& -N_{\alpha\beta}N_{\gamma\delta}(t_{\alpha\delta}\delta_{\beta\gamma} + t_{\alpha\gamma}\delta_{\beta\delta}
    +t_{\beta\gamma}\delta_{\alpha\delta}+t_{\beta\delta}\delta_{\alpha\gamma})
    \end{split}
    \label{eq:single_matrix}
\end{equation}
and for the interaction term as
\begin{equation}
    \begin{split}
    &H_{\mathrm{int},\alpha\beta\gamma\delta}\\
    =& N_{\alpha\beta}N_{\gamma\delta}
    (U_{\alpha\beta\gamma\delta}+U_{\beta\alpha\gamma\delta}
    +U_{\alpha\beta\delta\gamma}+U_{\beta\alpha\delta\gamma}).
    \end{split}
    \label{eq:int_matrix}
\end{equation}

In the site basis $\hat{b}_i,\hat{b}_i^\dagger,$ where $i$ is a site in an array, we consider site-local interaction $\hat{H}_{\mathrm{int}} = \sum_i U_i\hat{b}_i^\dagger\hat{b}_i^\dagger\hat{b}_i\hat{b}_i/2$, whose matrix elements in the two-particle basis are $H_{\mathrm{int},ijkl} = \delta_{ij}\delta_{ik}\delta_{il} U_i.$ It is useful to consider the interaction in the eigenbasis of the single-particle Hamiltonian instead, defined by $\hat{b}_{n} = \sum_iV_{in}^* \hat{b}_i$, where $V_{in}$ diagonalizes the single particle Hamiltonian matrix $H_{0,ij} = \sum_n V_{in}\Lambda_{nn}V_{jn}^*.$ In this basis, we have
\begin{equation}
    U_{nmop} = \sum_i U_i V_{in}^*V_{im}^*V_{io}V_{ip}/2~,
\end{equation}
which can be inserted in Eq. \eqref{eq:int_matrix} to obtain 
the two-particle state matrix elements.

\subsection{Effective two-particle models at small interactions}
\subsubsection{Stub lattice unit cell}
In Sec.~\ref{sec:three_sites_closed}, we have seen that for small enough $U$, the interaction acts as a perturbation to the non interacting eigenstates, given by $\ket{\mathrm{loc},\mathrm{loc}},\ket{+,-}$ at $E=0$, $\ket{\mathrm{loc},\pm}$ at $E=\pm t_{AC}\sqrt{r_{\mathrm{AB}}^2+1}$ and $\ket{\pm,\pm}$ at $E=\pm 2t_{AC}\sqrt{r_{\mathrm{AB}}^2+1}$. 
If $U\ll t_{AC}\sqrt{r_{\mathrm{AB}}^2+1}$, the states $\ket{+,-}$ and $\ket{\mathrm{loc},\mathrm{loc}}$ describe a two-level system, isolated from the other eigenstates, which remain stationary.
The interacting Hamiltonian can be projected onto the subspace spanned by the two states $\ket{+,-}$ and $\ket{\mathrm{loc},\mathrm{loc}}$ through the Schrieffer-Wolff transformation. In this way, we get the following effective two-level Hamiltonian\begin{equation}
    H^{\text{eff}U\ll t_{\mathrm{AC}}}_{\mathrm{Stub}}
    =
    \frac{U}{(1+r_{\mathrm{AB}}^2)^2}\begin{pmatrix}
        1+r_{\mathrm{AB}}^2+r_{\mathrm{AB}}^4 & \sqrt{2}r_{\mathrm{AB}}^2 \\
        \sqrt{2}r_{\mathrm{AB}}^2  & 1+r_{\mathrm{AB}}^4
    \end{pmatrix},
    \label{eq:stub_two_state_smallU}
\end{equation} 
where the energy off-set of the two states $\delta$ and the overlap between the states $t_{12}$ are
\begin{equation}
    \delta= \frac{U r_{AB}^2}{(1+r_{AB}^2)^2},\qquad
    t_{12}= \frac{\sqrt{2}U r_{AB}^2}{(1+r_{AB}^2)^2}.
    \label{eq:delta_and_t12_smallU}
\end{equation}

\subsubsection{Single rhombus diamond chain}
The eigenstates $\ket{L\pm,L\pm}$, $\ket{L\pm,R\pm}$ and
$\ket{R\pm,R\pm}$ are degenerate at energies $E=\pm \sqrt{2}t$ for $+$ and $-$, respectively. Also, the states $\ket{L+,L-}$, $\ket{R+,R-}$, and $\ket{R\pm,L\mp}$ are degenerate with each other at the energy $E=0$. One can directly show that the terms, which split a photon pair at an edge to two photons at the opposite edges are forbidden:  $\braket{L\pm,L\pm|\hat{H}_{\mathrm{int}}|L\pm,R\pm}= \braket{R\pm,R\pm|\hat{H}_{\mathrm{int}}|R\pm,L\pm}=0$ and $\braket{R\pm,L\mp|\hat{H}_{\mathrm{int}}|R+,R-}
=\braket{R\pm,L\mp|\hat{H}_{\mathrm{int}}|L+,L-}=0$. In other words, the interaction allows only pairs to move from one edge to the another.
Hence, at small interaction, Hamiltonian subspaces at $E=\pm\sqrt{2}t$ consists of two interaction-connected states $\ket{L\pm,L\pm}$ and $\ket{R\pm,R\pm}$ and a single interaction-disconnected state $\ket{L\pm,R\pm}.$ Similarly, the subspace at $E=0$ is formed by two mutually independent two-level systems: pairs of photons at the same edge: $\ket{L+,L-}$ and $\ket{R+,R-}$, and at the opposite edges: $\ket{L+,R-}$ and $\ket{L-,R+}.$
The effective two-level projections of the interaction Hamiltonian onto these subspaces are given by
\begin{equation}
H_{\mathrm{U\gg J,rhombus}}^{L\pm L\pm,R\pm R\pm}=
\begin{pmatrix}
    3U/8 & U/8 \\
    U/8 & 3U/8 \\
\end{pmatrix}
\label{eq:rhombus_small_U}
\end{equation}
and $H_{\mathrm{int,rhombus}}^{L+L-,R+R-} = H_{\mathrm{int,rhombus}}^{L+R-,L-R+} = 2 H_{\mathrm{int,rhombus}}^{L\pm L\pm,R\pm R\pm}$~.

\subsubsection{Two-rhombi diamond chain}
The single-particle eigenstates $\ket{L\pm L\pm}$ and $\ket{R\pm R\pm}$ are degenerate similar to the single-rhombus system, forming an effective two-level system at small interactions.
Since these states do not overlap directly, we need to calculate the second-order term in the Schrieffer-Wolff transformation to obtain an effective Hamiltonian. We note that, despite the state $\ket{L\pm,R\pm}$ being degenerate with $\ket{L\pm L\pm}$ and $\ket{R\pm R\pm}$, the former is disconnected from the latter two states, because the interaction prevents splitting of the photon pair.
The projected Hamiltonian is
\begin{equation}
H_{U\gg J,\text{2 rhombi}}^{\text{eff}}
= \frac{U^2}{256\sqrt{2} J}
\begin{pmatrix}
    -23& 6 \\
    6& -23
\end{pmatrix}~.
\label{eq:two_rhombi_small_U}
\end{equation}

\subsection{Large interaction limit of two-particle lattice model}
\label{sec:large_interaction}
Let us assume a lattice model with hopping elements 
$\hat{H}_{0} = \sum_{ij}t_{ij}\hat{b}_i^\dagger\hat{b}_j$ and onsite
Hubbard interaction $\hat{H}_{\mathrm{int}} = \sum_i U_i \hat{b}_i^\dagger\hat{b}_i^\dagger \hat{b}_i\hat{b}_i/2$. The two-particle eigenstates of the Hubbard term are the states $\ket{ij}$ where one particle is at the site $i$ and another at the site $j$ with energy $E=U_{i}\delta_{ij}.$ If the interaction is large in comparison
to the hopping, we utilize the Schrieffer-Wolff transformation as in Appendix~\ref{sec:sw_trans} to project
to the subspace spanned by states $\ket{ii}$, consisting of a doublon~\cite{bravyi2011schrieffer-wolff, cohen1998atom}.
Importantly, we observe that $\braket{ii|\hat{H}_{0}|ij} = \sqrt{2} t_{ij}$, where the factor $\sqrt{2}$ arises due to normalization.
The terms between different $\ket{ii}$, up to first order in $\hat{H}_0$, vanish since they require two-particle operations whereas there are single-particle terms.
The second-order terms in the projection are
\begin{equation}
    \hat{H}_{\mathrm{int},ii,jj}^{\mathrm{proj.}}
    = t_{ij}^2 \left(\frac{1}{U_i} +\frac{1}{U_j}\right)
\end{equation}
for $i\neq j$
and
\begin{equation}
        \hat{H}_{\mathrm{int},ii,ii}^{\mathrm{proj.}}
    = \sum_{j} 2\frac{|t_{ij}|^2}{U_j}~.
\end{equation}
At the lowest finite order, the Hamiltonian projected to the doublon states is the hopping Hamiltonian with modified hopping amplitudes and additional on-site potential.    

\subsubsection{Stub unit cell}
In the limit of large interaction, the hopping is a perturbation to the interaction Hamiltonian eigenstates. The interaction Hamiltonian for the three-site system has two sets of three degenerate eigenstates: $\ket{\mathrm{AA}},\ket{\mathrm{BB}}$ and $\ket{\mathrm{CC}}$ at energy $E=U$ and $\ket{\mathrm{AB}},\ket{\mathrm{AC}}$ and $\ket{\mathrm{BC}}$ at $E=0.$
The Hamiltonian of the full system can be projected to the subspace spanned by on-site pairs by using the second order Schrieffer-Wolff transformation.
Assuming further that $U \gg |r_{\mathrm{AB}}t_{AC}| \gg t_{AC}$, one obtains the effective two-state model
 \begin{equation}
    H^{\text{eff}}_{U\gg t_{\mathrm{AC}},\text{Stub}}
    =
    \begin{pmatrix}
        2t_{\mathrm{AC}}^2/U & \sqrt{2}t_{\mathrm{AC}}^2/U\\
        \sqrt{2}t_{\mathrm{AC}}^2/U & t_{\mathrm{AC}}^2/U\\
    \end{pmatrix}~.
    \label{eq:stub_two_state_largeU}
\end{equation}
In Eq.~\eqref{eq:stub_two_state_largeU} the energy off-set of the two states $\delta$ and the overlap between the states $t_{12}$ are
\begin{equation}
    \delta= \frac{t_{AC}^2}{U},\qquad
    t_{12}= \frac{\sqrt{2} t_{AC}^2}{U}
    \label{eq:delta_and_t12_largeU}
\end{equation}

\subsubsection{Single rhombus system}
At large interaction, a two-particle doublon moves in a lattice with the same geometry as the original one, but with modified hopping amplitudes $J\to 2J^2/U$ and on-site energies $\epsilon_i = \sum_j 2J^2/U$. While the on-site energies do not have any significant effect, the hopping amplitudes become real and positive so that the flux condition is lost.
The system has degenerate single-particle eigenstates that are localized on the opposite corners. 

\subsection{Closed-operation switching time from two-state model dynamics}
\label{sec:two_state_model}
With the localized states, we oftentimes find that the non-interacting two-photon states form collections of (nearly) degenerate states. In this work, especially two-state behavior is found to be prominent. Here we repeat the standard two-state system calculation in order to introduce our notation and for easy reference.

Let us consider a two-level system spanned by states $\ket{1}$ and $\ket{2}$ with Hamiltonian
\begin{equation}
    \hat{H} \doteq 
    \begin{pmatrix}
        \epsilon_1 & t_{12} \\
        t_{12}^* & \epsilon_2\\
    \end{pmatrix},
\end{equation}
whose eigenenergies are
\begin{equation}
    E_{\pm} 
    = \epsilon_1 + \delta/2 \pm \Omega~,
\end{equation}
where we have defined the 'detuning' $\delta = \epsilon_2-\epsilon_1$ and the Rabi frequency $\Omega = \sqrt{(\delta/2)^2+|t_{12}|^2}~.$ 
We label the eigenstates by $\ket{+},\ket{-}$ for the plus and minus energy, respectively. The eigenstates components $(\alpha_{\pm},\beta_{\pm})$ are obtained as
\begin{equation}
    \alpha_{\pm} = \frac{|t_{12}|}{\sqrt{|t_{12}|^2 + ( \delta/2 \pm \Omega)^2}}
\end{equation}
and
\begin{equation}
    \beta_{\pm} = \frac{|t_{12}|}{t_{12}} \frac{\delta/2 \pm \Omega}{\sqrt{|t_{12}|^2+ (\delta/2\pm\Omega)^2}}~.
\end{equation}

Let us assume that at time $t=0$, the system is in the state $\ket{1}$.
General time dependence is given by
\begin{equation}
    \begin{pmatrix}
        \braket{1|\psi(t)}\\
        \braket{2|\psi(t)}
    \end{pmatrix}
    = \exp[-i(\epsilon_1+\delta/2)t]
    \begin{pmatrix}
        \cos(\Omega t) + i\delta \sin(\Omega t)/2\Omega\\
        -i t_{12}^* \sin(\Omega t)/\Omega 
    \end{pmatrix}
\end{equation}
and the probabilities are
\begin{equation}
    \begin{pmatrix}
        P_1(t)\\
        P_2(t)
    \end{pmatrix}
    = 
    \begin{pmatrix}
        1- |t_{12}|^2\sin^2(\Omega t)/\Omega^2\\
        |t_{12}|^2\sin^2(\Omega t)/\Omega^2
    \end{pmatrix}~.
    \label{eq:Rabi_probability}
\end{equation}
The resulting behavior is Rabi oscillations with Rabi frequency $\Omega = \sqrt{(\delta/2)^2+|t_{12}|^2}$. The maximum probability of observing state $\ket{2}$, provided that the initial state is $\ket{1}$ is $|t_{12}|^2/\Omega^2 = 1/(1+(\delta/2)^2/|t_{12}|^2).$ 

The switching time can be extracted from the dynamics by noting that the Rabi oscillation separates photons
from the input to the output. In other words, the switching time is the half-period of the Rabi oscillation
$t_\text{switch} = \pi/2\Omega.$

\subsubsection{Stub unit cell}
At small interaction, the effective two-level model for Stub unit cell, given by Eq.~\eqref{eq:stub_two_state_smallU}, results in the
Rabi frequency $\Omega = 3/2 Ur_\text{AB}^2/(1+r_\text{AB}^2)^2$. The half-period of the oscillation gives
the switching time
\begin{equation}
    t_\text{switch} 
    = \frac{\pi}{3} \frac{(1+r_\text{AB}^2)^2}{Ur_\text{AB}^2}~. 
\end{equation}

The Rabi frequency of two-state model for Stub lattice unit cell at large interaction Eq. \eqref{eq:stub_two_state_largeU} is $\Omega = \frac{3}{2} t_{\mathrm{AC}}^2/U$, which sets the switching time at large interactions 
\begin{equation}
    t_{\mathrm{switch}} = \frac{\pi U}{3t_{\mathrm{AC}}^2}
\end{equation}
shown in Fig. \ref{fig:Fig2} (f) of the main text.
In contrast to the small interaction limit, the switching time at large interactions depends on the single-particle hopping amplitude $t_\mathrm{AC}$ but does not depend on the ratio $r_{\mathrm{AB}}.$ 

Combining the observations on the interaction strength dependence of the switching time, we see that the switching time is optimal around an intermediate $U$, the optimal value being dependent on $r_{\mathrm{AB}}.$

\subsubsection{Single-rhombus system with $\pi$ flux}
Solving this model with the initial state $\ket{\mathrm{AA}}$ leads to density time-evolution $n_{A}(t) = \cos^4(2J^2 \tau/U)$, $n_{D}(t) = \sin^4(2J^2t/U)$, and $n_B(t)=n_C(t)=\sin^2(2J^2t/U)\cos^2(2J^2t/U).$ The oscillation of particle-density between the opposite edges is Rabi-like but with fourth-order sine and cosine functions. This dynamics was considered in Ref. \cite{martinez2023interaction} for a circuit QED experiment, where simulations agree with our results.

\subsubsection{Two-rhombi diamond chain}
The Rabi frequency of the effective two-state model of two-rhombi diamond chain Eq. \eqref{eq:two_rhombi_small_U} is 
$\Omega = 3 U^2/(128\sqrt{2} J)$, giving the switching time
\begin{equation}
    t_\mathrm{switch} = \frac{64 \sqrt{2}\pi J}{3 U^2}~.
\end{equation}

\subsection{Open operation switching time from two-state model with losses}
\label{sec:two_state_losses}
Let us consider the aforementioned two-level system with an additional third level $\ket{3}$ into which the state $\ket{2}$ decays according with a decay constant $\gamma$. The Hamiltonian is represented by the matrix in such basis by
\begin{equation}
    \hat{H} = 
    \begin{pmatrix}
        \epsilon_{1} & t_{12} & 0 \\
        t_{12}^* & \epsilon_{2} & 0 \\
        0 & 0 & 0 \\
    \end{pmatrix}~.
\end{equation}
and, similarly, the jump operator describing the decay is
\begin{equation}
    \hat{L}
    =\begin{pmatrix}
        0 & 0 & 0 \\
        0 & 0 & 0 \\
        0 & 1 & 0 
    \end{pmatrix}~.
\end{equation}
The Lindblad master equation for the density matrix $\hat{\rho}$ is
\begin{equation}
    \frac{\mathrm{d}\hat{\rho}}{\mathrm{d}t} = -i[\hat{H},\hat{\rho}]
    + \gamma \left(\hat{L}\hat{\rho} \hat{L}^\dagger -\frac{1}{2}\left\{ \hat{L}^\dagger \hat{L},\hat{\rho}\right\}\right)~.
\end{equation}
The Hamiltonian term is 
\begin{equation}
\begin{split}
    &-i[\hat{H},\hat{\rho}]\\
    =&\begin{pmatrix}
        -it_{12}\rho_{21}+it_{12}^*\rho_{12} & -i\delta\rho_{12}+it_{12}(\rho_{11}-\rho_{22}) & 0 \\
        i\delta\rho_{21}-it_{12}^*(\rho_{11}-\rho_{22}) & -it_{12}^*\rho_{12}+it_{12}\rho_{21} & 0 \\
        0 & 0 & 0 \\
    \end{pmatrix}~,
    \end{split}
\end{equation}
where $\delta=\epsilon_1-\epsilon_2.$
The Lindblad term is
\begin{equation}
\hat{L}\rho\hat{L}^\dagger -\frac{1}{2}\left\{\hat{L}^\dagger\hat{L},\hat{\rho}\right\}
=
\begin{pmatrix}
    0 & -\rho_{12}/2 & 0\\
    -\rho_{21}/2 & -\rho_{22} & -\rho_{23}/2 \\
    0  & -\rho_{32}/2 & \rho_{22}  \\
\end{pmatrix}~.
\end{equation}
From these equations, we note that the terms $\rho_{13},\, \rho_{31},\, \rho_{23},$ and $\rho_{32}$ will remain zero if they are initially zero. The equations for the remaining density matrix elements
can be expressed in terms of the Liouville-Fock space consisting of vector representations of the density matrices
$\hat{\rho} \doteq (\rho_{11},\rho_{12},\rho_{21},\rho_{22},\rho_{33})^T$.
The Lindblad equation can be represented
in this basis by the matrix equation $\frac{\mathrm{d}}{\mathrm{d}t} \rho = L\rho$, where
\begin{equation}
    L   
    = 
    \begin{pmatrix}
        0 & it_{12}^* & -it_{12} & 0 & 0 \\
        it_{12} & -i\delta-\gamma/2 & 0 & -it_{12} & 0 \\
        -it_{12}^*  & 0 & i\delta-\gamma/2 & it_{12}^* & 0 \\
        0 & -it_{12}^* & it_{12} & -\gamma & 0 \\
        0 & 0 & 0 & \gamma & 0 \\ 
    \end{pmatrix}
\end{equation}

\subsubsection{Large \texorpdfstring{$\gamma$}{gamma} limit}
In the limit $\gamma \gg \Omega$, one can neglect the $\rho_{22}$ term since it decays fast and is initially zero. Similarly, we see from the above equation that (assuming $t_{12}$ to be real for simplicity, but similar argument can be made in general)
$\mathrm{d} \mathrm{Re}(\rho_{12})/\mathrm{d} t = \delta \mathrm{Im}(\rho_{12}) -\gamma/2 \mathrm{Re}(\rho_{12})$ and accordingly, due to the fast decay and small source term, $\mathrm{Re}(\rho_{12})$ will remain negligible since we assume that $\rho_{12}(t=0)=0.$ Effectively, we are left with two real variables, $\rho_{11}$ and $\mathrm{Im}(\rho_{12})$ with equations
\begin{equation}
    \frac{\mathrm{d}}{\mathrm{d} t}
    \begin{pmatrix}
        \rho_{11}\\
        \mathrm{Im}(\rho_{12})\\
    \end{pmatrix}
    =
    \begin{pmatrix}
        0 & - 2 t_{12} \\
        t_{12} & -\gamma/2
    \end{pmatrix}
      \begin{pmatrix}
        \rho_{11}\\
        \mathrm{Im}(\rho_{12})\\
    \end{pmatrix}.
\end{equation}
We obtain a second-order differential equation for 
$\mathrm{Im} (\rho_{12})$:
\begin{equation}
    \mathrm{Im}(\Ddot{\rho}_{12})+\frac{\gamma}{2}\mathrm{Im}(\dot{\rho}_{12})
    +2 t_{12}^2 \mathrm{Im}(\rho_{12}) = 0\,.
\end{equation}
With an ansatz $\mathrm{Im}(\rho_{12})= A\exp(r t)$, we get the equation for
$r$
\begin{equation}
    r^2 + \frac{\gamma}{2}r+ 2t_{12}^2 =0~, 
\end{equation}
which has the solution
\begin{equation}
    r_{\pm} = -\frac{\gamma}{4} \pm \sqrt{\frac{\gamma^2}{16}-2t_{12}^2}~.
\end{equation}
Thus, the solution for the element $\mathrm{Im}(\rho_{12})$ is, using the initial conditions,
\begin{equation}
\begin{split}
    &\mathrm{Im}(\rho_{12})(t)
    =
    t_{12}\exp\left(-\frac{\gamma}{4}t\right)\\
    &\begin{cases}
    \sinh\left[\sqrt{\frac{\gamma^2}{16}-2t_{12}^2}t\right]/\sqrt{\frac{\gamma^2}{16}-2t_{12}^2}, & \mathrm{if~} \gamma^2 > 32t_{12}^2 \\
        \sin\left[\sqrt{2t_{12}^2-\frac{\gamma^2}{16}}t\right]/\sqrt{32t_{12}^2-\frac{\gamma^2}{16}},& \mathrm{if~} \gamma^2 < 32t_{12}^2
    \end{cases}~.
    \end{split}
\end{equation}
Accordingly, we obtain the remaining density matrix element by $\rho_{11} = \gamma\mathrm{Im}(\dot{\rho}_{12})/2t_{12} + \mathrm{Im}(\rho_{12})/t_{12}$ as,
\begin{equation}
    \begin{split}
    \rho_{11}(t)
= \exp\left(-\frac{\gamma}{4}t\right)
    &\Bigg(\cosh\left[\sqrt{\frac{\gamma^2}{16}-2t_{12}^2}t\right] \\
    &+\frac{\gamma}{4}
    \frac{\sinh\left[\sqrt{\frac{\gamma^2}{16}-2t_{12}^2}t\right]}{\sqrt{\frac{\gamma^2}{16}-2t_{12}^2}}\Bigg)~,
    \end{split}
\end{equation}
and similarly with $\cos$ and $\sin$ if $\gamma^2 < 32t_{12}^2$. However,
since we assumed $\gamma \gg t_{12},\delta$, we can write $\sqrt{\gamma^2/16-2t_{12}^2} \approx \gamma/4-4t_{12}^2/\gamma.$ Thus, we have
\begin{equation}
\begin{split}
    &\exp(-\gamma t/4) \cosh(\sqrt{\gamma^2/16-2t_{12}^2} t)\\
    \approx&
\frac{1}{2}\exp(-4t_{12}^2t/\gamma) + \frac{1}{2}\exp(-\gamma t /2 + 4t_{12}^2t/\gamma)\,,
\end{split}
\end{equation}
and similarly for the $\sinh$ but with a minus sign in front of the second term.
Hence, we have the approximate form
\begin{equation}
    \rho_{11}(t)=\exp\left(-\frac{4t_{12}^2}{\gamma} t\right)\,.
\end{equation}
We note directly that, given a tolerance $\epsilon$, the decay time is proportional to $\gamma$ and inversely proportional to $t_{12}^2.$
In the context of the switch considered in the main text,
the switching time is given by
\begin{equation}
    t_\text{switch} = -\ln(\epsilon)\frac{\gamma}{4t_{12}^2}~.
\end{equation}

Based on the two-level model for Stub unit cell at small interaction $U \ll t_\text{AC}$, Eq.~\eqref{eq:stub_two_state_smallU},
the switching time becomes 
\begin{equation}
    t_\text{switch} = -\ln(\epsilon) \frac{\gamma(1+r_\text{AB}^2)^4}{8U^2r_\text{AB}^4}~.
\end{equation}
Similarly, at the large interaction $U \gg t_\text{AC}$, the Stub unit cell effective model Eq.~\eqref{eq:stub_two_state_largeU} gives the
switching time
\begin{equation}
    t_\text{switch} = -\ln(\epsilon) \frac{\gamma U^2}{8t_\text{AC}^4}~.
\end{equation}

For the single-rhombus diamond chain, we find at small interaction $U \ll J$, that
\begin{equation}
    t_{\text{switch}}
    = -\ln(\epsilon) \frac{16 \gamma }{U^2}
\end{equation}

\subsubsection{Small \texorpdfstring{$\gamma$}{gamma} limit}
Let us consider the limit where the Rabi frequency is substantially faster than the decay rate, that is, $\Omega \gg \gamma$, and we have damped Rabi oscillations.
The probability to find the system in the states $\ket{1}$ and $\ket{2}$ is $P(t) \equiv P_1(t)+P_2(t)$, which decays over time due to the sink. The decay of $P(t)$ has to be proportional to $P_2(t)$: $\frac{\mathrm{d}P}{\mathrm{d}t} = -\gamma P_ 2(t).$ 
Moreover, since the decay is slow, the evolution of the probabilities $P_1$ and $P_2$ still follows approximately the undamped dynamics given by Eq. \eqref{eq:Rabi_probability} but with the decay of the probability $P(t)$ taken into account. Then, the ratio $P_2(t)/P(t) = |t_{12}|^2\sin^2(\Omega t)/[(\delta/2)^2+|t_{12}|^2]$, leading to the equation 
\begin{equation}
    \frac{\mathrm{d}P}{\mathrm{d}t} = 
    -\frac{4 \gamma |t_{12}|^2}{\delta^2+4|t_{12}|^2}\sin^2(\Omega t)\,P~
\end{equation}
with the initial condition $P(t=0) = 1.$
Solution to this equation is
\begin{equation}
    P(t) = \exp\left[- \frac{2\gamma|t_{12}|^2}{\delta^2+4|t_{12}|^2}\left(t-\frac{\sin(2\Omega t)}{2\Omega}\right)\right]~.
\end{equation}
From this, the time it takes for the system to deplete beneath a given threshold $\epsilon$ is given as the solution of
\begin{equation}
    t -\frac{\sin(2\Omega t)}{2\Omega} = -\ln(\epsilon) \frac{\delta^2+4|t_{12}|^2}{2\gamma|t_{12}|^2}.
    \label{eq:time_small_gamma}
\end{equation}

For the switch considered in the main text, Eq. \eqref{eq:time_small_gamma} sets the switching time $t_\text{switch}$. Since the sine term on the left-hand side of the equation is small, we can neglect it giving the equation
\begin{equation}
    t_\text{switch} = -\ln(\epsilon) \frac{\delta^2+4|t_{12}|^2}{2\gamma|t_{12}|^2}~,
    \label{eq:switching_time_small_gamma}
\end{equation}
where now $\epsilon$ represents the switching threshold.
This equation specifies that the switching time is inversely proportional to $\gamma$ and not dependent on the interaction. In fact, both $t_{12}$ and $\delta$ scale similarly with the interaction strength $U$.

Using the two-state model for the Stub unit cell at small and large interactions strengths
$U \ll t_\text{AC}, U \gg t_\text{AC}$, Eqs.~\eqref{eq:stub_two_state_smallU}, \eqref{eq:stub_two_state_largeU}, respectively, we find the switching time
\begin{equation}
    t_\text{switch} = - \frac{9\ln(\epsilon)}{2\gamma}
\end{equation}
for both models. Since the switching occurs after many Rabi oscillation periods due to weak damping induced by the sink, only the average population at the sink determines the switching time. Since the average is independent of interaction strength $U$ and the parameter $r_{\mathrm{AB}}$, the switching time is also independent of these.

The switching time for the single rhombus diamond chain is given, according to the model
Eq.~\eqref{eq:rhombus_small_U} by 
\begin{equation}
    t_\text{switch} = - \frac{13\ln(\epsilon)}{2\gamma}~.
\end{equation}

\section{Sawtooth lattice}
\label{sec:sawtooth}
\begin{figure*}
    \centering
    \includegraphics[width=0.9\textwidth]{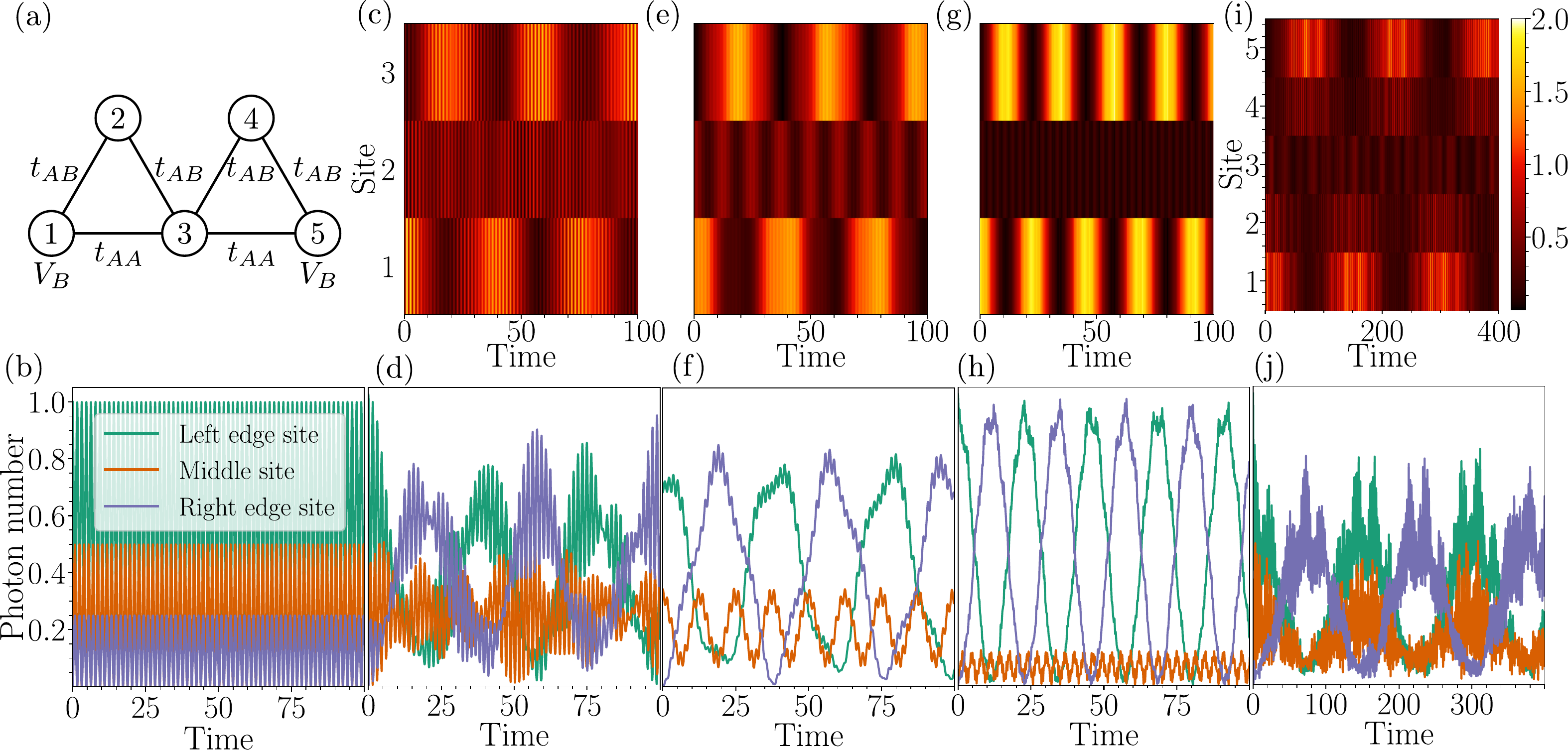}
    \caption{Switching using the sawtooth lattice. (a) The sawtooth lattice consists of sites of type A (1, 3, and 5) and B (2, and 4). On obtains the flat-band condition of the lattice by putting $t_{\mathrm{AB}} = \sqrt{2}t_{\mathrm{AA}}$. Furthermore, the edge-states localize if one sets the boundary potential $V_B=t_{\mathrm{AA}}$ at the edge sites. In the following, time is in units of $1/t_{\mathrm{AA}}$. The photon numbers are also normalized to 1. Panels (b) to (h) show the dynamics of a three-site sawtooth ladder (i.e., a triangle) with sites 1, 2, and 3, with $V_{B}$ at sites 1 and 3; panels (i) and (j) consider the complete system shown in panel (a). Panel (b) shows the single-photon dynamics starting at the site 1, showing partial delocalization to the site 3. The interaction is chosen as $U=t_{\mathrm{AA}}$ and photons are initially at the site 1. Population oscillation with some high-frequency noise occurs. Panels (e) and (f) show the same dynamics as (c) and (d) but with photons initially in the localized left-edge state, which reduces the high-frequency noise. Panels (g) and (h) show the large interaction two-photon dynamics at $U=10t_{\mathrm{AA}}$ starting with photons at site 1, but otherwise same parameters as panels (c) and (d). The oscillations have higher amplitude and occur at faster frequency than at the smaller interaction. Panels (i) and (j) show the two-unit cell dynamics with the same parameters and setting as panels (c) and (d). The population oscillation frequency is decreased.}
    \label{fig:sawtooth_dynamics}
\end{figure*}

Another prominent system that allows the switching concept proposed in this work is the sawtooth lattice. The sawtooth lattice belongs to the class of flat band systems where the tight-binding hopping amplitudes are purely positive while the destructive interference is due to the eigenstate itself having relative phases from one site to another. Its three-site version is
among the three-site models with localized states considered in Sec.~\ref{sec:three_site}. In general, the sawtooth lattice comprises of unit cells with two sites, labeled A and B, in a sawtooth formation
as shown in Fig. \ref{fig:sawtooth_dynamics} (a). If the direct hopping B sites vanishes and if $t_{AB}=\sqrt{2}t_{AA},$ the system possesses localized states. As a finite-size effect, the chain contains exponentially localized edge states, which become perfectly localized if one adds an edge potential $V_{B} = t_{\mathrm{AA}}$ on the edge A sites of the system.

If we use the sawtooth lattice flat-band condition $t_{AB}=t_{BC}=\sqrt{2}t_{AC}$ in Eq. \eqref{eq:three_loc_cond}, we find that $\epsilon_A = t_{AC}$, which is the boundary potential resonance condition \cite{pyykkonen2021}. Indeed, Eq. \eqref{eq:loc_energy} gives $E_{\mathrm{loc}}=2t_{AC}$, the flat band energy. The localized eigenstate is 
$\ket{\psi_{\mathrm{loc}}} \doteq (\sqrt{2},-1,0)^T/\sqrt{3}$. Assuming $\epsilon_C = t_{AC}$ for symmetry, we find $E_{\pm} = \pm 2 t_{AC}$, while the eigenvectors are $\ket{\psi_+} \doteq (1,\sqrt{2},-3)^T/2\sqrt{3}$ and $\ket{\psi_-} \doteq (1,\sqrt{2},1)^T/2.$ 

Figure \ref{fig:sawtooth_dynamics} presents the switching dynamics of the sawtooth lattice with three and five sites. In both cases, one observes similar population oscillation between the opposing edge sites of the system as we observed for the Stub unit cell and the diamond chain. The limiting feature is observed in the Fig. \ref{fig:sawtooth_dynamics} (b): the single-photon dynamics starting from the initial state with the photon the site 1 causes significant part of the population to delocalize, which is unwanted. This cannot remedied in other ways than what we essentially have done with the Stub lattice, that is, to optimize the hopping parameters to localize the left edge state more on the input site. Another mitigation could be to initialize the system in the edge state rather than edge site, which might be difficult in experiments due to relative phases of the edge state. However, the population oscillation to the opposite edge is not perfect even at this case. 

The oscillatory nature of the two-photon dynamics hint that one can understand the system by projecting to a subspace of the single-particle Hamiltonian eigenstates at small interaction or the interaction eigenstates at the large interaction.
In the sawtooth, the non-interacting two-particle states
are 
$\ket{\mathrm{loc},\mathrm{loc}}$, $\ket{\mathrm{loc},+}$, and 
$\ket{++}$ at $E= 4t_{AC}$, $\ket{\mathrm{loc},-}$ and $\ket{+-}$ at $E=0$
and $\ket{--}$ at $E=-4t_{AC}$. 
Gathering these results in a matrix, and assuming uniform interaction, we can project the Hamiltonian onto the basis $\ket{++},\ket{+\mathrm{loc}},\ket{\mathrm{loc},\mathrm{loc}}$ and obtain
\begin{equation}
    H_{\mathrm{int,sawtooth}}
    =
    \frac{U}{9}\begin{pmatrix}
    \frac{ 43}{8} & -\frac{1}{4} & 1\\
     -\frac{1}{4} & 2 & 1\\
     1 & 1 & 5
    \end{pmatrix}~.
\end{equation}
We have three states where two are approximately degenerate while the third one is off-set. This explains why the observed dynamics is more complicated in comparison to the diamond chain and the Stub unit cell, where one could understand the dynamics with a two-state model. However, one can extract the approximate switching time by considering the pair of states with equal diagonal elements, $\ket{++}$ and $\ket{\mathrm{loc,loc}}$ to obtain the Rabi frequqency. The large-interaction limit of the dynamics follows similarly.

\section{Sensitivity analysis}
\label{sec:sensitivity}
In the Stub unit cell and the diamond chain models, we have assumed specific properties for the parameters in order to have localized states. Specifically, for the Stub unit cell we required that $t_{AB}=0$, i.e. the A and B sites do not have mutual direct hopping, while for the diamond chain  we have included a $\pi$ flux. 
\begin{figure*}
    \centering
    \includegraphics[width=0.80\textwidth]{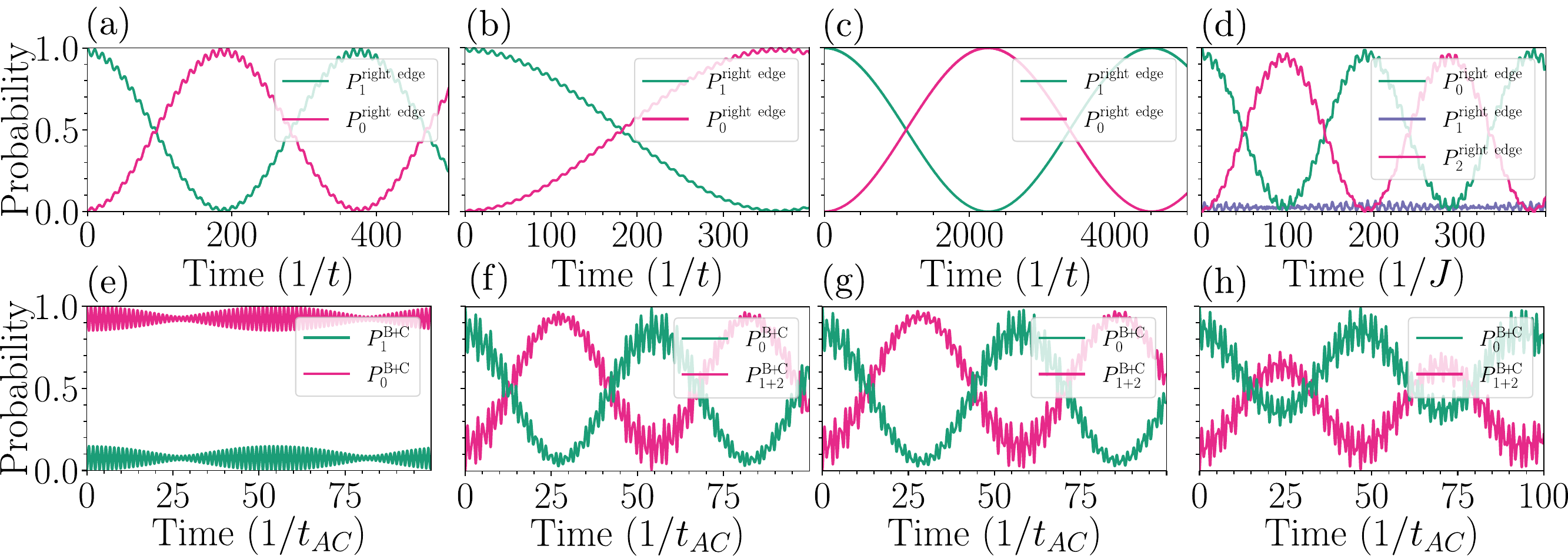}
    \caption{Sensitivity to parameter imperfections. Panels (a)-(d) demonstrate the effect of having off-set from the $\pi$-flux condition at the two-rhombi diamond chain. On panels (a), (b), and (c), the flux is put to $0.93\pi, 0.95\pi$, and $0.98\pi$, respectively, assuming single photon initially at the left edge state. One observes similar oscillations between the two edges as with finite interaction with two photons. Panel (d) shows the two-photon dynamics at $0.95\pi$ flux at $U=J$, which, together with Fig. \ref{fig:Fig3}, shows that the imperfection on the flux does not have effect on the interaction if the interaction delocalization time is faster than the flux off-set delocalization time. Panels (e)-(h) demonstrate the effect of having hopping between A and B sites of the Stub unit cell at $r_{\mathrm{AB}}=-5$. Panel (e) shows the effect on the single particle dynamics of $t_{\mathrm{AB}}=0.1t_{\mathrm{AC}}$, the effect being small oscillations in the populations but not increasing the average zero-photon probability. Panels (f), (g), and (h) show the effect of $t_{\mathrm{AB}} = 0,0.05t_{\mathrm{AC}}, 0.1t_{\mathrm{AC}}$ on the two-photon dynamics. At small $t_{\mathrm{AB}}$ the effect is very minute while suddenly at $0.1t_{\mathrm{AC}}$ the maximal probability of finding particles on the sites B and C reduces significantly. }
    \label{fig:sensitivity}
\end{figure*}

In Fig.~\ref{fig:sensitivity} (a)-(c) we note that the single-particle dynamics follows a Rabi-type oscillation between the edges if the flux is slightly off-set  from $\pi.$ For successful switching we need to prepare the flux so that this oscillation is slower than the interaction-induced delocalization. We show in Fig.~\ref{fig:sensitivity} (d) that the two-photon dynamics are not sensitive to slight inaccuracies in the flux if the interaction is strong enough to cause faster delocalization than the imperfection on the flux.

The flux-deviation effect can be understood by the following arguments.
For the diamond chain, we note that in a single rhombus, deviations from the $\pi$ flux directly couple the left and right edge states. If the difference to the $\pi$ flux is small enough, the perturbation mixes $\ket{L+}$ and $\ket{R+}$ together, for instance, but not the other states, up to first order in the deviation. With direct calculation, we find the coupling between the states to be $(\exp(i\phi)-1)J\braket{L+|\hat{b}_2^\dagger\hat{b}_4|R+} \approx |\phi|J \sqrt{2}/4$, 
which causes Rabi oscillations between the edges. Similarly, in the two rhombus case, since the left and right edge states are not directly coupled, we find that the coupling is given by the second-order Schrieffer-Wolff transformation as $\phi^2 J \sqrt{2}/8.$ 

For the Stub unit cell, the direct connection between sites A and B is found to be robust in terms of the single-particle localization, as shown in Fig.~\ref{fig:sensitivity} (e). However, we find that the two-particle dynamics delocalization to the other edge becomes imperfect with increasing hopping between A and B sites. For instance, at the considered ratio $r_{\mathrm{AB}}=-5$, deviation $t_{\mathrm{AB}}=0.1t_{\mathrm{AC}}$ significantly disturbs the delocalization, as shown in Fig.\ref{fig:sensitivity} (h). However, the change seems to be abrupt since $t_{AB}=0.05t_{\mathrm{AB}}$ is not causing large difference as shown in Fig.\ref{fig:sensitivity} (g).

\section{More than two photons}
\label{sec:multiphoton}
\begin{figure*}
    \centering
    \includegraphics[width=0.8\textwidth]{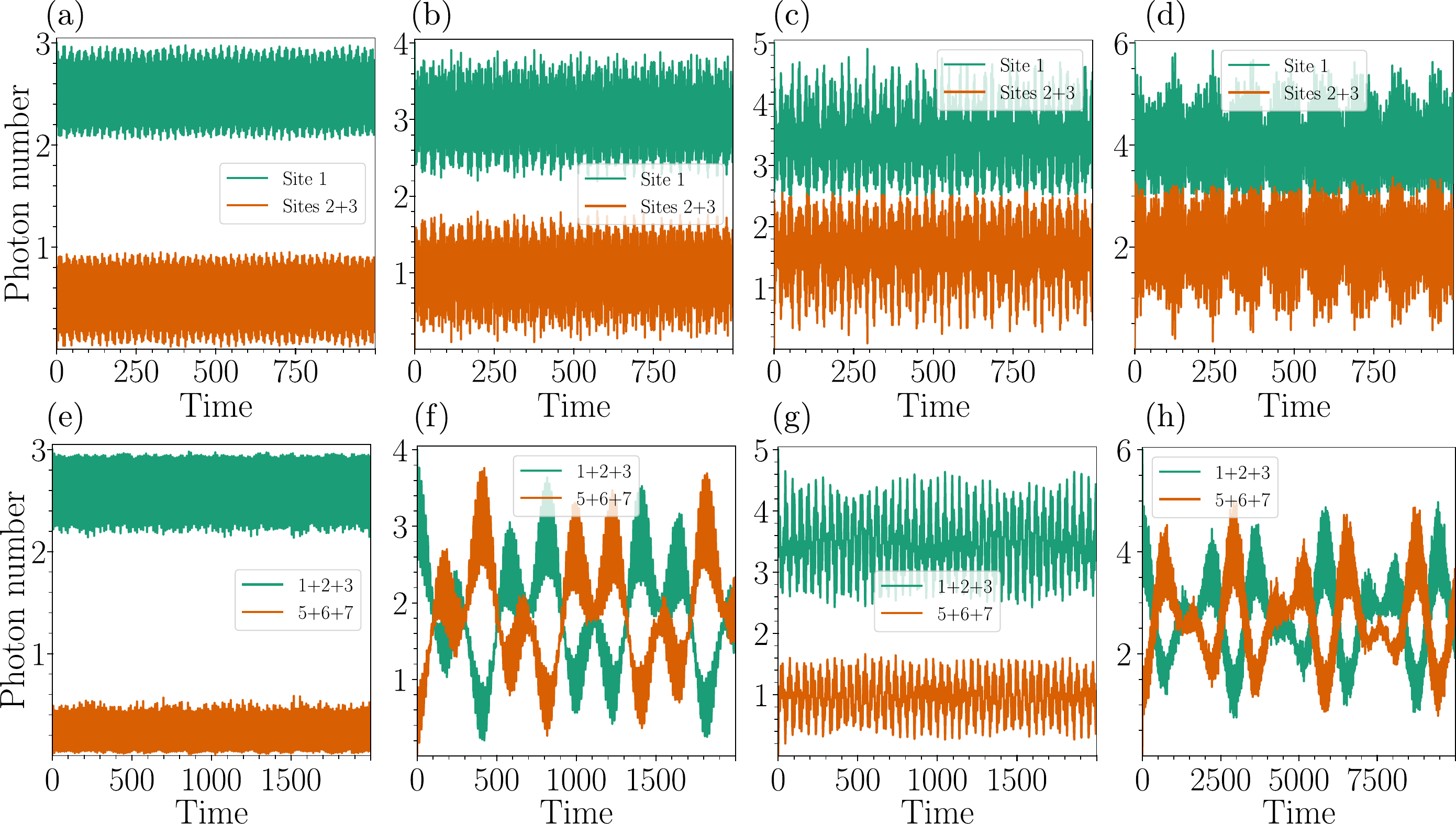}
    \caption{Multiphoton dynamics. The panels (a)-(d) present the time-evolution of 3, 4, 5, and 6 particle dynamics on the Stub unit cell, starting from all the photons on the A site, labeled site 1. The panels (e)-(h) present the 3, 4, 5, and 6 particle dynamics on the two-rhombi diamond chain. For the Stub unit cell we find that the coherent population oscillatory dynamics observed with two photons is not present. However, part of the photons delocalize: the photon number fluctuations seem to increase with photon number. In the case of the diamond chain, we find that the odd particle number does not result in particle oscillation from an edge to another while the even photon number does.   }
    \label{fig:more_photons}
\end{figure*}

Here we discuss the behavior of the switch if more than two photons enter the system. Fig.\ref{fig:more_photons} demonstrates the effect of having more than two photons at the initial state for the Stub unit cell and the two-rhombi diamond chain. For the Stub unit cell, the population delocalizes more than in the single-photon case. However, similar dynamics of all photons moving together, as found for the two-photon case, is not present. Nevertheless, the photon number fluctuations over time increase with photon number.

For the diamond chain, the oscillatory dynamics of photons moving together from an edge to the opposite one is observed for even photon number, while the odd photon number resembles the Stub unit cell shown above. However, the time-scale of the oscillation is substantially longer than with the two-photon oscillations. 

The difference between the diamond chain and Stub unit cell and the is that the former strictly allows only pair movement while in the latter single-particle processes can occur due to the presence of non-localized states. This, explains the diamond chain odd photon number different behavior.

The dynamics observed at the multiphoton initial states would allow switching since part of the photons can be found at the output sites of the systems. Also, the delocalization time is quick.
However, switching is difficult to arrange with more than one photon signal due to the fact that if there are multiple photons, they tend to delocalize even without the control photon. Thus, one would need to device another way to control the switch e.g. turn the interaction on and off, which is not experimentally practical. Another issue for the closed operation of the switching is the large fluctuations of the expected photon number over time, so one would need to be very precise when to deplete the system. This would not, however, be a problem in the open operation of the switch.

\section{Classical dynamics}
\label{sec:classical}
\begin{figure*}
    \centering
    \includegraphics[width=0.9\textwidth]{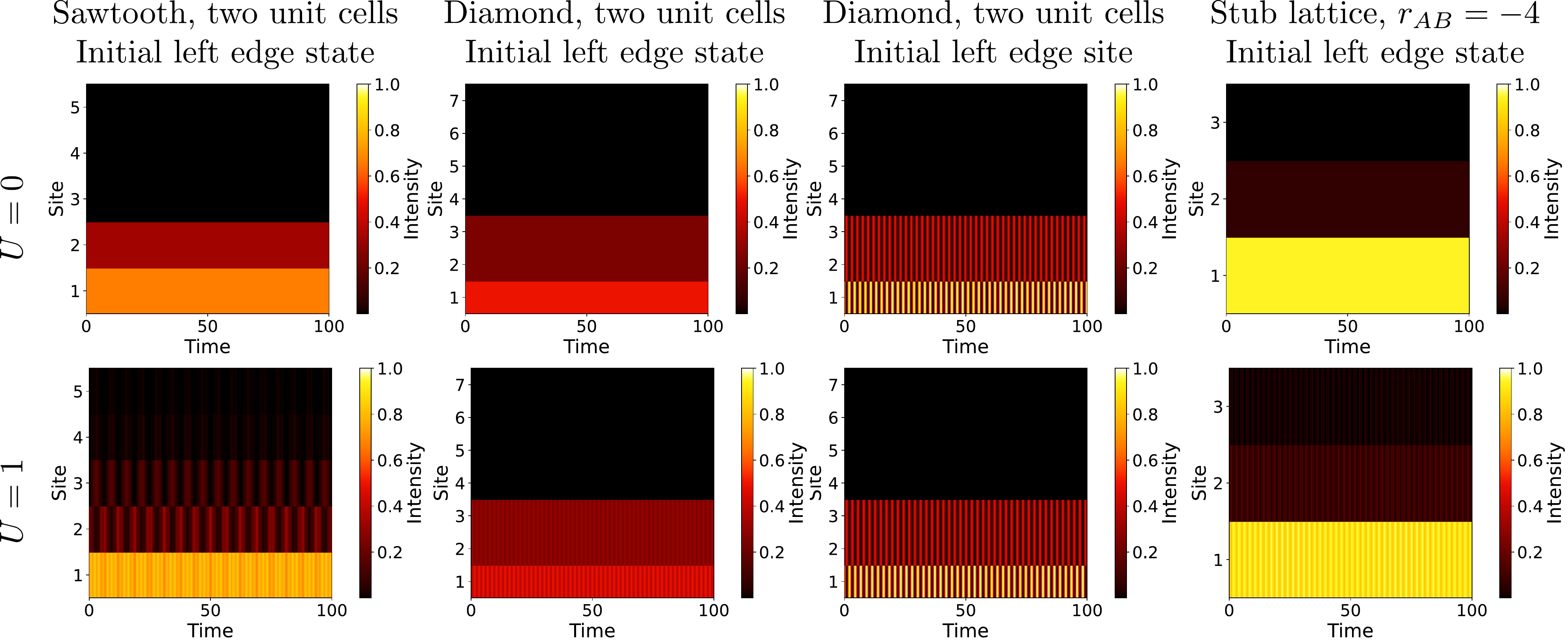}
    \caption{Closed-operation dynamics in the classical limit. The interaction does not delocalize the states at all in the either of the initial state configurations of the diamond lattice. For the sawtooth and stub lattices, the delocalization is exponential, resulting in a weak signal at the right edge. }
    \label{fig:classical_dynamics}
\end{figure*}

In the limit of many photons, the mean-field approximation becomes valid, where one can replace $\hat{b}_i\to b_i =\meanv{\hat{b}_i}$, that is, the operator with a classical number. 
The Heisenberg equation of motion for the operator $\hat{b}_i$ is
\begin{equation}
    -i\hbar \deriv{}{t}\hat{b}_i 
    = [\hat{H},\hat{b}_i]
    = \sum_j t_{ij}\hat{b}_j
    - U_i \hat{b}_i^\dagger\hat{b}_i\hat{b}_i
\end{equation}
where we obtain by the mean-field approximation 
\begin{equation}
    i\hbar\deriv{}{t}b_i
    = -\sum_{j}t_{ij} b_j
    + U_i |b_{i}|^2 b_{i}~,
\end{equation}
which is known as the nonlinear Schrödinger equation.
We solve this equation for the sawtooth edge system, diamond chain and the Stub unit cell with and without interaction $U_i$ in Fig.\ref{fig:classical_dynamics}. It turns out that, in contrast to the quantum limit, the interaction does not delocalize the light at the diamond chain, in accordance to Ref. \cite{diliberto2019nonlinear}, while the sawtooth and the Stub systems delocalizes weakly. Thus, the presence or absence of interaction does not lead to successful switching at the classical limit.


\begin{thebibliography}{87}%
\makeatletter
\providecommand \@ifxundefined [1]{%
 \@ifx{#1\undefined}
}%
\providecommand \@ifnum [1]{%
 \ifnum #1\expandafter \@firstoftwo
 \else \expandafter \@secondoftwo
 \fi
}%
\providecommand \@ifx [1]{%
 \ifx #1\expandafter \@firstoftwo
 \else \expandafter \@secondoftwo
 \fi
}%
\providecommand \natexlab [1]{#1}%
\providecommand \enquote  [1]{``#1''}%
\providecommand \bibnamefont  [1]{#1}%
\providecommand \bibfnamefont [1]{#1}%
\providecommand \citenamefont [1]{#1}%
\providecommand \href@noop [0]{\@secondoftwo}%
\providecommand \href [0]{\begingroup \@sanitize@url \@href}%
\providecommand \@href[1]{\@@startlink{#1}\@@href}%
\providecommand \@@href[1]{\endgroup#1\@@endlink}%
\providecommand \@sanitize@url [0]{\catcode `\\12\catcode `\$12\catcode
  `\&12\catcode `\#12\catcode `\^12\catcode `\_12\catcode `\%12\relax}%
\providecommand \@@startlink[1]{}%
\providecommand \@@endlink[0]{}%
\providecommand \url  [0]{\begingroup\@sanitize@url \@url }%
\providecommand \@url [1]{\endgroup\@href {#1}{\urlprefix }}%
\providecommand \urlprefix  [0]{URL }%
\providecommand \Eprint [0]{\href }%
\providecommand \doibase [0]{https://doi.org/}%
\providecommand \selectlanguage [0]{\@gobble}%
\providecommand \bibinfo  [0]{\@secondoftwo}%
\providecommand \bibfield  [0]{\@secondoftwo}%
\providecommand \translation [1]{[#1]}%
\providecommand \BibitemOpen [0]{}%
\providecommand \bibitemStop [0]{}%
\providecommand \bibitemNoStop [0]{.\EOS\space}%
\providecommand \EOS [0]{\spacefactor3000\relax}%
\providecommand \BibitemShut  [1]{\csname bibitem#1\endcsname}%
\let\auto@bib@innerbib\@empty
\bibitem [{\citenamefont {Almeida}\ \emph {et~al.}(2004)\citenamefont
  {Almeida}, \citenamefont {Barrios}, \citenamefont {Panepucci},\ and\
  \citenamefont {Lipson}}]{almeida2004all}%
  \BibitemOpen
  \bibfield  {author} {\bibinfo {author} {\bibfnamefont {V.~R.}\ \bibnamefont
  {Almeida}}, \bibinfo {author} {\bibfnamefont {C.~A.}\ \bibnamefont
  {Barrios}}, \bibinfo {author} {\bibfnamefont {R.~R.}\ \bibnamefont
  {Panepucci}},\ and\ \bibinfo {author} {\bibfnamefont {M.}~\bibnamefont
  {Lipson}},\ }\bibfield  {title} {\bibinfo {title} {{A}ll-optical control of
  light on a silicon chip},\ }\href {https://doi.org/10.1038/nature02921}
  {\bibfield  {journal} {\bibinfo  {journal} {Nature}\ }\textbf {\bibinfo
  {volume} {431}},\ \bibinfo {pages} {1081} (\bibinfo {year}
  {2004})}\BibitemShut {NoStop}%
\bibitem [{\citenamefont {Reed}\ \emph {et~al.}(2010)\citenamefont {Reed},
  \citenamefont {Mashanovich}, \citenamefont {Gardes},\ and\ \citenamefont
  {Thomson}}]{reed2010silicon}%
  \BibitemOpen
  \bibfield  {author} {\bibinfo {author} {\bibfnamefont {G.~T.}\ \bibnamefont
  {Reed}}, \bibinfo {author} {\bibfnamefont {G.}~\bibnamefont {Mashanovich}},
  \bibinfo {author} {\bibfnamefont {F.~Y.}\ \bibnamefont {Gardes}},\ and\
  \bibinfo {author} {\bibfnamefont {D.}~\bibnamefont {Thomson}},\ }\bibfield
  {title} {\bibinfo {title} {{S}ilicon optical modulators},\ }\href
  {https://doi.org/10.1038/nphoton.2010.179} {\bibfield  {journal} {\bibinfo
  {journal} {Nature Photonics}\ }\textbf {\bibinfo {volume} {4}},\ \bibinfo
  {pages} {518} (\bibinfo {year} {2010})}\BibitemShut {NoStop}%
\bibitem [{\citenamefont {Chai}\ \emph {et~al.}(2017)\citenamefont {Chai},
  \citenamefont {Hu}, \citenamefont {Wang}, \citenamefont {Niu}, \citenamefont
  {Xie},\ and\ \citenamefont {Gong}}]{chai2017ultrafast}%
  \BibitemOpen
  \bibfield  {author} {\bibinfo {author} {\bibfnamefont {Z.}~\bibnamefont
  {Chai}}, \bibinfo {author} {\bibfnamefont {X.}~\bibnamefont {Hu}}, \bibinfo
  {author} {\bibfnamefont {F.}~\bibnamefont {Wang}}, \bibinfo {author}
  {\bibfnamefont {X.}~\bibnamefont {Niu}}, \bibinfo {author} {\bibfnamefont
  {J.}~\bibnamefont {Xie}},\ and\ \bibinfo {author} {\bibfnamefont
  {Q.}~\bibnamefont {Gong}},\ }\bibfield  {title} {\bibinfo {title}
  {{U}ltrafast all-optical switching},\ }\href
  {https://doi.org/10.1002/adom.201600665} {\bibfield  {journal} {\bibinfo
  {journal} {Advanced Optical Materials}\ }\textbf {\bibinfo {volume} {5}},\
  \bibinfo {pages} {1600665} (\bibinfo {year} {2017})}\BibitemShut {NoStop}%
\bibitem [{\citenamefont {Sasikala}\ and\ \citenamefont
  {Chitra}(2018)}]{sasikala2018all}%
  \BibitemOpen
  \bibfield  {author} {\bibinfo {author} {\bibfnamefont {V.}~\bibnamefont
  {Sasikala}}\ and\ \bibinfo {author} {\bibfnamefont {K.}~\bibnamefont
  {Chitra}},\ }\bibfield  {title} {\bibinfo {title} {{A}ll optical switching
  and associated technologies: a review},\ }\href
  {https://doi.org/10.1007/s12596-018-0452-3} {\bibfield  {journal} {\bibinfo
  {journal} {Journal of Optics}\ }\textbf {\bibinfo {volume} {47}},\ \bibinfo
  {pages} {307} (\bibinfo {year} {2018})}\BibitemShut {NoStop}%
\bibitem [{\citenamefont {Chang}\ \emph {et~al.}(2014)\citenamefont {Chang},
  \citenamefont {Vuleti{\'c}},\ and\ \citenamefont {Lukin}}]{chang2014quantum}%
  \BibitemOpen
  \bibfield  {author} {\bibinfo {author} {\bibfnamefont {D.~E.}\ \bibnamefont
  {Chang}}, \bibinfo {author} {\bibfnamefont {V.}~\bibnamefont {Vuleti{\'c}}},\
  and\ \bibinfo {author} {\bibfnamefont {M.~D.}\ \bibnamefont {Lukin}},\
  }\bibfield  {title} {\bibinfo {title} {{{Q}uantum nonlinear optics—photon
  by photon}},\ }\href {https://doi.org/10.1038/nphoton.2014.192} {\bibfield
  {journal} {\bibinfo  {journal} {Nature Photonics}\ }\textbf {\bibinfo
  {volume} {8}},\ \bibinfo {pages} {685} (\bibinfo {year} {2014})}\BibitemShut
  {NoStop}%
\bibitem [{\citenamefont {Lukin}\ and\ \citenamefont
  {Imamo{\u{g}}lu}(2001)}]{lukin2001controlling}%
  \BibitemOpen
  \bibfield  {author} {\bibinfo {author} {\bibfnamefont {M.}~\bibnamefont
  {Lukin}}\ and\ \bibinfo {author} {\bibfnamefont {A.}~\bibnamefont
  {Imamo{\u{g}}lu}},\ }\bibfield  {title} {\bibinfo {title} {{Controlling
  photons using electromagnetically induced transparency}},\ }\href
  {https://doi.org/10.1038/35095000} {\bibfield  {journal} {\bibinfo  {journal}
  {Nature}\ }\textbf {\bibinfo {volume} {413}},\ \bibinfo {pages} {273}
  (\bibinfo {year} {2001})}\BibitemShut {NoStop}%
\bibitem [{\citenamefont {Pscherer}\ \emph {et~al.}(2021)\citenamefont
  {Pscherer}, \citenamefont {Meierhofer}, \citenamefont {Wang}, \citenamefont
  {Kelkar}, \citenamefont {Mart\'{\i}n-Cano}, \citenamefont {Utikal},
  \citenamefont {G\"otzinger},\ and\ \citenamefont
  {Sandoghdar}}]{pscherer2021single-molecule}%
  \BibitemOpen
  \bibfield  {author} {\bibinfo {author} {\bibfnamefont {A.}~\bibnamefont
  {Pscherer}}, \bibinfo {author} {\bibfnamefont {M.}~\bibnamefont
  {Meierhofer}}, \bibinfo {author} {\bibfnamefont {D.}~\bibnamefont {Wang}},
  \bibinfo {author} {\bibfnamefont {H.}~\bibnamefont {Kelkar}}, \bibinfo
  {author} {\bibfnamefont {D.}~\bibnamefont {Mart\'{\i}n-Cano}}, \bibinfo
  {author} {\bibfnamefont {T.}~\bibnamefont {Utikal}}, \bibinfo {author}
  {\bibfnamefont {S.}~\bibnamefont {G\"otzinger}},\ and\ \bibinfo {author}
  {\bibfnamefont {V.}~\bibnamefont {Sandoghdar}},\ }\bibfield  {title}
  {\bibinfo {title} {{S}ingle-{M}olecule {V}acuum {R}abi {S}plitting:
  {F}our-{W}ave {M}ixing and {O}ptical {S}witching at the {S}ingle-{P}hoton
  {L}evel},\ }\href {https://doi.org/10.1103/PhysRevLett.127.133603} {\bibfield
   {journal} {\bibinfo  {journal} {Physical Review Letters}\ }\textbf {\bibinfo
  {volume} {127}},\ \bibinfo {pages} {133603} (\bibinfo {year}
  {2021})}\BibitemShut {NoStop}%
\bibitem [{\citenamefont {Zasedatelev}\ \emph {et~al.}(2021)\citenamefont
  {Zasedatelev}, \citenamefont {Baranikov}, \citenamefont {Sannikov},
  \citenamefont {Urbonas}, \citenamefont {Scafirimuto}, \citenamefont
  {Shishkov}, \citenamefont {Andrianov}, \citenamefont {Lozovik}, \citenamefont
  {Scherf}, \citenamefont {St{\"o}ferle}, \citenamefont {Mahrt},\ and\
  \citenamefont {Lagoudakis}}]{zasedatelev2021}%
  \BibitemOpen
  \bibfield  {author} {\bibinfo {author} {\bibfnamefont {A.~V.}\ \bibnamefont
  {Zasedatelev}}, \bibinfo {author} {\bibfnamefont {A.~V.}\ \bibnamefont
  {Baranikov}}, \bibinfo {author} {\bibfnamefont {D.}~\bibnamefont {Sannikov}},
  \bibinfo {author} {\bibfnamefont {D.}~\bibnamefont {Urbonas}}, \bibinfo
  {author} {\bibfnamefont {F.}~\bibnamefont {Scafirimuto}}, \bibinfo {author}
  {\bibfnamefont {V.~Y.}\ \bibnamefont {Shishkov}}, \bibinfo {author}
  {\bibfnamefont {E.~S.}\ \bibnamefont {Andrianov}}, \bibinfo {author}
  {\bibfnamefont {Y.~E.}\ \bibnamefont {Lozovik}}, \bibinfo {author}
  {\bibfnamefont {U.}~\bibnamefont {Scherf}}, \bibinfo {author} {\bibfnamefont
  {T.}~\bibnamefont {St{\"o}ferle}}, \bibinfo {author} {\bibfnamefont {R.~F.}\
  \bibnamefont {Mahrt}},\ and\ \bibinfo {author} {\bibfnamefont {P.~G.}\
  \bibnamefont {Lagoudakis}},\ }\bibfield  {title} {\bibinfo {title}
  {{S}ingle-photon nonlinearity at room temperature},\ }\href
  {https://doi.org/10.1038/s41586-021-03866-9} {\bibfield  {journal} {\bibinfo
  {journal} {Nature}\ }\textbf {\bibinfo {volume} {597}},\ \bibinfo {pages}
  {493} (\bibinfo {year} {2021})}\BibitemShut {NoStop}%
\bibitem [{\citenamefont {Shomroni}\ \emph {et~al.}(2014)\citenamefont
  {Shomroni}, \citenamefont {Rosenblum}, \citenamefont {Lovsky}, \citenamefont
  {Bechler}, \citenamefont {Guendelman},\ and\ \citenamefont
  {Dayan}}]{shomroni2014all}%
  \BibitemOpen
  \bibfield  {author} {\bibinfo {author} {\bibfnamefont {I.}~\bibnamefont
  {Shomroni}}, \bibinfo {author} {\bibfnamefont {S.}~\bibnamefont {Rosenblum}},
  \bibinfo {author} {\bibfnamefont {Y.}~\bibnamefont {Lovsky}}, \bibinfo
  {author} {\bibfnamefont {O.}~\bibnamefont {Bechler}}, \bibinfo {author}
  {\bibfnamefont {G.}~\bibnamefont {Guendelman}},\ and\ \bibinfo {author}
  {\bibfnamefont {B.}~\bibnamefont {Dayan}},\ }\bibfield  {title} {\bibinfo
  {title} {{A}ll-optical routing of single photons by a one-atom switch
  controlled by a single photon},\ }\href
  {https://doi.org/10.1126/science.1254699} {\bibfield  {journal} {\bibinfo
  {journal} {Science}\ }\textbf {\bibinfo {volume} {345}},\ \bibinfo {pages}
  {903} (\bibinfo {year} {2014})}\BibitemShut {NoStop}%
\bibitem [{\citenamefont {Tiecke}\ \emph {et~al.}(2014)\citenamefont {Tiecke},
  \citenamefont {Thompson}, \citenamefont {de~Leon}, \citenamefont {Liu},
  \citenamefont {Vuleti{\'c}},\ and\ \citenamefont
  {Lukin}}]{tiecke2014nanophotonic}%
  \BibitemOpen
  \bibfield  {author} {\bibinfo {author} {\bibfnamefont {T.}~\bibnamefont
  {Tiecke}}, \bibinfo {author} {\bibfnamefont {J.~D.}\ \bibnamefont
  {Thompson}}, \bibinfo {author} {\bibfnamefont {N.~P.}\ \bibnamefont
  {de~Leon}}, \bibinfo {author} {\bibfnamefont {L.}~\bibnamefont {Liu}},
  \bibinfo {author} {\bibfnamefont {V.}~\bibnamefont {Vuleti{\'c}}},\ and\
  \bibinfo {author} {\bibfnamefont {M.~D.}\ \bibnamefont {Lukin}},\ }\bibfield
  {title} {\bibinfo {title} {{N}anophotonic quantum phase switch with a single
  atom},\ }\href {https://doi.org/10.1038/nature13188} {\bibfield  {journal}
  {\bibinfo  {journal} {Nature}\ }\textbf {\bibinfo {volume} {508}},\ \bibinfo
  {pages} {241} (\bibinfo {year} {2014})}\BibitemShut {NoStop}%
\bibitem [{\citenamefont {Hacker}\ \emph {et~al.}(2016)\citenamefont {Hacker},
  \citenamefont {Welte}, \citenamefont {Rempe},\ and\ \citenamefont
  {Ritter}}]{hacker2016photon}%
  \BibitemOpen
  \bibfield  {author} {\bibinfo {author} {\bibfnamefont {B.}~\bibnamefont
  {Hacker}}, \bibinfo {author} {\bibfnamefont {S.}~\bibnamefont {Welte}},
  \bibinfo {author} {\bibfnamefont {G.}~\bibnamefont {Rempe}},\ and\ \bibinfo
  {author} {\bibfnamefont {S.}~\bibnamefont {Ritter}},\ }\bibfield  {title}
  {\bibinfo {title} {{A} photon--photon quantum gate based on a single atom in
  an optical resonator},\ }\href {https://doi.org/10.1038/nature18592}
  {\bibfield  {journal} {\bibinfo  {journal} {Nature}\ }\textbf {\bibinfo
  {volume} {536}},\ \bibinfo {pages} {193} (\bibinfo {year}
  {2016})}\BibitemShut {NoStop}%
\bibitem [{\citenamefont {Volz}\ \emph {et~al.}(2012)\citenamefont {Volz},
  \citenamefont {Reinhard}, \citenamefont {Winger}, \citenamefont {Badolato},
  \citenamefont {Hennessy}, \citenamefont {Hu},\ and\ \citenamefont
  {Imamo{\u{g}}lu}}]{volz2012ultrafast}%
  \BibitemOpen
  \bibfield  {author} {\bibinfo {author} {\bibfnamefont {T.}~\bibnamefont
  {Volz}}, \bibinfo {author} {\bibfnamefont {A.}~\bibnamefont {Reinhard}},
  \bibinfo {author} {\bibfnamefont {M.}~\bibnamefont {Winger}}, \bibinfo
  {author} {\bibfnamefont {A.}~\bibnamefont {Badolato}}, \bibinfo {author}
  {\bibfnamefont {K.~J.}\ \bibnamefont {Hennessy}}, \bibinfo {author}
  {\bibfnamefont {E.~L.}\ \bibnamefont {Hu}},\ and\ \bibinfo {author}
  {\bibfnamefont {A.}~\bibnamefont {Imamo{\u{g}}lu}},\ }\bibfield  {title}
  {\bibinfo {title} {{U}ltrafast all-optical switching by single photons},\
  }\href {https://doi.org/10.1038/nphoton.2012.181} {\bibfield  {journal}
  {\bibinfo  {journal} {Nature Photonics}\ }\textbf {\bibinfo {volume} {6}},\
  \bibinfo {pages} {605} (\bibinfo {year} {2012})}\BibitemShut {NoStop}%
\bibitem [{\citenamefont {Giesz}\ \emph {et~al.}(2016)\citenamefont {Giesz},
  \citenamefont {Somaschi}, \citenamefont {Hornecker}, \citenamefont {Grange},
  \citenamefont {Reznychenko}, \citenamefont {De~Santis}, \citenamefont
  {Demory}, \citenamefont {Gomez}, \citenamefont {Sagnes}, \citenamefont
  {Lemaitre} \emph {et~al.}}]{giesz2016coherent}%
  \BibitemOpen
  \bibfield  {author} {\bibinfo {author} {\bibfnamefont {V.}~\bibnamefont
  {Giesz}}, \bibinfo {author} {\bibfnamefont {N.}~\bibnamefont {Somaschi}},
  \bibinfo {author} {\bibfnamefont {G.}~\bibnamefont {Hornecker}}, \bibinfo
  {author} {\bibfnamefont {T.}~\bibnamefont {Grange}}, \bibinfo {author}
  {\bibfnamefont {B.}~\bibnamefont {Reznychenko}}, \bibinfo {author}
  {\bibfnamefont {L.}~\bibnamefont {De~Santis}}, \bibinfo {author}
  {\bibfnamefont {J.}~\bibnamefont {Demory}}, \bibinfo {author} {\bibfnamefont
  {C.}~\bibnamefont {Gomez}}, \bibinfo {author} {\bibfnamefont
  {I.}~\bibnamefont {Sagnes}}, \bibinfo {author} {\bibfnamefont
  {A.}~\bibnamefont {Lemaitre}}, \emph {et~al.},\ }\bibfield  {title} {\bibinfo
  {title} {{C}oherent manipulation of a solid-state artificial atom with few
  photons},\ }\href {https://doi.org/10.1038/ncomms11986} {\bibfield  {journal}
  {\bibinfo  {journal} {Nature Communications}\ }\textbf {\bibinfo {volume}
  {7}},\ \bibinfo {pages} {11986} (\bibinfo {year} {2016})}\BibitemShut
  {NoStop}%
\bibitem [{\citenamefont {Dietrich}\ \emph {et~al.}(2016)\citenamefont
  {Dietrich}, \citenamefont {Fiore}, \citenamefont {Thompson}, \citenamefont
  {Kamp},\ and\ \citenamefont {H{\"o}fling}}]{dietrich2016gaas}%
  \BibitemOpen
  \bibfield  {author} {\bibinfo {author} {\bibfnamefont {C.~P.}\ \bibnamefont
  {Dietrich}}, \bibinfo {author} {\bibfnamefont {A.}~\bibnamefont {Fiore}},
  \bibinfo {author} {\bibfnamefont {M.~G.}\ \bibnamefont {Thompson}}, \bibinfo
  {author} {\bibfnamefont {M.}~\bibnamefont {Kamp}},\ and\ \bibinfo {author}
  {\bibfnamefont {S.}~\bibnamefont {H{\"o}fling}},\ }\bibfield  {title}
  {\bibinfo {title} {{G}a{A}s integrated quantum photonics: {T}owards compact
  and multi-functional quantum photonic integrated circuits},\ }\href
  {https://doi.org/10.1002/lpor.201500321} {\bibfield  {journal} {\bibinfo
  {journal} {Laser \& Photonics Reviews}\ }\textbf {\bibinfo {volume} {10}},\
  \bibinfo {pages} {870} (\bibinfo {year} {2016})}\BibitemShut {NoStop}%
\bibitem [{\citenamefont {Sun}\ \emph {et~al.}(2018)\citenamefont {Sun},
  \citenamefont {Kim}, \citenamefont {Luo}, \citenamefont {Solomon},\ and\
  \citenamefont {Waks}}]{sun2018}%
  \BibitemOpen
  \bibfield  {author} {\bibinfo {author} {\bibfnamefont {S.}~\bibnamefont
  {Sun}}, \bibinfo {author} {\bibfnamefont {H.}~\bibnamefont {Kim}}, \bibinfo
  {author} {\bibfnamefont {Z.}~\bibnamefont {Luo}}, \bibinfo {author}
  {\bibfnamefont {G.~S.}\ \bibnamefont {Solomon}},\ and\ \bibinfo {author}
  {\bibfnamefont {E.}~\bibnamefont {Waks}},\ }\bibfield  {title} {\bibinfo
  {title} {{A} single-photon switch and transistor enabled by a solid-state
  quantum memory},\ }\href {https://doi.org/10.1126/science.aat3581} {\bibfield
   {journal} {\bibinfo  {journal} {Science}\ }\textbf {\bibinfo {volume}
  {361}},\ \bibinfo {pages} {57} (\bibinfo {year} {2018})}\BibitemShut
  {NoStop}%
\bibitem [{\citenamefont {Peyronel}\ \emph {et~al.}(2012)\citenamefont
  {Peyronel}, \citenamefont {Firstenberg}, \citenamefont {Liang}, \citenamefont
  {Hofferberth}, \citenamefont {Gorshkov}, \citenamefont {Pohl}, \citenamefont
  {Lukin},\ and\ \citenamefont {Vuleti{\'c}}}]{peyronel2012quantum}%
  \BibitemOpen
  \bibfield  {author} {\bibinfo {author} {\bibfnamefont {T.}~\bibnamefont
  {Peyronel}}, \bibinfo {author} {\bibfnamefont {O.}~\bibnamefont
  {Firstenberg}}, \bibinfo {author} {\bibfnamefont {Q.-Y.}\ \bibnamefont
  {Liang}}, \bibinfo {author} {\bibfnamefont {S.}~\bibnamefont {Hofferberth}},
  \bibinfo {author} {\bibfnamefont {A.~V.}\ \bibnamefont {Gorshkov}}, \bibinfo
  {author} {\bibfnamefont {T.}~\bibnamefont {Pohl}}, \bibinfo {author}
  {\bibfnamefont {M.~D.}\ \bibnamefont {Lukin}},\ and\ \bibinfo {author}
  {\bibfnamefont {V.}~\bibnamefont {Vuleti{\'c}}},\ }\bibfield  {title}
  {\bibinfo {title} {{Q}uantum nonlinear optics with single photons enabled by
  strongly interacting atoms},\ }\href {https://doi.org/10.1038/nature11361}
  {\bibfield  {journal} {\bibinfo  {journal} {Nature}\ }\textbf {\bibinfo
  {volume} {488}},\ \bibinfo {pages} {57} (\bibinfo {year} {2012})}\BibitemShut
  {NoStop}%
\bibitem [{\citenamefont {Chen}\ \emph {et~al.}(2013)\citenamefont {Chen},
  \citenamefont {Beck}, \citenamefont {B{\"u}cker}, \citenamefont {Gullans},
  \citenamefont {Lukin}, \citenamefont {Tanji-Suzuki},\ and\ \citenamefont
  {Vuleti{\'c}}}]{chen2013all}%
  \BibitemOpen
  \bibfield  {author} {\bibinfo {author} {\bibfnamefont {W.}~\bibnamefont
  {Chen}}, \bibinfo {author} {\bibfnamefont {K.~M.}\ \bibnamefont {Beck}},
  \bibinfo {author} {\bibfnamefont {R.}~\bibnamefont {B{\"u}cker}}, \bibinfo
  {author} {\bibfnamefont {M.}~\bibnamefont {Gullans}}, \bibinfo {author}
  {\bibfnamefont {M.~D.}\ \bibnamefont {Lukin}}, \bibinfo {author}
  {\bibfnamefont {H.}~\bibnamefont {Tanji-Suzuki}},\ and\ \bibinfo {author}
  {\bibfnamefont {V.}~\bibnamefont {Vuleti{\'c}}},\ }\bibfield  {title}
  {\bibinfo {title} {{All-optical switch and transistor gated by one stored
  photon}},\ }\href {https://doi.org/10.1126/science.1238169} {\bibfield
  {journal} {\bibinfo  {journal} {Science}\ }\textbf {\bibinfo {volume}
  {341}},\ \bibinfo {pages} {768} (\bibinfo {year} {2013})}\BibitemShut
  {NoStop}%
\bibitem [{\citenamefont {Baur}\ \emph {et~al.}(2014)\citenamefont {Baur},
  \citenamefont {Tiarks}, \citenamefont {Rempe},\ and\ \citenamefont
  {D{\"u}rr}}]{baur2014single}%
  \BibitemOpen
  \bibfield  {author} {\bibinfo {author} {\bibfnamefont {S.}~\bibnamefont
  {Baur}}, \bibinfo {author} {\bibfnamefont {D.}~\bibnamefont {Tiarks}},
  \bibinfo {author} {\bibfnamefont {G.}~\bibnamefont {Rempe}},\ and\ \bibinfo
  {author} {\bibfnamefont {S.}~\bibnamefont {D{\"u}rr}},\ }\bibfield  {title}
  {\bibinfo {title} {{Single-photon switch based on Rydberg blockade}},\ }\href
  {https://doi.org/10.1103/PhysRevLett.112.073901} {\bibfield  {journal}
  {\bibinfo  {journal} {Physical Review Letters}\ }\textbf {\bibinfo {volume}
  {112}},\ \bibinfo {pages} {073901} (\bibinfo {year} {2014})}\BibitemShut
  {NoStop}%
\bibitem [{\citenamefont {Gorniaczyk}\ \emph {et~al.}(2014)\citenamefont
  {Gorniaczyk}, \citenamefont {Tresp}, \citenamefont {Schmidt}, \citenamefont
  {Fedder},\ and\ \citenamefont {Hofferberth}}]{gorniaczyk2014single}%
  \BibitemOpen
  \bibfield  {author} {\bibinfo {author} {\bibfnamefont {H.}~\bibnamefont
  {Gorniaczyk}}, \bibinfo {author} {\bibfnamefont {C.}~\bibnamefont {Tresp}},
  \bibinfo {author} {\bibfnamefont {J.}~\bibnamefont {Schmidt}}, \bibinfo
  {author} {\bibfnamefont {H.}~\bibnamefont {Fedder}},\ and\ \bibinfo {author}
  {\bibfnamefont {S.}~\bibnamefont {Hofferberth}},\ }\bibfield  {title}
  {\bibinfo {title} {{S}ingle-photon transistor mediated by interstate
  {R}ydberg interactions},\ }\href
  {https://doi.org/10.1103/PhysRevLett.113.053601} {\bibfield  {journal}
  {\bibinfo  {journal} {Physical Review Letters}\ }\textbf {\bibinfo {volume}
  {113}},\ \bibinfo {pages} {053601} (\bibinfo {year} {2014})}\BibitemShut
  {NoStop}%
\bibitem [{\citenamefont {Mu{\~n}oz-Matutano}\ \emph
  {et~al.}(2020)\citenamefont {Mu{\~n}oz-Matutano}, \citenamefont {Johnsson},
  \citenamefont {Mart{\'\i}nez-Pastor}, \citenamefont {Rivas~G{\'o}ngora},
  \citenamefont {Seravalli}, \citenamefont {Trevisi}, \citenamefont {Frigeri},
  \citenamefont {Volz},\ and\ \citenamefont {Gurioli}}]{munoz2020all}%
  \BibitemOpen
  \bibfield  {author} {\bibinfo {author} {\bibfnamefont {G.}~\bibnamefont
  {Mu{\~n}oz-Matutano}}, \bibinfo {author} {\bibfnamefont {M.}~\bibnamefont
  {Johnsson}}, \bibinfo {author} {\bibfnamefont {J.}~\bibnamefont
  {Mart{\'\i}nez-Pastor}}, \bibinfo {author} {\bibfnamefont {D.}~\bibnamefont
  {Rivas~G{\'o}ngora}}, \bibinfo {author} {\bibfnamefont {L.}~\bibnamefont
  {Seravalli}}, \bibinfo {author} {\bibfnamefont {G.}~\bibnamefont {Trevisi}},
  \bibinfo {author} {\bibfnamefont {P.}~\bibnamefont {Frigeri}}, \bibinfo
  {author} {\bibfnamefont {T.}~\bibnamefont {Volz}},\ and\ \bibinfo {author}
  {\bibfnamefont {M.}~\bibnamefont {Gurioli}},\ }\bibfield  {title} {\bibinfo
  {title} {{A}ll optical switching of a single photon stream by excitonic
  depletion},\ }\href {https://doi.org/10.1038/s42005-020-0292-8} {\bibfield
  {journal} {\bibinfo  {journal} {Communications Physics}\ }\textbf {\bibinfo
  {volume} {3}},\ \bibinfo {pages} {29} (\bibinfo {year} {2020})}\BibitemShut
  {NoStop}%
\bibitem [{\citenamefont {Liao}\ \emph {et~al.}(2009)\citenamefont {Liao},
  \citenamefont {Huang}, \citenamefont {Liu}, \citenamefont {Kuang},\ and\
  \citenamefont {Sun}}]{liao2009}%
  \BibitemOpen
  \bibfield  {author} {\bibinfo {author} {\bibfnamefont {J.-Q.}\ \bibnamefont
  {Liao}}, \bibinfo {author} {\bibfnamefont {J.-F.}\ \bibnamefont {Huang}},
  \bibinfo {author} {\bibfnamefont {Y.-x.}\ \bibnamefont {Liu}}, \bibinfo
  {author} {\bibfnamefont {L.-M.}\ \bibnamefont {Kuang}},\ and\ \bibinfo
  {author} {\bibfnamefont {C.~P.}\ \bibnamefont {Sun}},\ }\bibfield  {title}
  {\bibinfo {title} {{Q}uantum switch for single-photon transport in a coupled
  superconducting transmission-line-resonator array},\ }\href
  {https://doi.org/10.1103/PhysRevA.80.014301} {\bibfield  {journal} {\bibinfo
  {journal} {Physical Review A}\ }\textbf {\bibinfo {volume} {80}},\ \bibinfo
  {pages} {014301} (\bibinfo {year} {2009})}\BibitemShut {NoStop}%
\bibitem [{\citenamefont {Chang}\ \emph {et~al.}(2007)\citenamefont {Chang},
  \citenamefont {S{\o}rensen}, \citenamefont {Demler},\ and\ \citenamefont
  {Lukin}}]{chang2007single}%
  \BibitemOpen
  \bibfield  {author} {\bibinfo {author} {\bibfnamefont {D.~E.}\ \bibnamefont
  {Chang}}, \bibinfo {author} {\bibfnamefont {A.~S.}\ \bibnamefont
  {S{\o}rensen}}, \bibinfo {author} {\bibfnamefont {E.~A.}\ \bibnamefont
  {Demler}},\ and\ \bibinfo {author} {\bibfnamefont {M.~D.}\ \bibnamefont
  {Lukin}},\ }\bibfield  {title} {\bibinfo {title} {{A single-photon transistor
  using nanoscale surface plasmons}},\ }\href
  {https://doi.org/10.1038/nphys708} {\bibfield  {journal} {\bibinfo  {journal}
  {Nature physics}\ }\textbf {\bibinfo {volume} {3}},\ \bibinfo {pages} {807}
  (\bibinfo {year} {2007})}\BibitemShut {NoStop}%
\bibitem [{\citenamefont {Bermel}\ \emph {et~al.}(2006)\citenamefont {Bermel},
  \citenamefont {Rodriguez}, \citenamefont {Johnson}, \citenamefont
  {Joannopoulos},\ and\ \citenamefont {Solja\ifmmode \check{c}\else
  \v{c}\fi{}i\ifmmode~\acute{c}\else \'{c}\fi{}}}]{bermel2006}%
  \BibitemOpen
  \bibfield  {author} {\bibinfo {author} {\bibfnamefont {P.}~\bibnamefont
  {Bermel}}, \bibinfo {author} {\bibfnamefont {A.}~\bibnamefont {Rodriguez}},
  \bibinfo {author} {\bibfnamefont {S.~G.}\ \bibnamefont {Johnson}}, \bibinfo
  {author} {\bibfnamefont {J.~D.}\ \bibnamefont {Joannopoulos}},\ and\ \bibinfo
  {author} {\bibfnamefont {M.}~\bibnamefont {Solja\ifmmode \check{c}\else
  \v{c}\fi{}i\ifmmode~\acute{c}\else \'{c}\fi{}}},\ }\bibfield  {title}
  {\bibinfo {title} {{S}ingle-photon all-optical switching using
  waveguide-cavity quantum electrodynamics},\ }\href
  {https://doi.org/10.1103/PhysRevA.74.043818} {\bibfield  {journal} {\bibinfo
  {journal} {Physical Review A}\ }\textbf {\bibinfo {volume} {74}},\ \bibinfo
  {pages} {043818} (\bibinfo {year} {2006})}\BibitemShut {NoStop}%
\bibitem [{\citenamefont {Leykam}\ \emph {et~al.}(2018)\citenamefont {Leykam},
  \citenamefont {Andreanov},\ and\ \citenamefont {Flach}}]{leykam2018}%
  \BibitemOpen
  \bibfield  {author} {\bibinfo {author} {\bibfnamefont {D.}~\bibnamefont
  {Leykam}}, \bibinfo {author} {\bibfnamefont {A.}~\bibnamefont {Andreanov}},\
  and\ \bibinfo {author} {\bibfnamefont {S.}~\bibnamefont {Flach}},\ }\bibfield
   {title} {\bibinfo {title} {{A}rtificial flat band systems: from lattice
  models to experiments},\ }\href
  {https://doi.org/10.1080/23746149.2018.1473052} {\bibfield  {journal}
  {\bibinfo  {journal} {Advances in Physics: X}\ }\textbf {\bibinfo {volume}
  {3}},\ \bibinfo {pages} {1473052} (\bibinfo {year} {2018})}\BibitemShut
  {NoStop}%
\bibitem [{\citenamefont {Tovmasyan}\ \emph {et~al.}(2018)\citenamefont
  {Tovmasyan}, \citenamefont {Peotta}, \citenamefont {Liang}, \citenamefont
  {T\"orm\"a},\ and\ \citenamefont {Huber}}]{tovmasyan2018preformed}%
  \BibitemOpen
  \bibfield  {author} {\bibinfo {author} {\bibfnamefont {M.}~\bibnamefont
  {Tovmasyan}}, \bibinfo {author} {\bibfnamefont {S.}~\bibnamefont {Peotta}},
  \bibinfo {author} {\bibfnamefont {L.}~\bibnamefont {Liang}}, \bibinfo
  {author} {\bibfnamefont {P.}~\bibnamefont {T\"orm\"a}},\ and\ \bibinfo
  {author} {\bibfnamefont {S.~D.}\ \bibnamefont {Huber}},\ }\bibfield  {title}
  {\bibinfo {title} {{P}reformed pairs in flat {B}loch bands},\ }\href
  {https://doi.org/10.1103/PhysRevB.98.134513} {\bibfield  {journal} {\bibinfo
  {journal} {Physical Review B}\ }\textbf {\bibinfo {volume} {98}},\ \bibinfo
  {pages} {134513} (\bibinfo {year} {2018})}\BibitemShut {NoStop}%
\bibitem [{\citenamefont {T{\"o}rm{\"a}}\ \emph {et~al.}(2018)\citenamefont
  {T{\"o}rm{\"a}}, \citenamefont {Liang},\ and\ \citenamefont
  {Peotta}}]{torma2018quantum}%
  \BibitemOpen
  \bibfield  {author} {\bibinfo {author} {\bibfnamefont {P.}~\bibnamefont
  {T{\"o}rm{\"a}}}, \bibinfo {author} {\bibfnamefont {L.}~\bibnamefont
  {Liang}},\ and\ \bibinfo {author} {\bibfnamefont {S.}~\bibnamefont
  {Peotta}},\ }\bibfield  {title} {\bibinfo {title} {Quantum metric and
  effective mass of a two-body bound state in a flat band},\ }\href
  {https://doi.org/10.1103/PhysRevB.98.220511} {\bibfield  {journal} {\bibinfo
  {journal} {Physical Review B}\ }\textbf {\bibinfo {volume} {98}},\ \bibinfo
  {pages} {220511} (\bibinfo {year} {2018})}\BibitemShut {NoStop}%
\bibitem [{\citenamefont {Pyykk\"onen}\ \emph {et~al.}(2023)\citenamefont
  {Pyykk\"onen}, \citenamefont {Peotta},\ and\ \citenamefont
  {T\"orm\"a}}]{pyykkonen2023suppression}%
  \BibitemOpen
  \bibfield  {author} {\bibinfo {author} {\bibfnamefont {V.~A.~J.}\
  \bibnamefont {Pyykk\"onen}}, \bibinfo {author} {\bibfnamefont
  {S.}~\bibnamefont {Peotta}},\ and\ \bibinfo {author} {\bibfnamefont
  {P.}~\bibnamefont {T\"orm\"a}},\ }\bibfield  {title} {\bibinfo {title}
  {{S}uppression of {N}onequilibrium {Q}uasiparticle {T}ransport in
  {F}lat-{B}and {S}uperconductors},\ }\href
  {https://doi.org/10.1103/PhysRevLett.130.216003} {\bibfield  {journal}
  {\bibinfo  {journal} {Physical Review Letters}\ }\textbf {\bibinfo {volume}
  {130}},\ \bibinfo {pages} {216003} (\bibinfo {year} {2023})}\BibitemShut
  {NoStop}%
\bibitem [{\citenamefont {Kopnin}\ \emph {et~al.}(2011)\citenamefont {Kopnin},
  \citenamefont {Heikkil\"a},\ and\ \citenamefont {Volovik}}]{kopnin2011high}%
  \BibitemOpen
  \bibfield  {author} {\bibinfo {author} {\bibfnamefont {N.~B.}\ \bibnamefont
  {Kopnin}}, \bibinfo {author} {\bibfnamefont {T.~T.}\ \bibnamefont
  {Heikkil\"a}},\ and\ \bibinfo {author} {\bibfnamefont {G.~E.}\ \bibnamefont
  {Volovik}},\ }\bibfield  {title} {\bibinfo {title} {{H}igh-temperature
  surface superconductivity in topological flat-band systems},\ }\href
  {https://doi.org/10.1103/PhysRevB.83.220503} {\bibfield  {journal} {\bibinfo
  {journal} {Physical Review B}\ }\textbf {\bibinfo {volume} {83}},\ \bibinfo
  {pages} {220503} (\bibinfo {year} {2011})}\BibitemShut {NoStop}%
\bibitem [{\citenamefont {Peotta}\ and\ \citenamefont
  {T{\"o}rm{\"a}}(2015)}]{peotta2015superfluidity}%
  \BibitemOpen
  \bibfield  {author} {\bibinfo {author} {\bibfnamefont {S.}~\bibnamefont
  {Peotta}}\ and\ \bibinfo {author} {\bibfnamefont {P.}~\bibnamefont
  {T{\"o}rm{\"a}}},\ }\bibfield  {title} {\bibinfo {title} {{S}uperfluidity in
  topologically nontrivial flat bands},\ }\href
  {https://doi.org/10.1038/ncomms9944} {\bibfield  {journal} {\bibinfo
  {journal} {Nature Communications}\ }\textbf {\bibinfo {volume} {6}},\
  \bibinfo {pages} {8944} (\bibinfo {year} {2015})}\BibitemShut {NoStop}%
\bibitem [{\citenamefont {Julku}\ \emph {et~al.}(2016)\citenamefont {Julku},
  \citenamefont {Peotta}, \citenamefont {Vanhala}, \citenamefont {Kim},\ and\
  \citenamefont {T\"orm\"a}}]{julku2016geometric}%
  \BibitemOpen
  \bibfield  {author} {\bibinfo {author} {\bibfnamefont {A.}~\bibnamefont
  {Julku}}, \bibinfo {author} {\bibfnamefont {S.}~\bibnamefont {Peotta}},
  \bibinfo {author} {\bibfnamefont {T.~I.}\ \bibnamefont {Vanhala}}, \bibinfo
  {author} {\bibfnamefont {D.-H.}\ \bibnamefont {Kim}},\ and\ \bibinfo {author}
  {\bibfnamefont {P.}~\bibnamefont {T\"orm\"a}},\ }\bibfield  {title} {\bibinfo
  {title} {{G}eometric {O}rigin of {S}uperfluidity in the {L}ieb-{L}attice
  {F}lat {B}and},\ }\href {https://doi.org/10.1103/PhysRevLett.117.045303}
  {\bibfield  {journal} {\bibinfo  {journal} {Physical Review Letters}\
  }\textbf {\bibinfo {volume} {117}},\ \bibinfo {pages} {045303} (\bibinfo
  {year} {2016})}\BibitemShut {NoStop}%
\bibitem [{\citenamefont {Liang}\ \emph {et~al.}(2017)\citenamefont {Liang},
  \citenamefont {Vanhala}, \citenamefont {Peotta}, \citenamefont {Siro},
  \citenamefont {Harju},\ and\ \citenamefont {T\"orm\"a}}]{liang2017band}%
  \BibitemOpen
  \bibfield  {author} {\bibinfo {author} {\bibfnamefont {L.}~\bibnamefont
  {Liang}}, \bibinfo {author} {\bibfnamefont {T.~I.}\ \bibnamefont {Vanhala}},
  \bibinfo {author} {\bibfnamefont {S.}~\bibnamefont {Peotta}}, \bibinfo
  {author} {\bibfnamefont {T.}~\bibnamefont {Siro}}, \bibinfo {author}
  {\bibfnamefont {A.}~\bibnamefont {Harju}},\ and\ \bibinfo {author}
  {\bibfnamefont {P.}~\bibnamefont {T\"orm\"a}},\ }\bibfield  {title} {\bibinfo
  {title} {{B}and geometry, {B}erry curvature, and superfluid weight},\ }\href
  {https://doi.org/10.1103/PhysRevB.95.024515} {\bibfield  {journal} {\bibinfo
  {journal} {Physical Review B}\ }\textbf {\bibinfo {volume} {95}},\ \bibinfo
  {pages} {024515} (\bibinfo {year} {2017})}\BibitemShut {NoStop}%
\bibitem [{\citenamefont {Törmä}\ \emph {et~al.}(2022)\citenamefont
  {Törmä}, \citenamefont {Peotta},\ and\ \citenamefont
  {Bernevig}}]{torma2022}%
  \BibitemOpen
  \bibfield  {author} {\bibinfo {author} {\bibfnamefont {P.}~\bibnamefont
  {Törmä}}, \bibinfo {author} {\bibfnamefont {S.}~\bibnamefont {Peotta}},\
  and\ \bibinfo {author} {\bibfnamefont {B.~A.}\ \bibnamefont {Bernevig}},\
  }\bibfield  {title} {\bibinfo {title} {{S}uperconductivity, superfluidity and
  quantum geometry in twisted multilayer systems},\ }\href
  {https://doi.org/10.1038/s42254-022-00466-y} {\bibfield  {journal} {\bibinfo
  {journal} {Nature Reviews Physics}\ }\textbf {\bibinfo {volume} {4}},\
  \bibinfo {pages} {528} (\bibinfo {year} {2022})}\BibitemShut {NoStop}%
\bibitem [{\citenamefont {Vidal}\ \emph {et~al.}(1998)\citenamefont {Vidal},
  \citenamefont {Mosseri},\ and\ \citenamefont {Dou\ifmmode~\mbox{\c{c}}\else
  \c{c}\fi{}ot}}]{vidal1998aharonov}%
  \BibitemOpen
  \bibfield  {author} {\bibinfo {author} {\bibfnamefont {J.}~\bibnamefont
  {Vidal}}, \bibinfo {author} {\bibfnamefont {R.}~\bibnamefont {Mosseri}},\
  and\ \bibinfo {author} {\bibfnamefont {B.}~\bibnamefont
  {Dou\ifmmode~\mbox{\c{c}}\else \c{c}\fi{}ot}},\ }\bibfield  {title} {\bibinfo
  {title} {{A}haronov-{B}ohm {C}ages in {T}wo-{D}imensional {S}tructures},\
  }\href {https://doi.org/10.1103/PhysRevLett.81.5888} {\bibfield  {journal}
  {\bibinfo  {journal} {Physical Review Letters}\ }\textbf {\bibinfo {volume}
  {81}},\ \bibinfo {pages} {5888} (\bibinfo {year} {1998})}\BibitemShut
  {NoStop}%
\bibitem [{\citenamefont {Vidal}\ \emph {et~al.}(2000)\citenamefont {Vidal},
  \citenamefont {Dou{\c{c}}ot}, \citenamefont {Mosseri},\ and\ \citenamefont
  {Butaud}}]{vidal2000interaction}%
  \BibitemOpen
  \bibfield  {author} {\bibinfo {author} {\bibfnamefont {J.}~\bibnamefont
  {Vidal}}, \bibinfo {author} {\bibfnamefont {B.}~\bibnamefont {Dou{\c{c}}ot}},
  \bibinfo {author} {\bibfnamefont {R.}~\bibnamefont {Mosseri}},\ and\ \bibinfo
  {author} {\bibfnamefont {P.}~\bibnamefont {Butaud}},\ }\bibfield  {title}
  {\bibinfo {title} {{I}nteraction induced delocalization for two particles in
  a periodic potential},\ }\href {https://doi.org/10.1103/PhysRevLett.85.3906}
  {\bibfield  {journal} {\bibinfo  {journal} {Physical Review Letters}\
  }\textbf {\bibinfo {volume} {85}},\ \bibinfo {pages} {3906} (\bibinfo {year}
  {2000})}\BibitemShut {NoStop}%
\bibitem [{\citenamefont {Abilio}\ \emph {et~al.}(1999)\citenamefont {Abilio},
  \citenamefont {Butaud}, \citenamefont {Fournier}, \citenamefont {Pannetier},
  \citenamefont {Vidal}, \citenamefont {Tedesco},\ and\ \citenamefont
  {Dalzotto}}]{abilio1999magnetic}%
  \BibitemOpen
  \bibfield  {author} {\bibinfo {author} {\bibfnamefont {C.~C.}\ \bibnamefont
  {Abilio}}, \bibinfo {author} {\bibfnamefont {P.}~\bibnamefont {Butaud}},
  \bibinfo {author} {\bibfnamefont {T.}~\bibnamefont {Fournier}}, \bibinfo
  {author} {\bibfnamefont {B.}~\bibnamefont {Pannetier}}, \bibinfo {author}
  {\bibfnamefont {J.}~\bibnamefont {Vidal}}, \bibinfo {author} {\bibfnamefont
  {S.}~\bibnamefont {Tedesco}},\ and\ \bibinfo {author} {\bibfnamefont
  {B.}~\bibnamefont {Dalzotto}},\ }\bibfield  {title} {\bibinfo {title}
  {{M}agnetic {F}ield {I}nduced {L}ocalization in a {T}wo-{D}imensional
  {S}uperconducting {W}ire {N}etwork},\ }\href
  {https://doi.org/10.1103/PhysRevLett.83.5102} {\bibfield  {journal} {\bibinfo
   {journal} {Physical Review Letters}\ }\textbf {\bibinfo {volume} {83}},\
  \bibinfo {pages} {5102} (\bibinfo {year} {1999})}\BibitemShut {NoStop}%
\bibitem [{\citenamefont {Alaeian}\ \emph {et~al.}(2019)\citenamefont
  {Alaeian}, \citenamefont {Chang}, \citenamefont {Moghaddam}, \citenamefont
  {Wilson}, \citenamefont {Solano},\ and\ \citenamefont
  {Rico}}]{alaeian2019creating}%
  \BibitemOpen
  \bibfield  {author} {\bibinfo {author} {\bibfnamefont {H.}~\bibnamefont
  {Alaeian}}, \bibinfo {author} {\bibfnamefont {C.~W.~S.}\ \bibnamefont
  {Chang}}, \bibinfo {author} {\bibfnamefont {M.~V.}\ \bibnamefont
  {Moghaddam}}, \bibinfo {author} {\bibfnamefont {C.~M.}\ \bibnamefont
  {Wilson}}, \bibinfo {author} {\bibfnamefont {E.}~\bibnamefont {Solano}},\
  and\ \bibinfo {author} {\bibfnamefont {E.}~\bibnamefont {Rico}},\ }\bibfield
  {title} {\bibinfo {title} {{Creating lattice gauge potentials in circuit QED:
  The bosonic Creutz ladder}},\ }\href
  {https://doi.org/10.1103/PhysRevA.99.053834} {\bibfield  {journal} {\bibinfo
  {journal} {Physical Review A}\ }\textbf {\bibinfo {volume} {99}},\ \bibinfo
  {pages} {053834} (\bibinfo {year} {2019})}\BibitemShut {NoStop}%
\bibitem [{\citenamefont {Hung}\ \emph {et~al.}(2021)\citenamefont {Hung},
  \citenamefont {Busnaina}, \citenamefont {Chang}, \citenamefont {Vadiraj},
  \citenamefont {Nsanzineza}, \citenamefont {Solano}, \citenamefont {Alaeian},
  \citenamefont {Rico},\ and\ \citenamefont {Wilson}}]{hung2021quantum}%
  \BibitemOpen
  \bibfield  {author} {\bibinfo {author} {\bibfnamefont {J.~S.}\ \bibnamefont
  {Hung}}, \bibinfo {author} {\bibfnamefont {J.}~\bibnamefont {Busnaina}},
  \bibinfo {author} {\bibfnamefont {C.~S.}\ \bibnamefont {Chang}}, \bibinfo
  {author} {\bibfnamefont {A.}~\bibnamefont {Vadiraj}}, \bibinfo {author}
  {\bibfnamefont {I.}~\bibnamefont {Nsanzineza}}, \bibinfo {author}
  {\bibfnamefont {E.}~\bibnamefont {Solano}}, \bibinfo {author} {\bibfnamefont
  {H.}~\bibnamefont {Alaeian}}, \bibinfo {author} {\bibfnamefont
  {E.}~\bibnamefont {Rico}},\ and\ \bibinfo {author} {\bibfnamefont
  {C.}~\bibnamefont {Wilson}},\ }\bibfield  {title} {\bibinfo {title} {{Quantum
  simulation of the bosonic Creutz ladder with a parametric cavity}},\ }\href
  {https://doi.org/10.1103/PhysRevLett.127.100503} {\bibfield  {journal}
  {\bibinfo  {journal} {Physical Review Letters}\ }\textbf {\bibinfo {volume}
  {127}},\ \bibinfo {pages} {100503} (\bibinfo {year} {2021})}\BibitemShut
  {NoStop}%
\bibitem [{\citenamefont {Martinez}\ \emph {et~al.}(2023)\citenamefont
  {Martinez}, \citenamefont {Chiu}, \citenamefont {Smitham},\ and\
  \citenamefont {Houck}}]{martinez2023interaction}%
  \BibitemOpen
  \bibfield  {author} {\bibinfo {author} {\bibfnamefont {J.~G.}\ \bibnamefont
  {Martinez}}, \bibinfo {author} {\bibfnamefont {C.~S.}\ \bibnamefont {Chiu}},
  \bibinfo {author} {\bibfnamefont {B.~M.}\ \bibnamefont {Smitham}},\ and\
  \bibinfo {author} {\bibfnamefont {A.~A.}\ \bibnamefont {Houck}},\ }\bibfield
  {title} {\bibinfo {title} {{I}nteraction-induced escape from an
  {A}haronov-{B}ohm cage},\ }\bibfield  {journal} {\bibinfo  {journal}
  {arXiv:2303.02170}\ }\href {https://doi.org/10.48550/arXiv.2303.02170}
  {10.48550/arXiv.2303.02170} (\bibinfo {year} {2023})\BibitemShut {NoStop}%
\bibitem [{\citenamefont {Chase-Mayoral}\ \emph {et~al.}(2023)\citenamefont
  {Chase-Mayoral}, \citenamefont {English}, \citenamefont {Kim}, \citenamefont
  {Lee}, \citenamefont {Lape}, \citenamefont {Andreanov}, \citenamefont
  {Kevrekidis},\ and\ \citenamefont {Flach}}]{chase2023compact}%
  \BibitemOpen
  \bibfield  {author} {\bibinfo {author} {\bibfnamefont {C.}~\bibnamefont
  {Chase-Mayoral}}, \bibinfo {author} {\bibfnamefont {L.}~\bibnamefont
  {English}}, \bibinfo {author} {\bibfnamefont {Y.}~\bibnamefont {Kim}},
  \bibinfo {author} {\bibfnamefont {S.}~\bibnamefont {Lee}}, \bibinfo {author}
  {\bibfnamefont {N.}~\bibnamefont {Lape}}, \bibinfo {author} {\bibfnamefont
  {A.}~\bibnamefont {Andreanov}}, \bibinfo {author} {\bibfnamefont
  {P.}~\bibnamefont {Kevrekidis}},\ and\ \bibinfo {author} {\bibfnamefont
  {S.}~\bibnamefont {Flach}},\ }\bibfield  {title} {\bibinfo {title} {Compact
  {L}ocalized {S}tates in {E}lectric {C}ircuit {F}latband {L}attices},\
  }\bibfield  {journal} {\bibinfo  {journal} {arXiv:2307.15319}\ }\href
  {https://doi.org/10.48550/arXiv.2307.15319} {10.48550/arXiv.2307.15319}
  (\bibinfo {year} {2023})\BibitemShut {NoStop}%
\bibitem [{\citenamefont {Longhi}(2014)}]{longhi2014aharonov}%
  \BibitemOpen
  \bibfield  {author} {\bibinfo {author} {\bibfnamefont {S.}~\bibnamefont
  {Longhi}},\ }\bibfield  {title} {\bibinfo {title} {{A}haronov-{B}ohm photonic
  cages in waveguide and coupled resonator lattices by synthetic magnetic
  fields},\ }\href {https://doi.org/10.1364/OL.39.005892} {\bibfield  {journal}
  {\bibinfo  {journal} {Optics Letters}\ }\textbf {\bibinfo {volume} {39}},\
  \bibinfo {pages} {5892} (\bibinfo {year} {2014})}\BibitemShut {NoStop}%
\bibitem [{\citenamefont {Mukherjee}\ and\ \citenamefont
  {Thomson}(2015)}]{mukherjee2015observation}%
  \BibitemOpen
  \bibfield  {author} {\bibinfo {author} {\bibfnamefont {S.}~\bibnamefont
  {Mukherjee}}\ and\ \bibinfo {author} {\bibfnamefont {R.~R.}\ \bibnamefont
  {Thomson}},\ }\bibfield  {title} {\bibinfo {title} {{O}bservation of
  localized flat-band modes in a quasi-one-dimensional photonic rhombic
  lattice},\ }\href {https://doi.org/10.1364/OL.40.005443} {\bibfield
  {journal} {\bibinfo  {journal} {Optics Letters}\ }\textbf {\bibinfo {volume}
  {40}},\ \bibinfo {pages} {5443} (\bibinfo {year} {2015})}\BibitemShut
  {NoStop}%
\bibitem [{\citenamefont {Mukherjee}\ \emph {et~al.}(2018)\citenamefont
  {Mukherjee}, \citenamefont {Di~Liberto}, \citenamefont {{\"O}hberg},
  \citenamefont {Thomson},\ and\ \citenamefont
  {Goldman}}]{mukherjee2018experimental}%
  \BibitemOpen
  \bibfield  {author} {\bibinfo {author} {\bibfnamefont {S.}~\bibnamefont
  {Mukherjee}}, \bibinfo {author} {\bibfnamefont {M.}~\bibnamefont
  {Di~Liberto}}, \bibinfo {author} {\bibfnamefont {P.}~\bibnamefont
  {{\"O}hberg}}, \bibinfo {author} {\bibfnamefont {R.~R.}\ \bibnamefont
  {Thomson}},\ and\ \bibinfo {author} {\bibfnamefont {N.}~\bibnamefont
  {Goldman}},\ }\bibfield  {title} {\bibinfo {title} {{E}xperimental
  observation of {A}haronov-{B}ohm cages in photonic lattices},\ }\href
  {https://doi.org/10.1103/PhysRevLett.121.075502} {\bibfield  {journal}
  {\bibinfo  {journal} {Physical Review Letters}\ }\textbf {\bibinfo {volume}
  {121}},\ \bibinfo {pages} {075502} (\bibinfo {year} {2018})}\BibitemShut
  {NoStop}%
\bibitem [{\citenamefont {Di~Liberto}\ \emph {et~al.}(2019)\citenamefont
  {Di~Liberto}, \citenamefont {Mukherjee},\ and\ \citenamefont
  {Goldman}}]{diliberto2019nonlinear}%
  \BibitemOpen
  \bibfield  {author} {\bibinfo {author} {\bibfnamefont {M.}~\bibnamefont
  {Di~Liberto}}, \bibinfo {author} {\bibfnamefont {S.}~\bibnamefont
  {Mukherjee}},\ and\ \bibinfo {author} {\bibfnamefont {N.}~\bibnamefont
  {Goldman}},\ }\bibfield  {title} {\bibinfo {title} {{N}onlinear dynamics of
  {A}haronov-{B}ohm cages},\ }\href
  {https://doi.org/10.1103/PhysRevA.100.043829} {\bibfield  {journal} {\bibinfo
   {journal} {Physical Review A}\ }\textbf {\bibinfo {volume} {100}},\ \bibinfo
  {pages} {043829} (\bibinfo {year} {2019})}\BibitemShut {NoStop}%
\bibitem [{\citenamefont {C{\'a}ceres-Aravena}\ \emph
  {et~al.}(2022)\citenamefont {C{\'a}ceres-Aravena}, \citenamefont
  {Guzm{\'a}n-Silva}, \citenamefont {Salinas},\ and\ \citenamefont
  {Vicencio}}]{caceres2022controlled}%
  \BibitemOpen
  \bibfield  {author} {\bibinfo {author} {\bibfnamefont {G.}~\bibnamefont
  {C{\'a}ceres-Aravena}}, \bibinfo {author} {\bibfnamefont {D.}~\bibnamefont
  {Guzm{\'a}n-Silva}}, \bibinfo {author} {\bibfnamefont {I.}~\bibnamefont
  {Salinas}},\ and\ \bibinfo {author} {\bibfnamefont {R.~A.}\ \bibnamefont
  {Vicencio}},\ }\bibfield  {title} {\bibinfo {title} {{Controlled transport
  based on multiorbital {A}haronov-{B}ohm photonic caging}},\ }\href
  {https://doi.org/10.1103/PhysRevLett.128.256602} {\bibfield  {journal}
  {\bibinfo  {journal} {Physical Review Letters}\ }\textbf {\bibinfo {volume}
  {128}},\ \bibinfo {pages} {256602} (\bibinfo {year} {2022})}\BibitemShut
  {NoStop}%
\bibitem [{\citenamefont {Creffield}\ and\ \citenamefont
  {Platero}(2010)}]{creffield2010coherent}%
  \BibitemOpen
  \bibfield  {author} {\bibinfo {author} {\bibfnamefont {C.~E.}\ \bibnamefont
  {Creffield}}\ and\ \bibinfo {author} {\bibfnamefont {G.}~\bibnamefont
  {Platero}},\ }\bibfield  {title} {\bibinfo {title} {{Coherent control of
  interacting particles using dynamical and Aharonov-Bohm Phases}},\ }\href
  {https://doi.org/10.1103/PhysRevLett.105.086804} {\bibfield  {journal}
  {\bibinfo  {journal} {Physical Review Letters}\ }\textbf {\bibinfo {volume}
  {105}},\ \bibinfo {pages} {086804} (\bibinfo {year} {2010})}\BibitemShut
  {NoStop}%
\bibitem [{\citenamefont {Li}\ \emph {et~al.}(2022)\citenamefont {Li},
  \citenamefont {Dong}, \citenamefont {Longhi}, \citenamefont {Liang},
  \citenamefont {Xie},\ and\ \citenamefont {Yan}}]{li2022aharonov}%
  \BibitemOpen
  \bibfield  {author} {\bibinfo {author} {\bibfnamefont {H.}~\bibnamefont
  {Li}}, \bibinfo {author} {\bibfnamefont {Z.}~\bibnamefont {Dong}}, \bibinfo
  {author} {\bibfnamefont {S.}~\bibnamefont {Longhi}}, \bibinfo {author}
  {\bibfnamefont {Q.}~\bibnamefont {Liang}}, \bibinfo {author} {\bibfnamefont
  {D.}~\bibnamefont {Xie}},\ and\ \bibinfo {author} {\bibfnamefont
  {B.}~\bibnamefont {Yan}},\ }\bibfield  {title} {\bibinfo {title}
  {{A}haronov-{B}ohm {C}aging and {I}nverse {A}nderson {T}ransition in
  {U}ltracold {A}toms},\ }\href
  {https://doi.org/10.1103/PhysRevLett.129.220403} {\bibfield  {journal}
  {\bibinfo  {journal} {Physical Review Letters}\ }\textbf {\bibinfo {volume}
  {129}},\ \bibinfo {pages} {220403} (\bibinfo {year} {2022})}\BibitemShut
  {NoStop}%
\bibitem [{\citenamefont {C{\u{a}}lug{\u{a}}ru}\ \emph
  {et~al.}(2022)\citenamefont {C{\u{a}}lug{\u{a}}ru}, \citenamefont {Chew},
  \citenamefont {Elcoro}, \citenamefont {Xu}, \citenamefont {Regnault},
  \citenamefont {Song},\ and\ \citenamefont {Bernevig}}]{calugaru2022general}%
  \BibitemOpen
  \bibfield  {author} {\bibinfo {author} {\bibfnamefont {D.}~\bibnamefont
  {C{\u{a}}lug{\u{a}}ru}}, \bibinfo {author} {\bibfnamefont {A.}~\bibnamefont
  {Chew}}, \bibinfo {author} {\bibfnamefont {L.}~\bibnamefont {Elcoro}},
  \bibinfo {author} {\bibfnamefont {Y.}~\bibnamefont {Xu}}, \bibinfo {author}
  {\bibfnamefont {N.}~\bibnamefont {Regnault}}, \bibinfo {author}
  {\bibfnamefont {Z.-D.}\ \bibnamefont {Song}},\ and\ \bibinfo {author}
  {\bibfnamefont {B.~A.}\ \bibnamefont {Bernevig}},\ }\bibfield  {title}
  {\bibinfo {title} {{G}eneral construction and topological classification of
  crystalline flat bands},\ }\href {https://doi.org/10.1038/s41567-021-01445-3}
  {\bibfield  {journal} {\bibinfo  {journal} {Nature Physics}\ }\textbf
  {\bibinfo {volume} {18}},\ \bibinfo {pages} {185} (\bibinfo {year}
  {2022})}\BibitemShut {NoStop}%
\bibitem [{\citenamefont {Fleischhauer}\ \emph {et~al.}(2005)\citenamefont
  {Fleischhauer}, \citenamefont {Imamo{\u{g}}lu},\ and\ \citenamefont
  {Marangos}}]{fleischhauer2005electromagnetically}%
  \BibitemOpen
  \bibfield  {author} {\bibinfo {author} {\bibfnamefont {M.}~\bibnamefont
  {Fleischhauer}}, \bibinfo {author} {\bibfnamefont {A.}~\bibnamefont
  {Imamo{\u{g}}lu}},\ and\ \bibinfo {author} {\bibfnamefont {J.~P.}\
  \bibnamefont {Marangos}},\ }\bibfield  {title} {\bibinfo {title}
  {{E}lectromagnetically induced transparency: {O}ptics in coherent media},\
  }\href {https://doi.org/10.1103/RevModPhys.77.633} {\bibfield  {journal}
  {\bibinfo  {journal} {Review of Modern Physics}\ }\textbf {\bibinfo {volume}
  {77}},\ \bibinfo {pages} {633} (\bibinfo {year} {2005})}\BibitemShut
  {NoStop}%
\bibitem [{\citenamefont {Schmidt}\ and\ \citenamefont
  {Imamo{\u{g}}lu}(1996)}]{schmidt1996giant}%
  \BibitemOpen
  \bibfield  {author} {\bibinfo {author} {\bibfnamefont {H.}~\bibnamefont
  {Schmidt}}\ and\ \bibinfo {author} {\bibfnamefont {A.}~\bibnamefont
  {Imamo{\u{g}}lu}},\ }\bibfield  {title} {\bibinfo {title} {{Giant Kerr
  nonlinearities obtained by electromagnetically induced transparency}},\
  }\href {https://doi.org/10.1364/OL.21.001936} {\bibfield  {journal} {\bibinfo
   {journal} {Optics Letters}\ }\textbf {\bibinfo {volume} {21}},\ \bibinfo
  {pages} {1936} (\bibinfo {year} {1996})}\BibitemShut {NoStop}%
\bibitem [{\citenamefont {Gligori{\'c}}\ \emph {et~al.}(2019)\citenamefont
  {Gligori{\'c}}, \citenamefont {Beli{\v{c}}ev}, \citenamefont {Leykam},\ and\
  \citenamefont {Maluckov}}]{gligoric2019nonlinear}%
  \BibitemOpen
  \bibfield  {author} {\bibinfo {author} {\bibfnamefont {G.}~\bibnamefont
  {Gligori{\'c}}}, \bibinfo {author} {\bibfnamefont {P.~P.}\ \bibnamefont
  {Beli{\v{c}}ev}}, \bibinfo {author} {\bibfnamefont {D.}~\bibnamefont
  {Leykam}},\ and\ \bibinfo {author} {\bibfnamefont {A.}~\bibnamefont
  {Maluckov}},\ }\bibfield  {title} {\bibinfo {title} {{N}onlinear symmetry
  breaking of {A}haronov-{B}ohm cages},\ }\href
  {https://doi.org/10.1103/PhysRevA.99.013826} {\bibfield  {journal} {\bibinfo
  {journal} {Physical Review A}\ }\textbf {\bibinfo {volume} {99}},\ \bibinfo
  {pages} {013826} (\bibinfo {year} {2019})}\BibitemShut {NoStop}%
\bibitem [{\citenamefont {Kremer}\ \emph {et~al.}(2020)\citenamefont {Kremer},
  \citenamefont {Petrides}, \citenamefont {Meyer}, \citenamefont {Heinrich},
  \citenamefont {Zilberberg},\ and\ \citenamefont
  {Szameit}}]{kremer2020square}%
  \BibitemOpen
  \bibfield  {author} {\bibinfo {author} {\bibfnamefont {M.}~\bibnamefont
  {Kremer}}, \bibinfo {author} {\bibfnamefont {I.}~\bibnamefont {Petrides}},
  \bibinfo {author} {\bibfnamefont {E.}~\bibnamefont {Meyer}}, \bibinfo
  {author} {\bibfnamefont {M.}~\bibnamefont {Heinrich}}, \bibinfo {author}
  {\bibfnamefont {O.}~\bibnamefont {Zilberberg}},\ and\ \bibinfo {author}
  {\bibfnamefont {A.}~\bibnamefont {Szameit}},\ }\bibfield  {title} {\bibinfo
  {title} {{A} square-root topological insulator with non-quantized indices
  realized with photonic {A}haronov-{B}ohm cages},\ }\href
  {https://doi.org/10.1038/s41467-020-14692-4} {\bibfield  {journal} {\bibinfo
  {journal} {Nature Communications}\ }\textbf {\bibinfo {volume} {11}},\
  \bibinfo {pages} {1} (\bibinfo {year} {2020})}\BibitemShut {NoStop}%
\bibitem [{\citenamefont {Pelegr{\'\i}}\ \emph {et~al.}(2020)\citenamefont
  {Pelegr{\'\i}}, \citenamefont {Marques}, \citenamefont {Ahufinger},
  \citenamefont {Mompart},\ and\ \citenamefont
  {Dias}}]{pelegri2020interaction}%
  \BibitemOpen
  \bibfield  {author} {\bibinfo {author} {\bibfnamefont {G.}~\bibnamefont
  {Pelegr{\'\i}}}, \bibinfo {author} {\bibfnamefont {A.}~\bibnamefont
  {Marques}}, \bibinfo {author} {\bibfnamefont {V.}~\bibnamefont {Ahufinger}},
  \bibinfo {author} {\bibfnamefont {J.}~\bibnamefont {Mompart}},\ and\ \bibinfo
  {author} {\bibfnamefont {R.}~\bibnamefont {Dias}},\ }\bibfield  {title}
  {\bibinfo {title} {{Interaction-induced topological properties of two bosons
  in flat-band systems}},\ }\href
  {https://doi.org/10.1103/PhysRevResearch.2.033267} {\bibfield  {journal}
  {\bibinfo  {journal} {Physical Review Research}\ }\textbf {\bibinfo {volume}
  {2}},\ \bibinfo {pages} {033267} (\bibinfo {year} {2020})}\BibitemShut
  {NoStop}%
\bibitem [{\citenamefont {Danieli}\ \emph
  {et~al.}(2021{\natexlab{a}})\citenamefont {Danieli}, \citenamefont
  {Andreanov}, \citenamefont {Mithun},\ and\ \citenamefont
  {Flach}}]{danieli2021nonlinear}%
  \BibitemOpen
  \bibfield  {author} {\bibinfo {author} {\bibfnamefont {C.}~\bibnamefont
  {Danieli}}, \bibinfo {author} {\bibfnamefont {A.}~\bibnamefont {Andreanov}},
  \bibinfo {author} {\bibfnamefont {T.}~\bibnamefont {Mithun}},\ and\ \bibinfo
  {author} {\bibfnamefont {S.}~\bibnamefont {Flach}},\ }\bibfield  {title}
  {\bibinfo {title} {{{N}onlinear caging in all-bands-flat lattices}},\ }\href
  {https://doi.org/10.1103/PhysRevB.104.085131} {\bibfield  {journal} {\bibinfo
   {journal} {Physical Review B}\ }\textbf {\bibinfo {volume} {104}},\ \bibinfo
  {pages} {085131} (\bibinfo {year} {2021}{\natexlab{a}})}\BibitemShut
  {NoStop}%
\bibitem [{\citenamefont {Danieli}\ \emph
  {et~al.}(2021{\natexlab{b}})\citenamefont {Danieli}, \citenamefont
  {Andreanov}, \citenamefont {Mithun},\ and\ \citenamefont
  {Flach}}]{danieli2021quantum}%
  \BibitemOpen
  \bibfield  {author} {\bibinfo {author} {\bibfnamefont {C.}~\bibnamefont
  {Danieli}}, \bibinfo {author} {\bibfnamefont {A.}~\bibnamefont {Andreanov}},
  \bibinfo {author} {\bibfnamefont {T.}~\bibnamefont {Mithun}},\ and\ \bibinfo
  {author} {\bibfnamefont {S.}~\bibnamefont {Flach}},\ }\bibfield  {title}
  {\bibinfo {title} {{{Q}uantum caging in interacting many-body all-bands-flat
  lattices}},\ }\href {https://doi.org/10.1103/PhysRevB.104.085132} {\bibfield
  {journal} {\bibinfo  {journal} {Physical Review B}\ }\textbf {\bibinfo
  {volume} {104}},\ \bibinfo {pages} {085132} (\bibinfo {year}
  {2021}{\natexlab{b}})}\BibitemShut {NoStop}%
\bibitem [{\citenamefont {Kolovsky}\ \emph {et~al.}(2023)\citenamefont
  {Kolovsky}, \citenamefont {Muraev},\ and\ \citenamefont
  {Flach}}]{kolovsky2023conductance}%
  \BibitemOpen
  \bibfield  {author} {\bibinfo {author} {\bibfnamefont {A.~R.}\ \bibnamefont
  {Kolovsky}}, \bibinfo {author} {\bibfnamefont {P.~S.}\ \bibnamefont
  {Muraev}},\ and\ \bibinfo {author} {\bibfnamefont {S.}~\bibnamefont
  {Flach}},\ }\bibfield  {title} {\bibinfo {title} {{Conductance transition
  with interacting bosons in an Aharonov-Bohm cage}},\ }\bibfield  {journal}
  {\bibinfo  {journal} {arXiv:2303.00509}\ }\href
  {https://doi.org/10.48550/arXiv.2303.00509} {10.48550/arXiv.2303.00509}
  (\bibinfo {year} {2023})\BibitemShut {NoStop}%
\bibitem [{\citenamefont {Dudin}\ and\ \citenamefont
  {Kuzmich}(2012)}]{dudin2012strongly}%
  \BibitemOpen
  \bibfield  {author} {\bibinfo {author} {\bibfnamefont {Y.}~\bibnamefont
  {Dudin}}\ and\ \bibinfo {author} {\bibfnamefont {A.}~\bibnamefont
  {Kuzmich}},\ }\bibfield  {title} {\bibinfo {title} {Strongly interacting
  {R}ydberg excitations of a cold atomic gas},\ }\href
  {https://doi.org/10.1126/science.1217901} {\bibfield  {journal} {\bibinfo
  {journal} {Science}\ }\textbf {\bibinfo {volume} {336}},\ \bibinfo {pages}
  {887} (\bibinfo {year} {2012})}\BibitemShut {NoStop}%
\bibitem [{\citenamefont {Firstenberg}\ \emph {et~al.}(2016)\citenamefont
  {Firstenberg}, \citenamefont {Adams},\ and\ \citenamefont
  {Hofferberth}}]{firstenberg2016nonlinear}%
  \BibitemOpen
  \bibfield  {author} {\bibinfo {author} {\bibfnamefont {O.}~\bibnamefont
  {Firstenberg}}, \bibinfo {author} {\bibfnamefont {C.~S.}\ \bibnamefont
  {Adams}},\ and\ \bibinfo {author} {\bibfnamefont {S.}~\bibnamefont
  {Hofferberth}},\ }\bibfield  {title} {\bibinfo {title} {{Nonlinear quantum
  optics mediated by Rydberg interactions}},\ }\href
  {https://doi.org/10.1088/0953-4075/49/15/152003} {\bibfield  {journal}
  {\bibinfo  {journal} {Journal of Physics B: Atomic, Molecular and Optical
  Physics}\ }\textbf {\bibinfo {volume} {49}},\ \bibinfo {pages} {152003}
  (\bibinfo {year} {2016})}\BibitemShut {NoStop}%
\bibitem [{\citenamefont {Murray}\ and\ \citenamefont
  {Pohl}(2017)}]{murray2017coherent}%
  \BibitemOpen
  \bibfield  {author} {\bibinfo {author} {\bibfnamefont {C.~R.}\ \bibnamefont
  {Murray}}\ and\ \bibinfo {author} {\bibfnamefont {T.}~\bibnamefont {Pohl}},\
  }\bibfield  {title} {\bibinfo {title} {{Coherent photon manipulation in
  interacting atomic ensembles}},\ }\href
  {https://doi.org/10.1103/PhysRevX.7.031007} {\bibfield  {journal} {\bibinfo
  {journal} {Physical Review X}\ }\textbf {\bibinfo {volume} {7}},\ \bibinfo
  {pages} {031007} (\bibinfo {year} {2017})}\BibitemShut {NoStop}%
\bibitem [{\citenamefont {Bajcsy}\ \emph {et~al.}(2009)\citenamefont {Bajcsy},
  \citenamefont {Hofferberth}, \citenamefont {Balic}, \citenamefont {Peyronel},
  \citenamefont {Hafezi}, \citenamefont {Zibrov}, \citenamefont {Vuletic},\
  and\ \citenamefont {Lukin}}]{bajcsy2009efficient}%
  \BibitemOpen
  \bibfield  {author} {\bibinfo {author} {\bibfnamefont {M.}~\bibnamefont
  {Bajcsy}}, \bibinfo {author} {\bibfnamefont {S.}~\bibnamefont {Hofferberth}},
  \bibinfo {author} {\bibfnamefont {V.}~\bibnamefont {Balic}}, \bibinfo
  {author} {\bibfnamefont {T.}~\bibnamefont {Peyronel}}, \bibinfo {author}
  {\bibfnamefont {M.}~\bibnamefont {Hafezi}}, \bibinfo {author} {\bibfnamefont
  {A.~S.}\ \bibnamefont {Zibrov}}, \bibinfo {author} {\bibfnamefont
  {V.}~\bibnamefont {Vuletic}},\ and\ \bibinfo {author} {\bibfnamefont {M.~D.}\
  \bibnamefont {Lukin}},\ }\bibfield  {title} {\bibinfo {title} {{Efficient
  all-optical switching using slow light within a hollow fiber}},\ }\href
  {https://doi.org/10.1103/PhysRevLett.102.203902} {\bibfield  {journal}
  {\bibinfo  {journal} {Physical Review Letters}\ }\textbf {\bibinfo {volume}
  {102}},\ \bibinfo {pages} {203902} (\bibinfo {year} {2009})}\BibitemShut
  {NoStop}%
\bibitem [{\citenamefont {Vetsch}\ \emph {et~al.}(2010)\citenamefont {Vetsch},
  \citenamefont {Reitz}, \citenamefont {Sagu{\'e}}, \citenamefont {Schmidt},
  \citenamefont {Dawkins},\ and\ \citenamefont
  {Rauschenbeutel}}]{vetsch2010optical}%
  \BibitemOpen
  \bibfield  {author} {\bibinfo {author} {\bibfnamefont {E.}~\bibnamefont
  {Vetsch}}, \bibinfo {author} {\bibfnamefont {D.}~\bibnamefont {Reitz}},
  \bibinfo {author} {\bibfnamefont {G.}~\bibnamefont {Sagu{\'e}}}, \bibinfo
  {author} {\bibfnamefont {R.}~\bibnamefont {Schmidt}}, \bibinfo {author}
  {\bibfnamefont {S.}~\bibnamefont {Dawkins}},\ and\ \bibinfo {author}
  {\bibfnamefont {A.}~\bibnamefont {Rauschenbeutel}},\ }\bibfield  {title}
  {\bibinfo {title} {Optical interface created by laser-cooled atoms trapped in
  the evanescent field surrounding an optical nanofiber},\ }\href
  {https://doi.org/10.1103/PhysRevLett.104.203603} {\bibfield  {journal}
  {\bibinfo  {journal} {Physical review letters}\ }\textbf {\bibinfo {volume}
  {104}},\ \bibinfo {pages} {203603} (\bibinfo {year} {2010})}\BibitemShut
  {NoStop}%
\bibitem [{\citenamefont {{O’Shea}}\ \emph {et~al.}(2013)\citenamefont
  {{O’Shea}}, \citenamefont {Junge}, \citenamefont {Volz},\ and\
  \citenamefont {Rauschenbeutel}}]{oshea2013fiber}%
  \BibitemOpen
  \bibfield  {author} {\bibinfo {author} {\bibfnamefont {D.}~\bibnamefont
  {{O’Shea}}}, \bibinfo {author} {\bibfnamefont {C.}~\bibnamefont {Junge}},
  \bibinfo {author} {\bibfnamefont {J.}~\bibnamefont {Volz}},\ and\ \bibinfo
  {author} {\bibfnamefont {A.}~\bibnamefont {Rauschenbeutel}},\ }\bibfield
  {title} {\bibinfo {title} {Fiber-optical switch controlled by a single
  atom},\ }\href {https://doi.org/10.1103/PhysRevLett.107.193902} {\bibfield
  {journal} {\bibinfo  {journal} {Physical review letters}\ }\textbf {\bibinfo
  {volume} {111}},\ \bibinfo {pages} {193601} (\bibinfo {year}
  {2013})}\BibitemShut {NoStop}%
\bibitem [{\citenamefont {Yanay}\ \emph {et~al.}(2020)\citenamefont {Yanay},
  \citenamefont {Braum{\"u}ller}, \citenamefont {Gustavsson}, \citenamefont
  {Oliver},\ and\ \citenamefont {Tahan}}]{yanay2020two-dimensional}%
  \BibitemOpen
  \bibfield  {author} {\bibinfo {author} {\bibfnamefont {Y.}~\bibnamefont
  {Yanay}}, \bibinfo {author} {\bibfnamefont {J.}~\bibnamefont
  {Braum{\"u}ller}}, \bibinfo {author} {\bibfnamefont {S.}~\bibnamefont
  {Gustavsson}}, \bibinfo {author} {\bibfnamefont {W.~D.}\ \bibnamefont
  {Oliver}},\ and\ \bibinfo {author} {\bibfnamefont {C.}~\bibnamefont
  {Tahan}},\ }\bibfield  {title} {\bibinfo {title} {{T}wo-dimensional hard-core
  {B}ose--{H}ubbard model with superconducting qubits},\ }\href
  {https://doi.org/10.1038/s41534-020-0269-1} {\bibfield  {journal} {\bibinfo
  {journal} {npj Quantum Information}\ }\textbf {\bibinfo {volume} {6}},\
  \bibinfo {pages} {58} (\bibinfo {year} {2020})}\BibitemShut {NoStop}%
\bibitem [{\citenamefont {Amo}\ and\ \citenamefont
  {Bloch}(2016)}]{amo2016exciton}%
  \BibitemOpen
  \bibfield  {author} {\bibinfo {author} {\bibfnamefont {A.}~\bibnamefont
  {Amo}}\ and\ \bibinfo {author} {\bibfnamefont {J.}~\bibnamefont {Bloch}},\
  }\bibfield  {title} {\bibinfo {title} {{E}xciton-polaritons in lattices: {A}
  non-linear photonic simulator},\ }\href
  {https://doi.org/10.1016/j.crhy.2016.08.007} {\bibfield  {journal} {\bibinfo
  {journal} {Comptes Rendus Physique}\ }\textbf {\bibinfo {volume} {17}},\
  \bibinfo {pages} {934} (\bibinfo {year} {2016})}\BibitemShut {NoStop}%
\bibitem [{\citenamefont {Ozawa}\ \emph {et~al.}(2019)\citenamefont {Ozawa},
  \citenamefont {Price}, \citenamefont {Amo}, \citenamefont {Goldman},
  \citenamefont {Hafezi}, \citenamefont {Lu}, \citenamefont {Rechtsman},
  \citenamefont {Schuster}, \citenamefont {Simon}, \citenamefont {Zilberberg},\
  and\ \citenamefont {Carusotto}}]{ozawa2019topological}%
  \BibitemOpen
  \bibfield  {author} {\bibinfo {author} {\bibfnamefont {T.}~\bibnamefont
  {Ozawa}}, \bibinfo {author} {\bibfnamefont {H.~M.}\ \bibnamefont {Price}},
  \bibinfo {author} {\bibfnamefont {A.}~\bibnamefont {Amo}}, \bibinfo {author}
  {\bibfnamefont {N.}~\bibnamefont {Goldman}}, \bibinfo {author} {\bibfnamefont
  {M.}~\bibnamefont {Hafezi}}, \bibinfo {author} {\bibfnamefont
  {L.}~\bibnamefont {Lu}}, \bibinfo {author} {\bibfnamefont {M.~C.}\
  \bibnamefont {Rechtsman}}, \bibinfo {author} {\bibfnamefont {D.}~\bibnamefont
  {Schuster}}, \bibinfo {author} {\bibfnamefont {J.}~\bibnamefont {Simon}},
  \bibinfo {author} {\bibfnamefont {O.}~\bibnamefont {Zilberberg}},\ and\
  \bibinfo {author} {\bibfnamefont {I.}~\bibnamefont {Carusotto}},\ }\bibfield
  {title} {\bibinfo {title} {{T}opological photonics},\ }\href
  {https://doi.org/10.1103/RevModPhys.91.015006} {\bibfield  {journal}
  {\bibinfo  {journal} {Review of Modern Physics}\ }\textbf {\bibinfo {volume}
  {91}},\ \bibinfo {pages} {015006} (\bibinfo {year} {2019})}\BibitemShut
  {NoStop}%
\bibitem [{\citenamefont {Rey}(2015)}]{rey2015synthetic}%
  \BibitemOpen
  \bibfield  {author} {\bibinfo {author} {\bibfnamefont {A.~M.}\ \bibnamefont
  {Rey}},\ }\bibfield  {title} {\bibinfo {title} {{{S}ynthetic gauge fields for
  ultracold atoms}},\ }\href {https://doi.org/10.1093/nsr/nwv053} {\bibfield
  {journal} {\bibinfo  {journal} {National Science Review}\ }\textbf {\bibinfo
  {volume} {3}},\ \bibinfo {pages} {166} (\bibinfo {year} {2015})}\BibitemShut
  {NoStop}%
\bibitem [{\citenamefont {Cooper}\ \emph {et~al.}(2019)\citenamefont {Cooper},
  \citenamefont {Dalibard},\ and\ \citenamefont
  {Spielman}}]{cooper2019topological}%
  \BibitemOpen
  \bibfield  {author} {\bibinfo {author} {\bibfnamefont {N.~R.}\ \bibnamefont
  {Cooper}}, \bibinfo {author} {\bibfnamefont {J.}~\bibnamefont {Dalibard}},\
  and\ \bibinfo {author} {\bibfnamefont {I.~B.}\ \bibnamefont {Spielman}},\
  }\bibfield  {title} {\bibinfo {title} {{T}opological bands for ultracold
  atoms},\ }\href {https://doi.org/10.1103/RevModPhys.91.015005} {\bibfield
  {journal} {\bibinfo  {journal} {Review of Modern Physics}\ }\textbf {\bibinfo
  {volume} {91}},\ \bibinfo {pages} {015005} (\bibinfo {year}
  {2019})}\BibitemShut {NoStop}%
\bibitem [{\citenamefont {Roushan}\ \emph {et~al.}(2017)\citenamefont
  {Roushan}, \citenamefont {Neill}, \citenamefont {Megrant}, \citenamefont
  {Chen}, \citenamefont {Babbush}, \citenamefont {Barends}, \citenamefont
  {Campbell}, \citenamefont {Chen}, \citenamefont {Chiaro}, \citenamefont
  {Dunsworth} \emph {et~al.}}]{roushan2017chiral}%
  \BibitemOpen
  \bibfield  {author} {\bibinfo {author} {\bibfnamefont {P.}~\bibnamefont
  {Roushan}}, \bibinfo {author} {\bibfnamefont {C.}~\bibnamefont {Neill}},
  \bibinfo {author} {\bibfnamefont {A.}~\bibnamefont {Megrant}}, \bibinfo
  {author} {\bibfnamefont {Y.}~\bibnamefont {Chen}}, \bibinfo {author}
  {\bibfnamefont {R.}~\bibnamefont {Babbush}}, \bibinfo {author} {\bibfnamefont
  {R.}~\bibnamefont {Barends}}, \bibinfo {author} {\bibfnamefont
  {B.}~\bibnamefont {Campbell}}, \bibinfo {author} {\bibfnamefont
  {Z.}~\bibnamefont {Chen}}, \bibinfo {author} {\bibfnamefont {B.}~\bibnamefont
  {Chiaro}}, \bibinfo {author} {\bibfnamefont {A.}~\bibnamefont {Dunsworth}},
  \emph {et~al.},\ }\bibfield  {title} {\bibinfo {title} {{C}hiral ground-state
  currents of interacting photons in a synthetic magnetic field},\ }\href
  {https://doi.org/10.1038/nphys3930} {\bibfield  {journal} {\bibinfo
  {journal} {Nature Physics}\ }\textbf {\bibinfo {volume} {13}},\ \bibinfo
  {pages} {146} (\bibinfo {year} {2017})}\BibitemShut {NoStop}%
\bibitem [{\citenamefont {Eisaman}\ \emph {et~al.}(2011)\citenamefont
  {Eisaman}, \citenamefont {Fan}, \citenamefont {Migdall},\ and\ \citenamefont
  {Polyakov}}]{eisaman2011invited}%
  \BibitemOpen
  \bibfield  {author} {\bibinfo {author} {\bibfnamefont {M.~D.}\ \bibnamefont
  {Eisaman}}, \bibinfo {author} {\bibfnamefont {J.}~\bibnamefont {Fan}},
  \bibinfo {author} {\bibfnamefont {A.}~\bibnamefont {Migdall}},\ and\ \bibinfo
  {author} {\bibfnamefont {S.~V.}\ \bibnamefont {Polyakov}},\ }\bibfield
  {title} {\bibinfo {title} {Invited review article: {S}ingle-photon sources
  and detectors},\ }\href {https://doi.org/10.1063/1.3610677} {\bibfield
  {journal} {\bibinfo  {journal} {Review of Scientific Instruments}\ }\textbf
  {\bibinfo {volume} {82}},\ \bibinfo {pages} {071101} (\bibinfo {year}
  {2011})}\BibitemShut {NoStop}%
\bibitem [{\citenamefont {Meyer-Scott}\ \emph {et~al.}(2020)\citenamefont
  {Meyer-Scott}, \citenamefont {Silberhorn},\ and\ \citenamefont
  {Migdall}}]{meyer2020single}%
  \BibitemOpen
  \bibfield  {author} {\bibinfo {author} {\bibfnamefont {E.}~\bibnamefont
  {Meyer-Scott}}, \bibinfo {author} {\bibfnamefont {C.}~\bibnamefont
  {Silberhorn}},\ and\ \bibinfo {author} {\bibfnamefont {A.}~\bibnamefont
  {Migdall}},\ }\bibfield  {title} {\bibinfo {title} {{S}ingle-photon sources:
  {A}pproaching the ideal through multiplexing},\ }\href
  {https://doi.org/10.1063/5.0003320} {\bibfield  {journal} {\bibinfo
  {journal} {Review of Scientific Instruments}\ }\textbf {\bibinfo {volume}
  {91}},\ \bibinfo {pages} {041101} (\bibinfo {year} {2020})}\BibitemShut
  {NoStop}%
\bibitem [{\citenamefont {Tomadin}\ and\ \citenamefont
  {Fazio}(2010)}]{tomadin2010manybody}%
  \BibitemOpen
  \bibfield  {author} {\bibinfo {author} {\bibfnamefont {A.}~\bibnamefont
  {Tomadin}}\ and\ \bibinfo {author} {\bibfnamefont {R.}~\bibnamefont
  {Fazio}},\ }\bibfield  {title} {\bibinfo {title} {{M}any-body phenomena in
  {QED}-cavity array},\ }\href {https://doi.org/10.1364/JOSAB.27.00A130}
  {\bibfield  {journal} {\bibinfo  {journal} {Journal of the Optical Society of
  America B}\ }\textbf {\bibinfo {volume} {27}},\ \bibinfo {pages} {A130}
  (\bibinfo {year} {2010})}\BibitemShut {NoStop}%
\bibitem [{\citenamefont {Kang}\ \emph {et~al.}(2023)\citenamefont {Kang},
  \citenamefont {Wei}, \citenamefont {Zhang},\ and\ \citenamefont
  {Dong}}]{kang2023topological}%
  \BibitemOpen
  \bibfield  {author} {\bibinfo {author} {\bibfnamefont {J.}~\bibnamefont
  {Kang}}, \bibinfo {author} {\bibfnamefont {R.}~\bibnamefont {Wei}}, \bibinfo
  {author} {\bibfnamefont {Q.}~\bibnamefont {Zhang}},\ and\ \bibinfo {author}
  {\bibfnamefont {G.}~\bibnamefont {Dong}},\ }\bibfield  {title} {\bibinfo
  {title} {{T}opological {P}hotonic {S}tates in {W}aveguide {A}rrays},\ }\href
  {https://doi.org/10.1002/apxr.202200053} {\bibfield  {journal} {\bibinfo
  {journal} {Advanced Physics Research}\ }\textbf {\bibinfo {volume} {2}},\
  \bibinfo {pages} {2200053} (\bibinfo {year} {2023})}\BibitemShut {NoStop}%
\bibitem [{\citenamefont {Amo}\ \emph {et~al.}(2010)\citenamefont {Amo},
  \citenamefont {Liew}, \citenamefont {Adrados}, \citenamefont {Houdr{\'e}},
  \citenamefont {Giacobino}, \citenamefont {Kavokin},\ and\ \citenamefont
  {Bramati}}]{amo2010exciton}%
  \BibitemOpen
  \bibfield  {author} {\bibinfo {author} {\bibfnamefont {A.}~\bibnamefont
  {Amo}}, \bibinfo {author} {\bibfnamefont {T.}~\bibnamefont {Liew}}, \bibinfo
  {author} {\bibfnamefont {C.}~\bibnamefont {Adrados}}, \bibinfo {author}
  {\bibfnamefont {R.}~\bibnamefont {Houdr{\'e}}}, \bibinfo {author}
  {\bibfnamefont {E.}~\bibnamefont {Giacobino}}, \bibinfo {author}
  {\bibfnamefont {A.}~\bibnamefont {Kavokin}},\ and\ \bibinfo {author}
  {\bibfnamefont {A.}~\bibnamefont {Bramati}},\ }\bibfield  {title} {\bibinfo
  {title} {Exciton--polariton spin switches},\ }\href
  {https://doi.org/10.1038/nphoton.2010.79} {\bibfield  {journal} {\bibinfo
  {journal} {Nature Photonics}\ }\textbf {\bibinfo {volume} {4}},\ \bibinfo
  {pages} {361} (\bibinfo {year} {2010})}\BibitemShut {NoStop}%
\bibitem [{\citenamefont {T\"{o}rm\"{a}}\ and\ \citenamefont
  {Sengstock}(2014)}]{torma2014quantum}%
  \BibitemOpen
  \bibfield  {author} {\bibinfo {author} {\bibfnamefont {P.}~\bibnamefont
  {T\"{o}rm\"{a}}}\ and\ \bibinfo {author} {\bibfnamefont {K.}~\bibnamefont
  {Sengstock}},\ }\href@noop {} {\emph {\bibinfo {title} {{Q}uantum {G}as
  {E}xperiments: {E}xploring {M}any-{B}ody {S}tates}}},\ Vol.~\bibinfo {volume}
  {3}\ (\bibinfo  {publisher} {World Scientific},\ \bibinfo {year}
  {2014})\BibitemShut {NoStop}%
\bibitem [{\citenamefont {Gross}\ and\ \citenamefont
  {Bloch}(2017)}]{gross2017quantum}%
  \BibitemOpen
  \bibfield  {author} {\bibinfo {author} {\bibfnamefont {C.}~\bibnamefont
  {Gross}}\ and\ \bibinfo {author} {\bibfnamefont {I.}~\bibnamefont {Bloch}},\
  }\bibfield  {title} {\bibinfo {title} {{Q}uantum simulations with ultracold
  atoms in optical lattices},\ }\href {https://doi.org/10.1126/science.aal3837}
  {\bibfield  {journal} {\bibinfo  {journal} {Science}\ }\textbf {\bibinfo
  {volume} {357}},\ \bibinfo {pages} {995} (\bibinfo {year}
  {2017})}\BibitemShut {NoStop}%
\bibitem [{\citenamefont {Sch{\"a}fer}\ \emph {et~al.}(2020)\citenamefont
  {Sch{\"a}fer}, \citenamefont {Fukuhara}, \citenamefont {Sugawa},
  \citenamefont {Takasu},\ and\ \citenamefont {Takahashi}}]{schafer2020tools}%
  \BibitemOpen
  \bibfield  {author} {\bibinfo {author} {\bibfnamefont {F.}~\bibnamefont
  {Sch{\"a}fer}}, \bibinfo {author} {\bibfnamefont {T.}~\bibnamefont
  {Fukuhara}}, \bibinfo {author} {\bibfnamefont {S.}~\bibnamefont {Sugawa}},
  \bibinfo {author} {\bibfnamefont {Y.}~\bibnamefont {Takasu}},\ and\ \bibinfo
  {author} {\bibfnamefont {Y.}~\bibnamefont {Takahashi}},\ }\bibfield  {title}
  {\bibinfo {title} {{T}ools for quantum simulation with ultracold atoms in
  optical lattices},\ }\href {https://doi.org/10.1038/s42254-020-0195-3}
  {\bibfield  {journal} {\bibinfo  {journal} {Nature Reviews Physics}\ }\textbf
  {\bibinfo {volume} {2}},\ \bibinfo {pages} {411} (\bibinfo {year}
  {2020})}\BibitemShut {NoStop}%
\bibitem [{\citenamefont {Bakr}\ \emph {et~al.}(2009)\citenamefont {Bakr},
  \citenamefont {Gillen}, \citenamefont {Peng}, \citenamefont {F{\"o}lling},\
  and\ \citenamefont {Greiner}}]{bakr2009quantum}%
  \BibitemOpen
  \bibfield  {author} {\bibinfo {author} {\bibfnamefont {W.~S.}\ \bibnamefont
  {Bakr}}, \bibinfo {author} {\bibfnamefont {J.~I.}\ \bibnamefont {Gillen}},
  \bibinfo {author} {\bibfnamefont {A.}~\bibnamefont {Peng}}, \bibinfo {author}
  {\bibfnamefont {S.}~\bibnamefont {F{\"o}lling}},\ and\ \bibinfo {author}
  {\bibfnamefont {M.}~\bibnamefont {Greiner}},\ }\bibfield  {title} {\bibinfo
  {title} {{A} quantum gas microscope for detecting single atoms in a
  {H}ubbard-regime optical lattice},\ }\href
  {https://doi.org/10.1038/nature08482} {\bibfield  {journal} {\bibinfo
  {journal} {Nature}\ }\textbf {\bibinfo {volume} {462}},\ \bibinfo {pages}
  {74} (\bibinfo {year} {2009})}\BibitemShut {NoStop}%
\bibitem [{\citenamefont {Sherson}\ \emph {et~al.}(2010)\citenamefont
  {Sherson}, \citenamefont {Weitenberg}, \citenamefont {Endres}, \citenamefont
  {Cheneau}, \citenamefont {Bloch},\ and\ \citenamefont
  {Kuhr}}]{sherson2010single}%
  \BibitemOpen
  \bibfield  {author} {\bibinfo {author} {\bibfnamefont {J.~F.}\ \bibnamefont
  {Sherson}}, \bibinfo {author} {\bibfnamefont {C.}~\bibnamefont {Weitenberg}},
  \bibinfo {author} {\bibfnamefont {M.}~\bibnamefont {Endres}}, \bibinfo
  {author} {\bibfnamefont {M.}~\bibnamefont {Cheneau}}, \bibinfo {author}
  {\bibfnamefont {I.}~\bibnamefont {Bloch}},\ and\ \bibinfo {author}
  {\bibfnamefont {S.}~\bibnamefont {Kuhr}},\ }\bibfield  {title} {\bibinfo
  {title} {Single-atom-resolved fluorescence imaging of an atomic {M}ott
  insulator},\ }\href {https://doi.org/10.1038/nature09378} {\bibfield
  {journal} {\bibinfo  {journal} {Nature}\ }\textbf {\bibinfo {volume} {467}},\
  \bibinfo {pages} {68} (\bibinfo {year} {2010})}\BibitemShut {NoStop}%
\bibitem [{\citenamefont {Carusotto}\ and\ \citenamefont
  {Ciuti}(2013)}]{carusotto2013}%
  \BibitemOpen
  \bibfield  {author} {\bibinfo {author} {\bibfnamefont {I.}~\bibnamefont
  {Carusotto}}\ and\ \bibinfo {author} {\bibfnamefont {C.}~\bibnamefont
  {Ciuti}},\ }\bibfield  {title} {\bibinfo {title} {{Q}uantum fluids of
  light},\ }\href {https://doi.org/10.1103/RevModPhys.85.299} {\bibfield
  {journal} {\bibinfo  {journal} {Review of Modern Physics}\ }\textbf {\bibinfo
  {volume} {85}},\ \bibinfo {pages} {299} (\bibinfo {year} {2013})}\BibitemShut
  {NoStop}%
\bibitem [{\citenamefont {Miller}\ \emph {et~al.}(1983)\citenamefont {Miller},
  \citenamefont {Chemla}, \citenamefont {Eilenberger}, \citenamefont {Smith},
  \citenamefont {Gossard},\ and\ \citenamefont
  {Wiegmann}}]{miller1983degenerate}%
  \BibitemOpen
  \bibfield  {author} {\bibinfo {author} {\bibfnamefont {D.~A.~B.}\
  \bibnamefont {Miller}}, \bibinfo {author} {\bibfnamefont {D.~S.}\
  \bibnamefont {Chemla}}, \bibinfo {author} {\bibfnamefont {D.~J.}\
  \bibnamefont {Eilenberger}}, \bibinfo {author} {\bibfnamefont {P.~W.}\
  \bibnamefont {Smith}}, \bibinfo {author} {\bibfnamefont {A.~C.}\ \bibnamefont
  {Gossard}},\ and\ \bibinfo {author} {\bibfnamefont {W.}~\bibnamefont
  {Wiegmann}},\ }\bibfield  {title} {\bibinfo {title} {{{D}egenerate
  four‐wave mixing in room‐temperature {G}a{A}s/{G}a{A}l{A}s multiple
  quantum well structures}},\ }\href {https://doi.org/10.1063/1.93802}
  {\bibfield  {journal} {\bibinfo  {journal} {Applied Physics Letters}\
  }\textbf {\bibinfo {volume} {42}},\ \bibinfo {pages} {925} (\bibinfo {year}
  {1983})}\BibitemShut {NoStop}%
\bibitem [{\citenamefont {Zhang}\ \emph {et~al.}(2022)\citenamefont {Zhang},
  \citenamefont {Bai}, \citenamefont {Cai}, \citenamefont {Hao}, \citenamefont
  {Wang}, \citenamefont {Zhang}, \citenamefont {Zhao}, \citenamefont {Teng},
  \citenamefont {Sui}, \citenamefont {Du},\ and\ \citenamefont
  {Wang}}]{zhang2022nonlinear}%
  \BibitemOpen
  \bibfield  {author} {\bibinfo {author} {\bibfnamefont {Q.}~\bibnamefont
  {Zhang}}, \bibinfo {author} {\bibfnamefont {Q.}~\bibnamefont {Bai}}, \bibinfo
  {author} {\bibfnamefont {E.}~\bibnamefont {Cai}}, \bibinfo {author}
  {\bibfnamefont {L.}~\bibnamefont {Hao}}, \bibinfo {author} {\bibfnamefont
  {M.}~\bibnamefont {Wang}}, \bibinfo {author} {\bibfnamefont {S.}~\bibnamefont
  {Zhang}}, \bibinfo {author} {\bibfnamefont {Q.}~\bibnamefont {Zhao}},
  \bibinfo {author} {\bibfnamefont {L.}~\bibnamefont {Teng}}, \bibinfo {author}
  {\bibfnamefont {N.}~\bibnamefont {Sui}}, \bibinfo {author} {\bibfnamefont
  {F.}~\bibnamefont {Du}},\ and\ \bibinfo {author} {\bibfnamefont
  {X.}~\bibnamefont {Wang}},\ }\bibfield  {title} {\bibinfo {title}
  {{N}onlinear optical properties of graphdiyne/graphene van der {W}aals
  heterostructure for laser modulations},\ }\href
  {https://doi.org/https://doi.org/10.1016/j.rinp.2022.105654} {\bibfield
  {journal} {\bibinfo  {journal} {Results in Physics}\ }\textbf {\bibinfo
  {volume} {38}},\ \bibinfo {pages} {105654} (\bibinfo {year}
  {2022})}\BibitemShut {NoStop}%
\bibitem [{\citenamefont {Tomm}\ \emph {et~al.}(2021)\citenamefont {Tomm},
  \citenamefont {Javadi}, \citenamefont {Antoniadis}, \citenamefont {Najer},
  \citenamefont {L{\"o}bl}, \citenamefont {Korsch}, \citenamefont {Schott},
  \citenamefont {Valentin}, \citenamefont {Wieck}, \citenamefont {Ludwig} \emph
  {et~al.}}]{tomm2021bright}%
  \BibitemOpen
  \bibfield  {author} {\bibinfo {author} {\bibfnamefont {N.}~\bibnamefont
  {Tomm}}, \bibinfo {author} {\bibfnamefont {A.}~\bibnamefont {Javadi}},
  \bibinfo {author} {\bibfnamefont {N.~O.}\ \bibnamefont {Antoniadis}},
  \bibinfo {author} {\bibfnamefont {D.}~\bibnamefont {Najer}}, \bibinfo
  {author} {\bibfnamefont {M.~C.}\ \bibnamefont {L{\"o}bl}}, \bibinfo {author}
  {\bibfnamefont {A.~R.}\ \bibnamefont {Korsch}}, \bibinfo {author}
  {\bibfnamefont {R.}~\bibnamefont {Schott}}, \bibinfo {author} {\bibfnamefont
  {S.~R.}\ \bibnamefont {Valentin}}, \bibinfo {author} {\bibfnamefont {A.~D.}\
  \bibnamefont {Wieck}}, \bibinfo {author} {\bibfnamefont {A.}~\bibnamefont
  {Ludwig}}, \emph {et~al.},\ }\bibfield  {title} {\bibinfo {title} {A bright
  and fast source of coherent single photons},\ }\href
  {https://doi.org/10.1038/s41565-020-00831-x} {\bibfield  {journal} {\bibinfo
  {journal} {Nature Nanotechnology}\ }\textbf {\bibinfo {volume} {16}},\
  \bibinfo {pages} {399} (\bibinfo {year} {2021})}\BibitemShut {NoStop}%
\bibitem [{\citenamefont {{O'Faolain}}\ \emph {et~al.}(2010)\citenamefont
  {{O'Faolain}}, \citenamefont {Beggs}, \citenamefont {White}, \citenamefont
  {Kampfrath}, \citenamefont {Kuipers},\ and\ \citenamefont
  {Krauss}}]{o2010compact}%
  \BibitemOpen
  \bibfield  {author} {\bibinfo {author} {\bibfnamefont {L.}~\bibnamefont
  {{O'Faolain}}}, \bibinfo {author} {\bibfnamefont {D.~M.}\ \bibnamefont
  {Beggs}}, \bibinfo {author} {\bibfnamefont {T.~P.}\ \bibnamefont {White}},
  \bibinfo {author} {\bibfnamefont {T.}~\bibnamefont {Kampfrath}}, \bibinfo
  {author} {\bibfnamefont {K.}~\bibnamefont {Kuipers}},\ and\ \bibinfo {author}
  {\bibfnamefont {T.~F.}\ \bibnamefont {Krauss}},\ }\bibfield  {title}
  {\bibinfo {title} {{C}ompact optical switches and modulators based on
  dispersion engineered photonic crystals},\ }\href
  {https://doi.org/10.1109/JPHOT.2010.2047918} {\bibfield  {journal} {\bibinfo
  {journal} {IEEE Photonics Journal}\ }\textbf {\bibinfo {volume} {2}},\
  \bibinfo {pages} {404} (\bibinfo {year} {2010})}\BibitemShut {NoStop}%
\bibitem [{\citenamefont {Sanders}\ and\ \citenamefont
  {Milburn}(1992{\natexlab{a}})}]{sanders1992quantumPRA}%
  \BibitemOpen
  \bibfield  {author} {\bibinfo {author} {\bibfnamefont {B.~C.}\ \bibnamefont
  {Sanders}}\ and\ \bibinfo {author} {\bibfnamefont {G.~J.}\ \bibnamefont
  {Milburn}},\ }\bibfield  {title} {\bibinfo {title} {{Q}uantum limits to
  all-optical phase shifts in a {K}err nonlinear medium},\ }\href
  {https://doi.org/10.1103/PhysRevA.45.1919} {\bibfield  {journal} {\bibinfo
  {journal} {Physical Review A}\ }\textbf {\bibinfo {volume} {45}},\ \bibinfo
  {pages} {1919} (\bibinfo {year} {1992}{\natexlab{a}})}\BibitemShut {NoStop}%
\bibitem [{\citenamefont {Sanders}\ and\ \citenamefont
  {Milburn}(1992{\natexlab{b}})}]{sanders1992quantum}%
  \BibitemOpen
  \bibfield  {author} {\bibinfo {author} {\bibfnamefont {B.~C.}\ \bibnamefont
  {Sanders}}\ and\ \bibinfo {author} {\bibfnamefont {G.~J.}\ \bibnamefont
  {Milburn}},\ }\bibfield  {title} {\bibinfo {title} {{Q}uantum limits to
  all-optical switching in the nonlinear {M}ach--{Z}ehnder interferometer},\
  }\href {https://doi.org/10.1364/JOSAB.9.000915} {\bibfield  {journal}
  {\bibinfo  {journal} {Journal of the Optical Society of America B}\ }\textbf
  {\bibinfo {volume} {9}},\ \bibinfo {pages} {915} (\bibinfo {year}
  {1992}{\natexlab{b}})}\BibitemShut {NoStop}%
\bibitem [{\citenamefont {Bravyi}\ \emph {et~al.}(2011)\citenamefont {Bravyi},
  \citenamefont {DiVincenzo},\ and\ \citenamefont
  {Loss}}]{bravyi2011schrieffer-wolff}%
  \BibitemOpen
  \bibfield  {author} {\bibinfo {author} {\bibfnamefont {S.}~\bibnamefont
  {Bravyi}}, \bibinfo {author} {\bibfnamefont {D.~P.}\ \bibnamefont
  {DiVincenzo}},\ and\ \bibinfo {author} {\bibfnamefont {D.}~\bibnamefont
  {Loss}},\ }\bibfield  {title} {\bibinfo {title} {{S}chrieffer–{W}olff
  transformation for quantum many-body systems},\ }\href
  {https://doi.org/https://doi.org/10.1016/j.aop.2011.06.004} {\bibfield
  {journal} {\bibinfo  {journal} {Annals of Physics}\ }\textbf {\bibinfo
  {volume} {326}},\ \bibinfo {pages} {2793} (\bibinfo {year}
  {2011})}\BibitemShut {NoStop}%
\bibitem [{\citenamefont {Cohen-Tannoudji}\ \emph {et~al.}(1998)\citenamefont
  {Cohen-Tannoudji}, \citenamefont {Dupont-Roc},\ and\ \citenamefont
  {Grynberg}}]{cohen1998atom}%
  \BibitemOpen
  \bibfield  {author} {\bibinfo {author} {\bibfnamefont {C.}~\bibnamefont
  {Cohen-Tannoudji}}, \bibinfo {author} {\bibfnamefont {J.}~\bibnamefont
  {Dupont-Roc}},\ and\ \bibinfo {author} {\bibfnamefont {G.}~\bibnamefont
  {Grynberg}},\ }\href@noop {} {\emph {\bibinfo {title} {{A}tom-photon
  interactions: basic processes and applications}}}\ (\bibinfo  {publisher}
  {John Wiley \& Sons},\ \bibinfo {year} {1998})\BibitemShut {NoStop}%
\bibitem [{\citenamefont {Pyykk\"onen}\ \emph {et~al.}(2021)\citenamefont
  {Pyykk\"onen}, \citenamefont {Peotta}, \citenamefont {Fabritius},
  \citenamefont {Mohan}, \citenamefont {Esslinger},\ and\ \citenamefont
  {T\"orm\"a}}]{pyykkonen2021}%
  \BibitemOpen
  \bibfield  {author} {\bibinfo {author} {\bibfnamefont {V.~A.~J.}\
  \bibnamefont {Pyykk\"onen}}, \bibinfo {author} {\bibfnamefont
  {S.}~\bibnamefont {Peotta}}, \bibinfo {author} {\bibfnamefont
  {P.}~\bibnamefont {Fabritius}}, \bibinfo {author} {\bibfnamefont
  {J.}~\bibnamefont {Mohan}}, \bibinfo {author} {\bibfnamefont
  {T.}~\bibnamefont {Esslinger}},\ and\ \bibinfo {author} {\bibfnamefont
  {P.}~\bibnamefont {T\"orm\"a}},\ }\bibfield  {title} {\bibinfo {title}
  {{F}lat-band transport and {J}osephson effect through a finite-size sawtooth
  lattice},\ }\href {https://doi.org/10.1103/PhysRevB.103.144519} {\bibfield
  {journal} {\bibinfo  {journal} {Physical Review B}\ }\textbf {\bibinfo
  {volume} {103}},\ \bibinfo {pages} {144519} (\bibinfo {year}
  {2021})}\BibitemShut {NoStop}%
\end{thebibliography}
\end{document}